\documentclass[3p,12pt]{elsarticle}
\usepackage{lineno}
\modulolinenumbers[5]
\usepackage{graphicx}%,subfloat}
\usepackage{amssymb}
\usepackage{bm,amsmath}
\usepackage[caption=false]{subfig}
\usepackage{hyperref}
\usepackage{caption}
\usepackage{subeqnarray}
\usepackage{multirow}
\usepackage{lscape}
\usepackage{rotating}
\usepackage{graphics}
\usepackage{lineno}

\usepackage{color,ulem}

\usepackage{amsthm}

\journal{Journal of Computational Physics}

\begin{document}

%\title{ An ultra-fast scheme to solve the Boltzmann equation for  steady-state solutions of general rarefied gas flows}
\title{ Can we find steady-state solutions to multiscale rarefied gas flows within dozens of iterations?  } 
	\author{Wei Su\fnref{label1}}%
	\author{Lianhua Zhu\fnref{label1}}%
		\author{Peng Wang\fnref{label1}}%
	\author{Yonghao Zhang\fnref{label1}}%
	\author{Lei Wu\corref{cor1}\fnref{label1}}%
	\ead{lei.wu.100@strath.ac.uk}
	\address[label1]{James Weir Fluids Laboratory, Department of Mechanical and Aerospace Engineering, University of Strathclyde, Glasgow G1 1XJ, UK}
	\cortext[cor1]{Wei Su and Lianhua Zhu contribute equally.\\Corresponding author:}
	\date{\today}

	%(especially the singular behavior of the velocity gradient near the wall)

\begin{abstract}
	One of the central problems in the study of rarefied gas dynamics is to find the steady-state solution of the Boltzmann equation quickly. When the Knudsen number is large, i.e. the system is highly rarefied, the conventional iteration scheme can lead to convergence within a few iterations. However, when the Knudsen number is small, i.e. the flow falls in the near-continuum regime, hundreds of thousands iterations are needed, and yet the ``converged'' solutions are prone to be contaminated by accumulated error and large numerical dissipation. Recently, based on the gas kinetic models, the implicit unified gas kinetic scheme (UGKS) and its variants have significantly reduced the iterations in the near-continuum flow regime, but still much higher than that of the highly rarefied gas flows. In this paper, we put forward a general synthetic iteration scheme (GSIS) to find the steady-state solutions of general rarefied gas flows within dozens of iterations at any Knudsen number. The key ingredient of our scheme is that the macroscopic equations, which are solved together with the Boltzmann equation and help to adjust the velocity distribution function, not only asymptotically preserves the Navier-Stokes limit in the framework of Chapman-Enskog expansion, but also contain Newton's law for stress and Fourier's law for heat conduction explicitly. For this reason, like implicit UGKS, the constraint that the numerical cell size should be smaller than the mean free path of gas molecules is removed, but we do not need the complex evaluation of numerical flux at the cell interface. What's more, as the GSIS does not rely on the specific kinetic model/collision operator, it can be naturally extended to quickly find converged solutions for mixture flows and even flows involving chemical reactions. These two superior advantages are also expected to accelerate the slow convergence in simulation of near-continuum flows via the direct simulation Monte Carlo method and its low-variance version. 

\end{abstract}

	\maketitle

\section{Introduction}

Multiscale rarefied gas flows involving a wide range of Knudsen number have been encountered in massive engineering problems, e.g.  high-altitude aerothermodynamics of space vehicles, microelectromechanical systems, and gas transportation in ultra-tight shale strata. A gas flow can be described by either the macroscopic or the microscopic model. At the macroscopic level, the gas is assumed as a continuous medium and the evolution of gas system is described in terms of the spatial and temporal variations of the familiar flow properties such as density, velocity, pressure and temperature. The mathematical description of any macroscopic model is grounded in two primary aspects: 1) the conservation laws that describe how the mass, momentum and energy must be conserved during transport processes; 2) the constitutive equations that describe how the fluxes of mass dissipation, momentum diffusion and heat conduction response to various stimuli such as pressure difference, gradients of temperature and velocity, and external force. The Navier-Stokes equations provide the conventional mathematical model for a gas as a continuum, in which the conservation laws are closed by the famous constitutive equations of the Newton's law of viscosity and Fourier's law of heat conduction. Since the transport terms are expressed in terms of the first-order macroscopic quantities, the Navier-Stokes equations are only valid when the length scale of the gradients of the macroscopic variables is much larger than the mean free path of gas molecules, i.e. the Knudsen number is far smaller than one~\cite{CE}.

The microscopic model postulates that the gas is not continuous but is composed of a finite number of molecules. The molecules rush hither and thither at large speed, and strike with boundary and collide with each other. Actually, the macroscopic transport phenomena stem no other than the random motions of the gas molecules. The mathematical model at the microscopic level is the Boltzmann equation, which governs the evolution of the one-particle velocity distribution function providing information on the state of every molecule at all times~\cite{Bird1970}. Then, the macroscopic flow properties are identified with average values of the molecular quantities. Note that the Boltzmann equation is applicable for the entire range of Knudsen number. 

The Boltzmann equation can be numerically solved either in discretized molecular velocity space via the discrete velocity method (DVM)\cite{Aristov2001}, or by applying the direct simulation Monte Carlo (DSMC) method that uses a collection of particles to represent random points in the molecular velocity space~\cite{Bird1970}. Compared to the traditional computational fluid dynamic (CFD) techniques for solving macroscopic equations, the Boltzmann equation (or simplified kinetic model equations) is much more expensive to solve in terms of computation time and memory. This is mainly due to the following facts. First of all, additional dimensions of the molecular velocity space are required to be discretized in DVM and particles are required to generate in DSMC. Second, since the random behaviors of gas molecules are modeled on length and time scales comparable to the cell size and simulation time interval, respectively, in order to suppress numerical diffusion errors it is suggested that the size of grid cell and the time interval should be smaller than the molecule's mean free path and the mean collision time, respectively~\cite{BURT20084653}. As a consequence, the computational cost dramatically increases as the gas flow approaches the near-continuum flow regime. Finally, in DVM, the conventional iteration scheme (CIS) to find steady-state solution converges extremely slowly for flows at low Knudsen numbers, since the exchange of information (e.g. perturbation in the flow field) through molecular steaming becomes very inefficient when binary collisions dominate~\cite{LeiJCP2017}. Worse still, the ``converged'' solutions are prone to be contaminated by numerical errors, e.g. the accumulated error from finite discrete molecular velocities~\cite{Luc2000JCP} and error stemming from the evaluation of molecular collisions, say, by the  projection method~\cite{Tcheremissine2005} and the fast spectral method~\cite{Mouhot2006Math}. In DSMC, the simulation time also increases significantly due to this inefficient information exchange process in the near-continuum flow. Note that unified gas-kinetic scheme (UGKS)~\cite{Kun_Huang2011,Xu2012,guo2013discrete,guo2015discrete,zhuyajun2016} can remove the restrictions on cell size and time step by simultaneously handling free streaming and collision of gas molecules during transport processes. However, as information exchanging relays on the evolution of velocity distribution function, UGKS still needs a large number of iterations to obtained steady-state solutions in near-continuum flows~\cite{CHEN201552,WANG201833}.

There has been a tremendous growth of researches on multiscale hybrid numerical methods that combine multiple models defined at fundamentally different length and time scales within the same overall spatial and temporal domain. Specifically for the flow of interest, the continuum CFD methods are used in regions where the Navier-Stokes equations are valid, while methods based on gas kinetic theory are applied in regions where the continuum equations fail~\cite{Roveda1998,Wang2003,boyd_1,Wijesinghe2004,SCHWARTZENTRUBER20071159,KOLOBOV2007589,Tang2014}. However, intrinsic difficulties arises when coupling the two different models. First, the mechanism for continuum breakdown is unclear and the criterion to determine where the continuum model is valid relies on empirical parameters that varies for different flow conditions~\cite{SCHWARTZENTRUBER20071159}. Second, the continuum-kinetic coupling is strictly required to lie in the region that can be accurately modeled by Navier-Stokes equations, so that the Boltzmann equation is still employed in low-Knudsen-number regions. Therefore, the CIS still needs lots of iterations to achieve convergence; also, DSMC still needs small cell size and time step and hence large evolution steps to find the converged solutions.

In recent years, the synthetic iterative scheme (SIS), which is initially developed for the radiation transport processes~\cite{DSA2002}, has been extended to achieve high efficiency and accuracy in DVM, in particular with fast convergence property across the whole gas flow regimes~\cite{Valougeorgis2003,Lihnaropoulos2007}. In this scheme, the gas kinetic equations and macroscopic equations are solved simultaneously on the same grids in the entire domain. Since the velocity distribution function is guided by the macroscopic flow quantities solved from diffusion-type equations at each iterative step, information propagates accurately and fast even when Knudsen number is small. When the Knudsen number is small, the synthetic macroscopic equations reduce to the Navier-Stokes equations. However, the macroscopic equation contains high-order terms to take into account rarefaction effects, thus the SIS also preserves accuracy in high Knudsen number regimes. The SIS has been successfully applied to Poiseuille flow in channels of arbitrary shapes using the Bhatnagar-Gross-Krook kinetic model for single-species gases~\cite{SZALMAS20104315}, and flows of binary and ternary gas mixtures driven by local pressure, temperature and concentration gradients using the McCormak model~\cite{NARIS2004629,NARIS2004294,Naris2005Pof,SZALMAS20132430,SZALMAS201691}. The SIS has also been extended to solve the linearized Boltzmann equation, where the role of realistic intermolecular potentials in Poiseuille, Couette and thermal transpiration flows has been analyzed~\cite{LeiJCP2017,SU2019573}.

It is interesting to note that the similar idea of SIS has also been used in DSMC, that is, in addition to traditional DSMC, macroscopic variables are solved and updated according to macroscopic rules/equations. For instances, in the information preservation (IP)-DSMC, the information velocity is introduced to compute macroscopic velocity and shear stress, with the aim of removing ``the statistical fluctuation source inherent in the DSMC method that results from the randomness of the thermal velocity''~\cite{Fan2001,ZhangJun2011JCP,Fei2013}, although the rule of updating the information velocity and/or other macroscopic variables is not exactly derived from the Boltzmann equation. On the other hand, the moment guided DSMC is also proposed to reduce the statistical error, where the density, velocity and temperature are updated by five exact macroscopic equations from the conservation law, but with the pressure tensor and heat flux calculated from the DSMC~\cite{Degond2011}.

In DVM, the SIS can not only asymptotically achieve the Navier-Stokes limit with fast convergence rate, but also preserve accuracy in high Knudsen number regimes. The critical point to develop this scheme is that the macroscopic equations must explicitly contain both the constitutive relations predicting the transport phenomena at the continuum level, and high-order terms taking into account rarefaction effects. To the author's awareness, the SIS is still limited to simple rarefied gas flows such as the Poiseuille, Couette and thermal transpiration flows, where the flow velocity is perpendicular to the computational domain, we refer to~\cite{Valougeorgis2003} for example. In this paper, we intend to put forward the general SIS (GSIS) with the aim to find the steady-state solutions of general rarefied gas flows within dozens of iterations at any Knudsen number. For simplicity we considered linearized flows but the methodology can be extended to nonlinear flows easily.

The remainder of the paper is organized as follows. In Section~\ref{sectionII}, the linearized Boltzmann equation (LBE) is introduced. In Section~\ref{secIII}, the GSIS for general rarefied gas flow is proposed. Numerical tests to assess the efficiency and accuracy of the proposed scheme are presented for stationary problems in Section~\ref{sec:results1} and for periodic oscillatory problems in Section~\ref{sec:results2}.  The paper closes with some finial comments in Section~\ref{sec:summary}.

\section{The linearized Boltzmann equation }\label{sectionII}

In kinetic theory, the state of a gas system is described by the one-particle velocity distribution $f\left(t,\bm{x},\bm{v}\right)$. Evolution of the velocity distribution function to the independent variables, i.e. time $t$, spatial position $\bm{x}=\left(x_1,x_2,x_3\right)$ and molecular velocity $\bm{v}=\left(v_1,v_2,v_3\right)$, is governed by the Boltzmann equation~\cite{CE}. When the system deviates slightly from the global equilibrium described by
\begin{equation}
f_{eq}(\bm{v})={\pi^{-3/2}}{\exp(-|\bm{v}|^2)},
\end{equation}
the velocity distribution function of gas molecules can be linearized around $f_{eq}$ as:
\begin{equation}\label{eq:1}
f(t,\bm{x},\bm{v})=f_{eq}(\bm{v})+\alpha h(t,\bm{x},\bm{v}), 
\end{equation}
where $\alpha h(t,\bm{x},\bm{v})$ is the small perturbance satisfying $|\alpha h/f_{eq}|\ll1$ with $\alpha$ being a small constant related to the amplitude of perturbation. The velocity distribution function $h(t,\bm{x},\bm{v})$, however, is not necessary smaller than the equilibrium distribution function $f_{eq}$. The LBE for $h(t,\bm{x},\bm{v})$ is:
\begin{equation}\label{LBE}
\frac{\partial {h}}{\partial t}+\bm{v}\cdot\frac{\partial
	{h}}{\partial{\bm{x}}}=L(h,f_{eq}),
\end{equation}
where the linearized Boltzmann collision operator is~\cite{lei_Jfm}:
\begin{equation}\label{LBE_collision}
L=\underbrace{\iint B(\theta,|\bm{u}|)  [f_{eq}(\bm{v}')h({\bm{v}}'_{\ast})+f_{eq}(\bm{v}'_\ast)h({\bm{v}}')-f_{eq}(\bm{v})h({\bm{v}}_\ast)]\mathrm{d}\Omega\mathrm{d}{\bm{v}}_\ast}_{L^+}-\nu_{eq}(\bm{v})h(\bm{v}),
\end{equation}
and the equilibrium collision frequency is
\begin{equation}
\nu_{eq}(\bm{v})=\iint B(|\bm{u}|,\theta) f_{eq}(\bm{v}_\ast)\mathrm{d}\Omega{}\mathrm{d}{\bm{v}}_\ast.
\end{equation}
Note that the relative velocity of the two molecules before binary collision is $\bm{u}=\bm{v}-\bm{v}_\ast$, and $\Omega$ is a unit vector along the relative post-collision velocity $\bm{v}'-\bm{v}'_\ast$. The deflection angle $\theta$ between the pre- and post-collision relative velocities satisfies $\cos\theta=\Omega\cdot{\bm{u}}/|\bm{u}|$, $0\le\theta\le\pi$. Finally, $B(\theta,|\bm{u}|)=|\bm{u}|\sigma$ is the collision kernel, with $\sigma$ being the differential cross-section that is determined by the intermolecular potential. In the present paper, we consider the inverse power-law potentials, where the collision kernels are modeled as~\cite{Lei2013,lei_Jfm}
\begin{equation}\label{collision_kernel}
B(|{\bm{u}}|,\theta)=\frac{|{\bm{u}}|^{2(1-\omega)}}{K}	{\sin^{\frac{1}{2}-\omega}\left(\frac{\theta}{2}\right)\cos^{\frac{1}{2}-\omega}\left(\frac{\theta}{2}\right)},
\end{equation}
with $\omega$ being the viscosity index (i.e. the shear viscosity $\mu$ of the gas is proportional to $T^\omega$) and $K$ some normalization constants~\cite{lei_Jfm}. HS and Maxwell molecules have $\omega=0.5$ and 1, respectively. The details of implementation of the Lennard-Jones potentials in fast spectral method can be found in Ref.~\cite{Wu:2015yu}, but here only the collision kernel~\eqref{collision_kernel} will be used to demonstrated that the GSIS works for the LBE.

Note that we present the governing system in terms of dimensionless variables. The coordinate $\bm{x}$ is normalized by the characteristic flow length $H$, the molecular velocity $\mathbf{v}$ is normalized by the most probable speed ${v_m}=\sqrt{2k_BT_0/m}$, the time $t$ is normalized by $H/v_m$, and velocity distribution functions $f_{eq}$ and $h$ are normalized by $n_0/v_m^3$, where $n_0$ is the average number density of the gas molecules, $T_0$ is the reference temperature, $k_B$ is the Boltzmann constant, and $m$ is the mass of gas molecules.
 % $\epsilon$ is a potential depth, and

To fully determine the gas dynamics in spatially-inhomogeneous problems, the gas-surface boundary condition should be specified.  In this paper, the Maxwell diffuse boundary condition will be used: the velocity distribution function $f(t,\bm{x},\bm{v})$ of the reflected gas molecules at the solid surface satisfies the following equation:
\begin{equation}\label{label_general}
f(t,\bm{x},\bm{v})=\frac{2\int_{v_n'<0} |v_n'| f(t,\bm{x},\bm{v}')\mathrm{d}\bm{v}'}{\pi{}T_w^2} \exp\left(-\frac{  |\bm{v}-\bm{U}_w|^2 }{T_w}\right), 
\end{equation}
where  $T_w$ is the wall temperature normalized by the reference temperature $T_0$, $\bm{U_w}$ is the wall velocity normalized by the most probable speed $v_m$, and $v_n$ is the normal component of the peculiar velocity $\bm{v}-\bm{U}_w$ redirected into the gas.

The macroscopic quantities of interest including the number density $\rho$, bulk velocity $\bm U$, temperature $T$, pressure $p$, stress tensor $\sigma_{ij}$ and heat flux $\bm{q}$, which are further normalized by the dimensionless constant $\alpha$, can be calculated as
\begin{eqnarray}\label{MP}
\rho=\int{h}\mathrm{d}\bm{v}, \quad \bm{U}=\int{\bm{v}h}\mathrm{d}\bm{v}, \quad 
T=\frac{2}{3}\int{|\bm{v}|^2}h\mathrm{d}\bm{v}-\rho, \quad p=\rho+T\label{nuT} \\
\sigma_{ij}=2\int{\left(v_iv_j-\frac{|\bm{v}|^2}{3}\delta_{ij}\right)h}\mathrm{d}\bm{v}, \quad
\bm{q}=\int{\bm{v}|\bm{v}|^2h}\mathrm{d}\bm{v}-\frac{5}{2}\bm{U}, \label{sigmaQ}
\end{eqnarray}
where $\delta$ is the Kronecker delta function, and $i,j=1,2,3$ represent the three orthogonal spatial directions in the Cartesian coordinates.

\section{The general synthetic iteration scheme}\label{secIII}

The steady state solution of the integro-differential system~\eqref{LBE} is usually solved by the CIS. Given the value of $h^{(k)}(\bm{x},\bm{v})$ at the $k$-th iteration step, the velocity distribution function at the next iteration step is calculated by solving the following equation~\cite{ohwada1989numerical,Lei2013,Wu:2015yu}:
\begin{equation}\label{LBE_iteration}
\nu_{eq}^{(k)}h^{(k+1)}+
\bm{v}\cdot\frac{\partial
	{h}^{(k+1)}}{\partial{\bm{x}}}=L^+(h^{(k)},f_{eq}),
\end{equation}
where the derivative with respect to $\textbf{x}$ can be approximated by any conventional CFD schemes such as the finite difference, finite volume, or Discontinuous Galerkin (DG) methods~\cite{WeiSuJCP1,Su2019IDG}, and the collision operator in Eq.~\eqref{LBE_collision} can be calculated by the fast spectral method~\cite{lei_Jfm,Wu:2015yu} based on the velocity distribution function at the $k$-th iteration step. The process is repeated until relative differences between successive estimates of macroscopic quantities are less than a convergence criterion $\epsilon$.

A key parameter in the rarefied gas flow is the rarefaction parameter, which is defined as
\begin{equation}\label{delta}
\delta_{rp}=\frac{H}{\lambda}, ~\lambda=\frac{\mu{(T_0)}v_m}{n_0k_BT_0},
\end{equation}
where $\mu(T_0)$ is the shear viscosity of the gas at the reference temperature, and $\lambda$ is the mean free path of the gas molecules. Alternatively, the Knudsen number is defined as
\begin{equation}
\mathrm{Kn}=\frac{\sqrt{\pi}}{2\delta_{rp}}.
\end{equation}

The CIS is efficient for highly rarefied gas flows when $\delta_{rp}$ is very small, where converged solutions can be quickly found after several iterations. However, the number of iteration increases significantly with the rarefaction parameter~\cite{Valougeorgis:2003zr,LeiJCP2017}. This is due to the frequent collisions of gas molecules, which quickly smear the perturbance and hinder the fluid information exchange. 
In order to enhance the information exchange across the whole computational domain, synthetic equations for the evolution of macroscopic flow variables that are asymptotic preserving the Navier-Stokes limit should be developed~\cite{LeiJCP2017}.

To this end, we first multiply Eq.~\eqref{LBE} by 1, 2$\bm{v}$, and $|\bm{v}|^2-\frac{3}{2}$, respectively, and integrate the resultant equations with respect to $\bm{v}$; we obtain the following equations for the evolution of the density, velocity, and temperature: 
\begin{equation}\label{eq123}
\begin{aligned}
\frac{\partial {\rho}}{\partial{t}}+\frac{\partial {U_i}}{\partial{x_i}}=0, \\
2\frac{\partial {U_i}}{\partial{t}}+\frac{\partial {\rho}}{\partial{x_i}}+\frac{\partial {T}}{\partial{x_i}}+\frac{\partial {\sigma_{ij}}}{\partial{x_j}}=0, \\
\frac{3}{2}\frac{\partial {T}}{\partial{t}}+\frac{\partial {q_j}}{\partial{x_j}}+\frac{\partial {U_j}}{\partial{x_j}}=0,
\end{aligned}
\end{equation}
which are not closed, since expressions for the shear stress $\sigma_{ij}$ and heat flux $\bm{q}$ are not known. One way to close Eq.~\eqref{eq123} is to use the Chapman-Enskog expansion, where the distribution function is expressed in the power series of $\mathrm{Kn}$~\citep{CE}: $h=\mathrm{Kn} h^{(1)}+\mathrm{Kn}^2 h^{(2)}+\cdots$. When $f=f^{(0)}$, we have $\sigma_{ij}= q_i=0$, and Euler equations are recovered. When the distribution function is truncated at the first-order of $\mathrm{Kn}$, that is, $
h=\mathrm{Kn} h^{(1)}$,
we have 
\begin{equation}\label{GTMNSF}
\sigma_{ij} =-\delta_{rp}^{-1}\left(\frac{\partial U_{i}}{\partial x_{j}}+\frac{\partial U_{j}}{\partial x_{i}}-\frac{2}{3}\frac{\partial U_{k}}{\partial x_{k}}\delta_{ij}\right)\equiv-2\delta_{rp}^{-1}\frac{\partial U_{<i}}{\partial {x_{j>}}}, \quad
q_i = -\frac{5}{4\mathrm{Pr}}\delta_{rp}^{-1} \frac{\partial T}{\partial x_i},
\end{equation}
and Eq.~\eqref{eq123} reduces to Navier-Stokes equations with $\mathrm{Pr}$ being the Prandtl number. Higher-order macroscopic equations can be obtained successively but they are not stable. On the other hand, even the obtained high-order macroscopic equations are stable, they are only the approximate solutions of the Boltzmann equation, rather than the exact solutions.

It should be noted that in implicit UGKS~\cite{Zhu2019JCP} and other variants~\cite{yang2018PoF,yang2018PRE}, both the gas kinetic equation and  macroscopic equations~\eqref{eq123} are solved, where $\sigma_{ij}$ and $\bm{q}$ are obtained according to Eq.~\eqref{sigmaQ}. These methods are efficient when the Knudsen number is large, like the CIS. However, in the near-continuum flow regime, the number of iterations are still large, at the order of thousands iterations. The reason for the relative slow convergence is that, if the iteration starts from the global equilibrium state where $\sigma_{ij}$ and $\bm{q}$ are zero, in most of the time the Euler equations, rather than the Navier-Stokes equations that dominates the steady-state flow dynamics, are solved, due to the fact that perturbance from the wall boundary takes a long time to reach the bulk region for near-continuum flows.	Even when the shear stress and heat flux are non-zero, solutions of Eq.~\eqref{eq123} deviate from that of the Navier-Stokes equations in the near-continuum flow regime unless they nearly converge to the steady-state solutions. As a matter of fact, the authors have checked, in the linearized Poiseuille flow, that Eq.~\eqref{eq123} cannot boost  convergence~\cite{LeiJCP2017};

Bearing this in mind, to develop an ultra-fast convergence scheme, the macroscopic equations must reduce to the Navier-Stokes equation in the near-continuum flow regime, and must contain the Newton's law for stress and Fourier's law for heat conduction explicitly to recover the macroscopic transport mechanism; that is, the shear stress and heat flux should be expressed as follows:
\begin{eqnarray}
\sigma_{ij} =-2\delta_{rp}^{-1}\frac{\partial U_{<i}}{\partial {x_{j>}}}+\text{HoT}_{\sigma_{ij}}, \label{sigma_HoT}\\
q_i =-\frac{5}{4\mathrm{Pr}}\delta_{rp}^{-1} \frac{\partial T}{\partial x_i}+\text{HoT}_{q_i}, \label{q_HoT}
\end{eqnarray}
where $\text{HoT}_{\sigma_{ij}}$ and $\text{HoT}_{q_i}$ are the high-order terms containing contributions of all the orders $O(Kn^\alpha)$ with $\alpha=2,3,\cdots,\infty$.

To obtain~\eqref{sigma_HoT}, we multiply Eq.~\eqref{LBE} by $2(v_iv_j-\delta_{ij}|\bm{v}|^2/3)$ and integrate the resultant equation with respect to $\bm{v}$, and obtain 
\begin{equation}\label{HoT_sigma}
\frac{\partial \sigma_{ij}}{\partial {t}}+\text{HoT}_{\sigma_{ij}}
+\underline{2\frac{\partial{U_{<i}}}{\partial {x_{j>}}}=-\delta_{rp}\sigma_{ij}}+2\int{(L-L_s)v_iv_j}\mathrm{d}\bm{v},
\end{equation}
where
\begin{equation}\label{LBE_shakhov}
 L_{s}=\delta_{rp}\left\{\left[\rho+2\bm{U}\cdot\bm{v}+T\left(|\bm{v}|^2-\frac{3}{2}\right)+\frac{4\left(1-\mathrm{Pr}\right)}{5}\bm{q}\cdot{\bm{v}}\left(|\bm{v}|^2-\frac{5}{2}\right)\right]f_{eq}-h\right\}
 \end{equation} is the linearized collision operator of the Shakhov kinetic model equation~\cite{Shakhov1968}, and
\begin{equation}\label{HoT_sigma2}
\text{HoT}_{\sigma_{ij}}=\left\{
\begin{array}{lr}
& \frac{\partial}{\partial x_i}\int{}(2v_i^2-1)v_jh\mathrm{d}\bm{v}
+\frac{\partial}{\partial x_j}\int{}(2v_j^2-1)v_ih\mathrm{d}\bm{v}
+\frac{\partial}{\partial x_k}\int{}2v_1v_2v_3h\mathrm{d}\bm{v}, \\
& \text{for~} i\neq{j}, k\neq{i}, k\neq{j}, \\
& \frac{\partial}{\partial x_i}\int{}2(v_i^2-\frac{|\bm{v}|^2}{3}-\frac{2}{3})v_ih\mathrm{d}\bm{v}
+\sum_{k}\frac{\partial}{\partial x_k}\int{}2(v_i^2-\frac{|\bm{v}|^2}{3}+\frac{1}{3})v_kh\mathrm{d}\bm{v}, \\
& \text{for~} i=j, k\neq{i}. 
\end{array}
\right.
\end{equation}
Note that this derivation is rather simple as we just separate the underlined term in Eq.~\eqref{HoT_sigma} from high-order moments $\int{}2(v_iv_j-\delta_{ij}|\bm{v}|^2/3)v_khd\mathbf{v}$, and the purpose of introducing $L_s$ is only to recover the term
$\delta_{rp}\sigma_{ij}$, so that the Newton's law of stress is recovered explicitly. It should also be noted that, for the linearized Boltzmann collision operator, the term $2\int{(L-L_s)v_iv_j} d\mathbf{v}$ is negligible small when compared to $\delta_{rp}\sigma_{ij}$. For instances, for the Maxwell model, this term is zero, while for the HS molecular model, this term is less than 2\% of $\delta_{rp}\sigma_{ij}$, see page no.~169 in the third edition of the book~\cite{CE}.

Similarly, to obtain Eq.~\eqref{q_HoT}, we multiply Eq.~\eqref{LBE} by $v_i(|\bm{v}|^2-5/2)$ and integrate the resultant equation with respect to $\bm{v}$; we obtain 
\begin{equation}\label{HoT_q}
\frac{\partial q_{i}}{\partial {t}}+\text{HoT}_{q_i}
+\underline{\frac{3C_q}{2}\frac{\partial{T}}{\partial {x_{i}}}=-\frac{2}{3}\delta_{rp}q_{i}}+\int{(L-L_s)v_i|\bm{v}|^2} \mathrm{d}\bm{v},
\end{equation}
where 
\begin{equation}\label{HoT_q2}
\text{HoT}_{q_i}=\frac{\partial}{\partial{x_i}}\int\left[(v_i^2-C_q)\left(|\bm{v}|^2-\frac{3}{2}\right)-v_i^2\right]h\mathrm{d}\bm{v}+\sum_{j\neq{i}}\frac{\partial}{\partial{x_j}}\int{}v_iv_j\left(|\bm{v}|^2-\frac{5}{2}\right)h\mathrm{d}\bm{v},
\end{equation}
and for the linearized Boltzmann collision operator, the term $\int{(L-L_s)v_i|\bm{v}|^2} d\bm{v}$ is negligible small when compared to $\delta_{rp}q_{i}$, i.e. within 3\% of $\delta_{rp}q_{i}$~\cite{CE}. If we choose $C_q=5/9\mathrm{Pr}$, then the under-braced term in Eq.~\eqref{HoT_q} recovers the Fourier's heat conduction law in Eq.~\eqref{GTMNSF}. Since for monatomic gas the Prandtl number is very close to $2/3$, in the following paper we choose $C_q=5/6$.

Note that the macroscopic equations~~\eqref{eq123}, \eqref{HoT_sigma} and~\eqref{HoT_q} resemble the Grad 13 moment equations~\cite{Grad1949,henning}. However, since the higher-order terms~\eqref{HoT_sigma2} and~\eqref{HoT_q2}  are computed directly from the velocity distribution function, no approximations are introduced here. If the velocity distribution function is approximated by the Gauss-Hermite polynomials to the third order, where the coefficients before those polynomials are determined by the first 13 moments of the velocity distribution function, then G13 moment equations will be recovered.  Since the first-order Chapman-Enskog expansion to G13 equations leads to Eqs.~\eqref{eq123} and~\eqref{GTMNSF}, that is, only the underlined terms in Eqs.~\eqref{HoT_sigma} and~\eqref{HoT_q} are retained, the derived synthetic equations~\eqref{eq123}, \eqref{HoT_sigma} and~\eqref{HoT_q} are asymptotic preserving the Navier-Stokes limit. Thus, they should be able to boost the convergence to the steady-state solution of the LBE significantly, as in the bulk region (a few mean free path of gas molecules away from  solid surfaces) we are effectively solving the Navier-Stokes equations.

With these macroscopic equations to update the macroscopic quantities and the velocity distribution function, we devise the following  iteration scheme to find the steady-state solution of the LBE~\eqref{LBE} efficiently:
\begin{itemize}

	\item Step 1. When the velocity distribution function $h^{(k)}$ and the corresponding macroscopic quantities in Eqs.~\eqref{nuT} and~\eqref{sigmaQ} are known at the $k$-th iteration, we calculate $2\int{(L-L_s)v_iv_j}\mathrm{d}\bm{v}$ in Eq.~\eqref{HoT_sigma} and $\int{(L-L_s)v_i|\bm{v}|^2}\mathrm{d}\bm{v}$ in Eq.~\eqref{HoT_q}. We also calculate the velocity distribution function $h^{(k+1/2)}$ according to the conventional iteration scheme~\eqref{LBE_iteration}, that is, we solve the following equation:
	\begin{equation}\label{syn_LBE0}
	{\nu_{eq}^{(k)}}h^{(k+1/2)}+\bm{v}\cdot\frac{\partial
		{h}^{(k+1/2)}}{\partial{\bm{x}}}=L^+(h^{(k)},f_{eq}),
	\end{equation}
	by a second-order upwind finite difference in the bulk and a first-order upwind scheme at the solid surface~\cite{ohwada1989numerical} or the DG method~\cite{WeiSuJCP1,Su2019IDG}.

	\item Step 2. From $h^{(k+1/2)}$, we calculate the density $\rho^{(k+1/2)}(\bm{x})$, flow velocity $\bm{U}^{(k+1/2)}(\bm{x})$, the temperature $T^{(k+1/2)}(\bm{x})$, the shear stress $\sigma_{ij}^{(k+1/2)}(\bm{x})$, the heat flux $\bm{q}^{(k+1/2)}(\bm{x})$, and the high-order terms $\text{HoT}_{\sigma_{ij}}$ and $\text{HoT}_{q_i}$ defined in Eqs.~\eqref{HoT_sigma2} and~\eqref{HoT_q2}, respectively.

	\item Step 3. We obtain the macroscopic quantities $\rho^{(k+1)}(\bm{x})$,  $\bm{U}^{(k+1)}(\bm{x})$, $T^{(k+1)}(\bm{x})$, $\sigma_{ij}^{(k+1)}(\bm{x})$, and $\bm{q}^{(k+1)}(\bm{x})$ by solving the synthetic equations~\eqref{eq123}, \eqref{HoT_sigma} and~\eqref{HoT_q}, That is, for the steady-state problems the shear stress and heat flux can be solved from Eq.~\eqref{HoT_sigma} and~\eqref{HoT_q}, which will then be substituted to Eq.~\eqref{eq123} to form the Navier-Stokes equations with source terms related to the higher-order terms defined in Eqs.~\eqref{HoT_sigma2} and~\eqref{HoT_q2}. These equation can be solved by the SIMPLE algorithm and/or DG method easily in the bulk region, where the boundary values in the vicinity of the walls for the density, velocity, temperature are obtained from the step 2. The detailed DG algorithm to solve the synthetic equations can be found in the Appendix.

	\item  Step 4. The velocity distribution function $h$ is modified to incorporate the change of macroscopic quantities. That is,
	\begin{equation}\label{guided0}
	\begin{aligned}[b]
	h^{(k+1)}(\bm{x},\bm{v})=&h^{(k+1/2)}(\bm{x},\bm{v})
	+\left[2\lambda_{\bm{U}}(\bm{x})\cdot{\bm{v}}+\frac{4}{5}{\lambda_{\bm q}}(\bm{x})\cdot\bm{v}\left(|\bm{v}|^2-\frac{5}{2}\right)\right]f_{eq}\\
	&+\left[\lambda_{\rho}(\bm{x})+\lambda_T(\bm{x})\left(|\bm{v}|^2-\frac{3}{2}\right)
	+\lambda_{\sigma_{ij}}(\bm{x})\left(v_iv_j-\frac{|\bm{v}|^2}{3}\delta_{ij}\right)\right]f_{eq},
	\end{aligned}
	\end{equation}
	where $\lambda_{\bm{U}}(\bm{x})=\bm{U}^{(k+1)}(\bm{x})-\bm{U}^{(k+1/2)}(\bm{x})$, $\lambda_{\bm{q}}(\bm{x})=\bm{q}^{(k+1)}(\bm{x})-\bm{q}^{(k+1/2)}(\bm{x})$, $\lambda_{\rho}(\bm{x})=\rho^{(k+1)}(\bm{x})-\rho^{(k+1/2)}(\bm{x})$,  $\lambda_T(\bm{x})=T^{(k+1)}(\bm{x})-T^{(k+1/2)}(\bm{x})$, and $\lambda_{\sigma_{ij}}(\bm{x})=B\sigma_{ij}^{(k+1)}(\bm{x})-B\sigma_{ij}^{(k+1/2)}(\bm{x})$, with $B=3/2$ when $i=j$ and $B=2$ otherwise.

	\item Step 5. The above steps are repeated until convergence.\\
\end{itemize}

Since the gas kinetic equation is solved together with the macroscopic equations~\eqref{eq123}, \eqref{HoT_sigma} and~\eqref{HoT_q} for general rarefied gas flows, the above scheme is called the GSIS. Note that although the SIS has been widely applied to the radiation transport processes~\cite{DSA2002} and rarefied gas flows driven by local pressure, temperature, and concentration gradients~\cite{Valougeorgis:2003zr,Naris2005Pof,CircularSIS2013,szalmas2010,WeiSuJCP1} to overcome the slow convergence in the near-continuum flow regime, it is the first time that the GSIS is developed for general rarefied gas flows described by the LBE.  Also, it is with no doubt that such a methodology can be directly applied to construct the GSIS for the nonlinear Boltzmann equation.

\section{Numerical results for stationary problems} \label{sec:results1}

Numerical simulations are carried out to assess the efficiency and accuracy of the GSIS. To this end, we consider the one-dimensional heat transfer between two parallel plates, two-dimensional lid-driven cavity flow and shear-driven flow between two eccentric cylinders. The reason is that in previous cases the special SIS is only applicable for rarefied gas flows~\cite{Valougeorgis:2003zr,Naris2005Pof,CircularSIS2013,szalmas2010,WeiSuJCP1,Su2019JCP2}, where the flow velocity is perpendicular to the computational domain. Here we investigate the performance of the GSIS for typical general rarefied gas flows, where the flow velocity (or other macroscopic variables) varies within the computational domain.

\subsection{Heat transfer between two parallel plates}

Consider the steady Fourier flow of a gas between two infinite parallel plates with a distance $H$, located at $x_2=0$ and $x_2=1$. The two plates are stationary, the one at $x_2=0$ has a temperature $T_0-\Delta{T}/2$, while that at $x_2=H$ has a temperature $T_0+\Delta{T}/2$. We assume that the temperature difference $\Delta{T}$ is negligible compared to $T_0$, so that the problem is symmetrical around $x_2=1/2$. Therefore, in numerical simulations only the region $x_2\in[0,1/2]$ is considered. The Boltzmann equation is linearized by choosing $\alpha=\Delta{T}/{T_0}$ in Eq.~\eqref{eq:1}. The boundary condition at $x_2=0$, as according to Eqs.~\eqref{eq:1} and~\eqref{label_general}, is 
\begin{equation}
h(x_2=0,\bm{v})= \left[ 1-\frac{|\bm{v}|^2}{2}-2\sqrt{\pi}\int_{v_2<0} v_2h(x_2=0,\bm{v})\mathrm{d}v_2\right] f_{eq}, \ \  \text{when~}  {v_2>0},
\end{equation}
while that at $x_2=0.5$ is
\begin{equation}\label{Fourier_symm}
h(v_1,v_2,v_3)=-h(v_1,-v_2,v_3),
\end{equation}
due to the symmetry of this linearized problem.

From the synthetic equations~\eqref{eq123}, \eqref{HoT_sigma} and~\eqref{HoT_q}, as well as the symmetry condition~\eqref{Fourier_symm}, we know
\begin{equation}
\bm{U}=0, \quad \sigma_{ij}=0~ \text{when}~ i\neq{j}, \quad q_1=q_3=0,
\end{equation}
the heat flux perpendicular to the two plates $q_2$ is a constant, and
the variation of the perturbed temperature satisfies
\begin{equation}\label{HoT_q_Fourier}
\frac{\partial T}{\partial x_2}=-\frac{4\delta_{rp}}{9C_q}q_{2}+\underbrace{\frac{2}{3C_q}\int{}v_2|\bm{v}|^2(L-L_{s})\mathrm{d}\bm{v}}_{H_1^{(k)}(x_2)}-\underbrace{\frac{2}{3C_q}\frac{\partial }{\partial x_2}\int{}(v_2^2-C_q)\left(|\bm{v}|^2-\frac{3}{2}\right)h\mathrm{d}\bm{v}}_{H_2^{(k+1/2)}(x_2)},
\end{equation}
whose solution at the $(k+1)$-th iteration step is given by 
\begin{equation}
T^{(k+1)}(x_2)=-\frac{4\delta_{rp}{q}_2}{9C_q}\left(x_2-\frac{1}{2}\right)+\int_{1/2}^{x_2}H_1^{(k)}(x_2)\mathrm{d}x_2-H_2^{(k+1/2)}(x_2),
\end{equation}
where the constant heat flux $q_2$ is
\begin{equation}
q_2=\frac{9C_q}{2\delta_{rp}}\left[
T^{(k+1/2)}(x_2=0)+H_2^{(k+1/2)}(x_2=0)-H_1^{(k)}(x_2=0) \right].
\end{equation}

When the temperature is known, the density variation can be easily obtained by solving the following equation
\begin{equation}\label{synthetic44}
\rho+T+\sigma_{22}=\int{2v_2^2}hd\mathbf{v},
\end{equation}
where the term at the right-hand-side of Eq.~\eqref{synthetic44} is zero due to the symmetry condition~\eqref{Fourier_symm}, and according to Eq.~\eqref{HoT_sigma} the stress $\sigma_{22}$ can be calculated as
\begin{equation}\label{HoT_sigma_Fourier}
\sigma_{22} 
=-\frac{\frac{\partial}{\partial x_2}\int{}2\left(v_2^2-\frac{|\bm{v}|^2}{3}\right)v_2h\mathrm{d}\bm{v}}{\delta_{rp}}+\frac{2}{\delta_{rp}}\int{(L-L_s)v_2^2}\mathrm{d}\bm{v}.
\end{equation}

\begin{figure}[t]
	\centering
	\includegraphics[scale=0.68,viewport=20 10 540 420,clip=true]{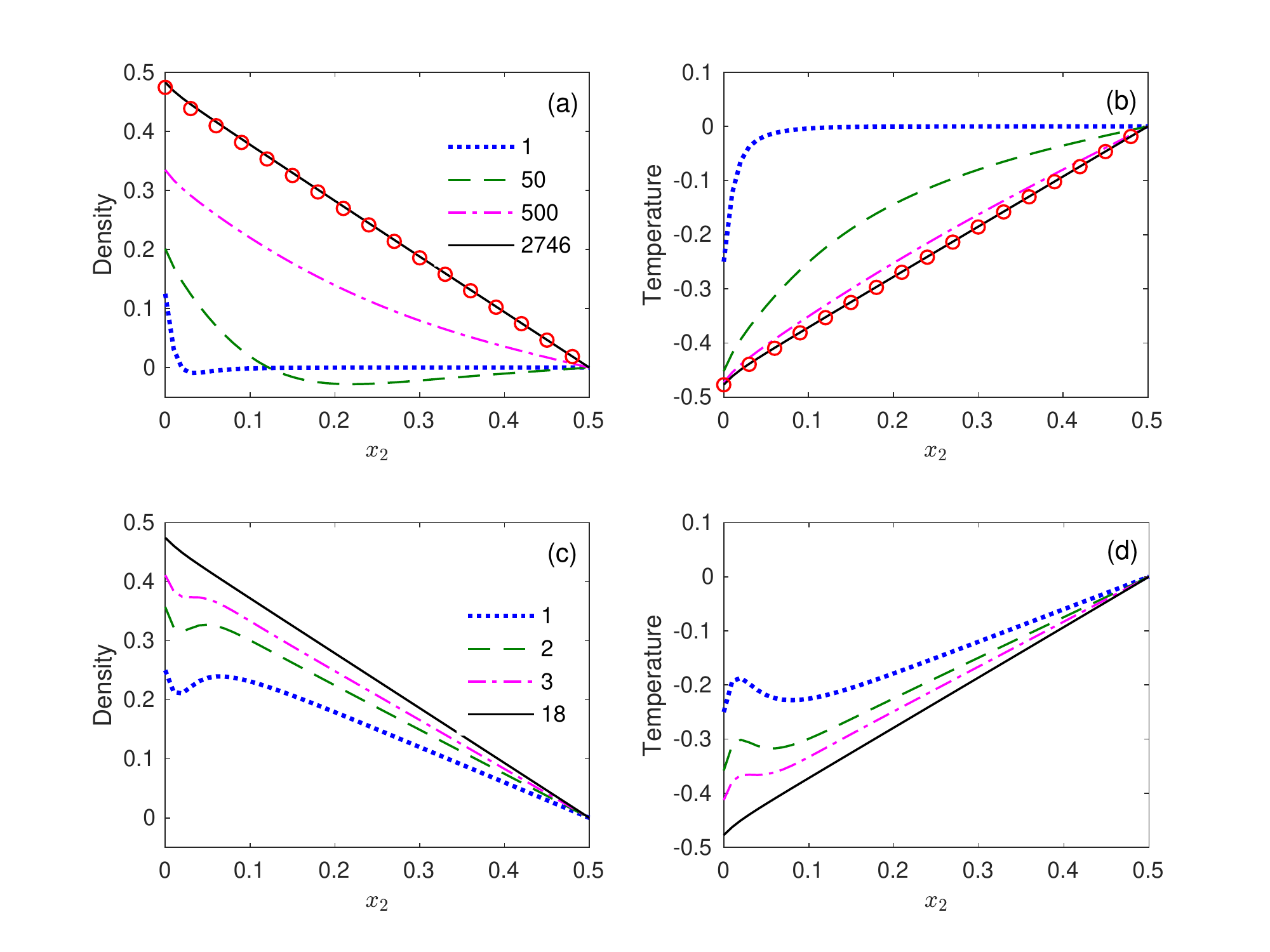}
	\caption{The density and temperature profiles at different iteration steps obtained from the CIS (a, b) and GSIS (c, d), when $\delta_{rp} = 50$. Circles show the converged solution obtained from the GSIS. The linearized Shakhov model is used with the initial condition  $h(x_2,\bm{v})=0$. The spatial region is discretized by $N_2=51$ equidistant points. The iteration stops when $\epsilon$ in Eq.~\eqref{epsilon_Fourier} is less than $10^{-5}$. Data in the legends are the iteration steps.}
	\label{fig:Fourier_histroy}
\end{figure}

We first test the efficiency of the GSIS based on the Shakhov model, that is, in Eq.~\eqref{LBE} we let the linearized Boltzmann collision operator equal to that of the linearized Shakhov model~\eqref{LBE_shakhov}. We choose the rarefaction parameter $\delta_{rp}=50$ and discretize the half spatial space into $N_2$ even-spaced points, where the derivative with respect to $x_2$ is approximated by a second-order upwind finite difference. The molecular velocity space in the $v_1$ and $v_3$ directions is truncated to the region $[-6, 6]$ by $24\times24$ equidistant points, while the molecular velocity $v_2$ is truncated to $[-6,6]$ and approximated by the non-uniform points~\cite{lei_Jfm,SuWeiPRE2017}:
\begin{equation}\label{nonuniform_v}
v_{2}=\frac{6}{(N_{v}-1)^\imath}[(-N_{v}+1)^\imath,(-N_{v}+3)^\imath,\cdots,(N_{v}-1)^\imath],
\end{equation}
which is useful to capture the discontinuity in the velocity distribution function near $v_2\sim0$. In this test we take $\imath=3$ and $N_v=64$. The iterations in both CIS and GSIS are terminated when 
\begin{equation}\label{epsilon_Fourier}
\epsilon= \max\left\{
\int{}\left|\frac{\rho^{(k+1)}}{\rho^{(k)}}-1\right|\mathrm{d}x_2, 
\int{}\left|\frac{T^{(k+1)}}{T^{(k)}}-1\right|\mathrm{d}x_2,
\int{}\left|\frac{q_2^{(k+1)}}{q_2^{(k)}}-1\right|\mathrm{d}x_2
\right\}
\end{equation}
is less than a certain value. Note that since $\rho$ and $T$  at $x_2=1/2$ are excluded in the above equation since they are zero.

\begin{figure}[t]
	\centering
	\includegraphics[scale=0.55,viewport=20 10 540 420,clip=true]{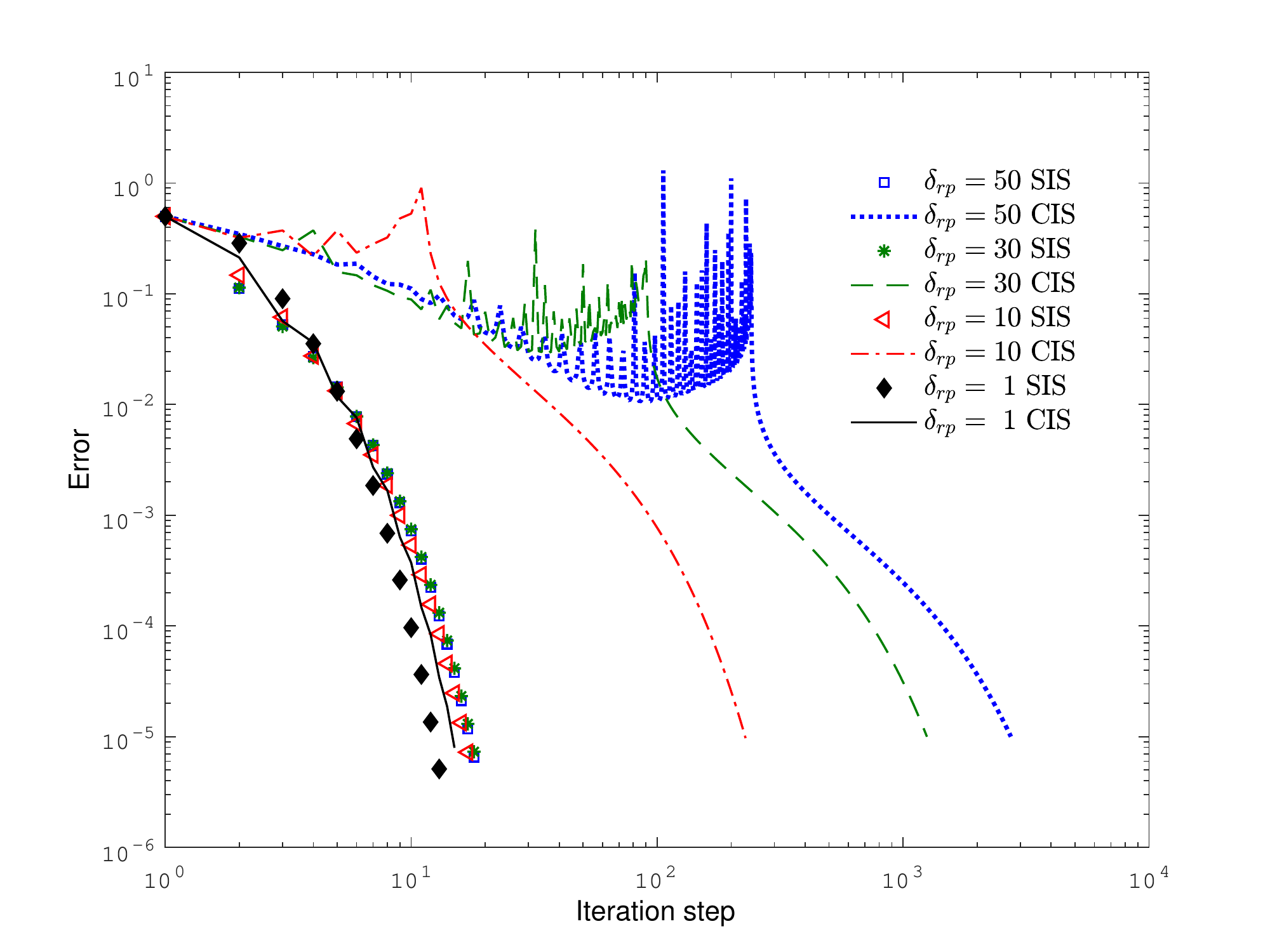}
	\caption{The decay of the error $\epsilon$ as a function of the iteration step, for the Fourier flow between two parallel plates described by the linearized Shakhov model. The spatial region is discretized by $N_2=51$ equidistant points.  }
	\label{fig:Fourier_speed}
\end{figure}

Figure~\ref{fig:Fourier_histroy} compares the convergence history of the GSIS and CIS when the rarefaction parameter is $\delta_{rp}=50$, that is, the flow is in the near-continuum regime. Starting from the initial guess $h(x_2,\bm{v})=0$, the perturbance from the solid surface quickly changes the density and temperature near the solid surface in the CIS (about one molecular mean free path away from the wall). However, due to the frequent collision between gas molecules, those in the bulk region takes a long time (i.e. iteration steps) to feel this change. From example, from Fig.~\ref{fig:Fourier_histroy}(b) we see that it takes about 50 iteration steps for the temperature at $x_2=0.5$ to feel this change. Moreover, such a change does not necessary lead to the final converged state monotonically, but it could be deviate further away from the final steady state: from Fig.~\ref{fig:Fourier_histroy}(a) we see that the density perturbance in the bulk region is even negative after 50 iterations, while the final steady state the density is always non-negative in the region of $x_2\in[0, 0.5]$. This is also evidenced in Fig.~\ref{fig:Fourier_speed} that the error does not decay monotonically but oscillates several times. Such a slow convergence is completely changed in the GSIS, where the temperature and density are corrected according to the synthetic equations~\eqref{HoT_q_Fourier} and~\eqref{synthetic44}, which the dominated parts are respectively $\frac{\partial T}{\partial x_2}=-\frac{4\delta_{rp}}{9C_q}q_{2}$ and $\rho=-T$ when $\delta_{rp}$ is large. This means that the temperature and density in the bulk region are corrected to be nearly linear immediately. As we can see from Fig.~\ref{fig:Fourier_histroy}(d), after the first iteration, the temperature from the GSIS at $x_2=0$ is the same as that from the CIS, but the temperature from the GSIS in the bulk region varies linearly, while that from the CIS is still zero. From Fig.~\ref{fig:Fourier_histroy}(c) we see that the density also varies linearly in the bulk, while at the solid surface it is more close to the final state than that obtained from the CIS.  Since the diffusion-type  macroscopic equation~\eqref{HoT_q_Fourier} allows the efficient exchange of information, fast convergence is realized in the whole computational domain, see Fig.~\ref{fig:Fourier_histroy}(c) and (d).

%Since the symmetry condition always guarantees $n(x_2=1/2)=T(x_2=1/2)=0$, a large number of iterations are needed to alter the density and temperature profiles between the solid surface and the center of the channel to be nearly linear.

Figure~\ref{fig:Fourier_speed} demonstrates how fast the solution is converged at different values of rarefaction parameter. When $\delta_{rp}$ is small, the errors in both the CIS and GSIS decays at the same rate, which means that the two schemes are as efficient as each other. As $\delta_{rp}$ increases so that the flow enters the transition and near-continuum regimes, the error in the CIS oscillates several times before it decays monotonically. As a consequence, the iteration number of CIS increases rapidly with the rarefaction parameter, which nearly scales as $\delta_{rp}^2$. For the GSIS, however, the error is monotonically decreasing, and the rarefaction parameter does not influence the error decay rate, where the converged solutions are obtained within the same number of iterations (here 20 iterations) for each rarefaction parameter from the free molecular to continuum flow regimes. At $\delta=50$, the GSIS is about 100 times more efficient than the CIS, and it can be expected that the gain of using GSIS becomes larger and larger as $\delta_{rp}$ further increases.

\begin{figure}[t]
	\centering
	\includegraphics[scale=0.44,viewport=20 0 540 420,clip=true]{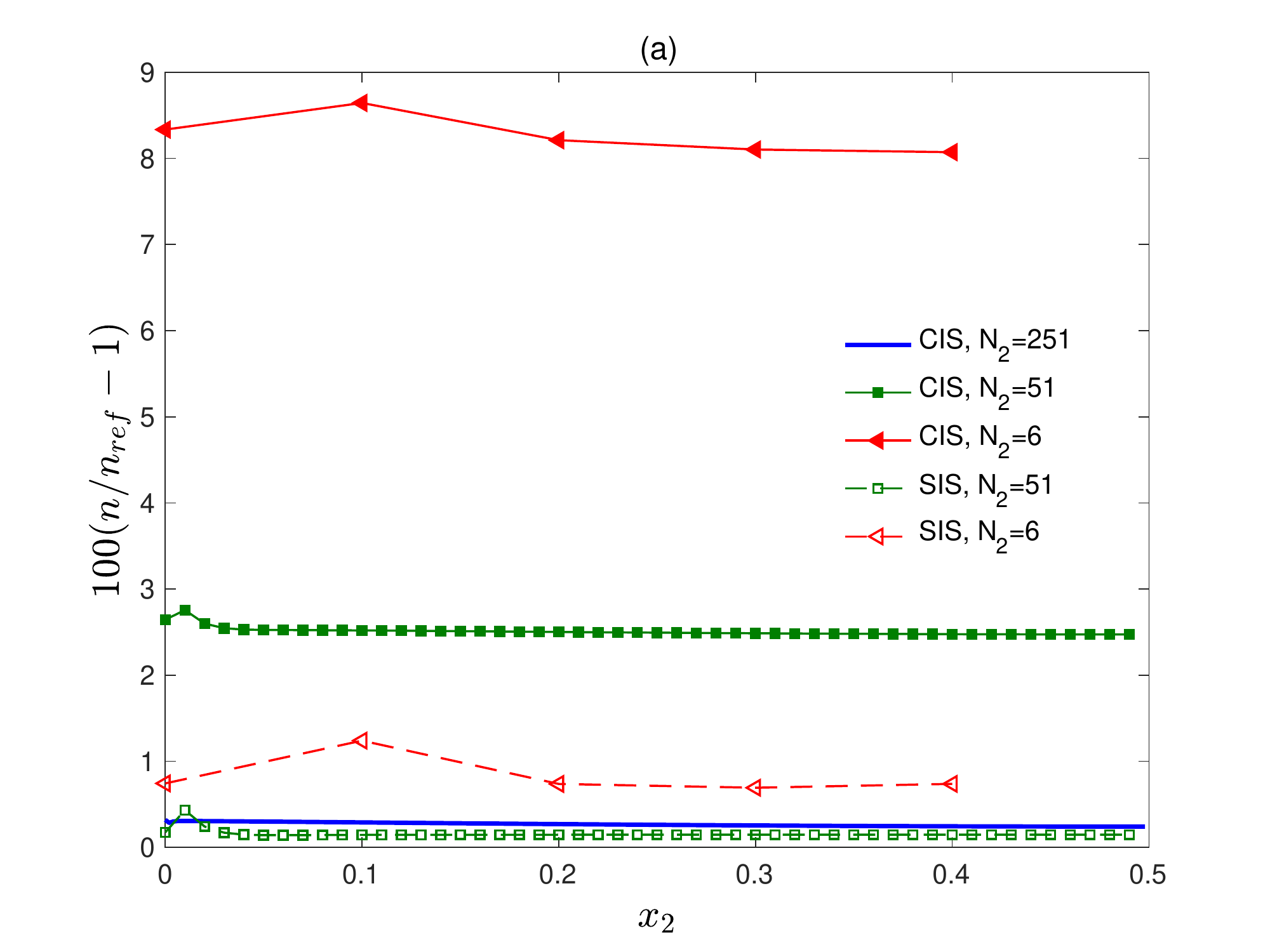}
	\includegraphics[scale=0.44,viewport=20 0 540 420,clip=true]{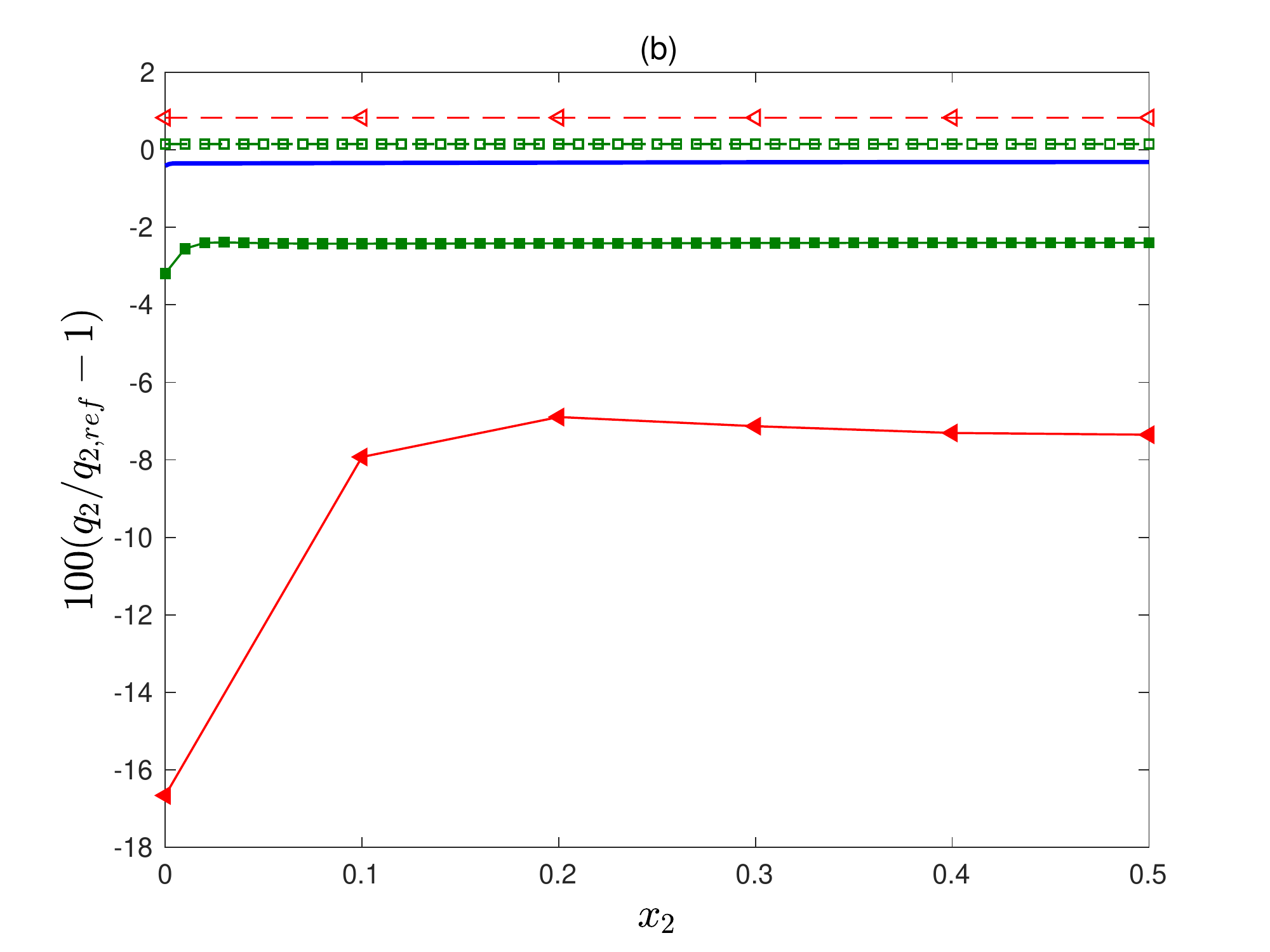}
	\caption{The influence of the spatial discretization on the accuracy of both the CIS and GSIS, for the Fourier flow between two parallel plates described by the linearized Shakhov model with $\delta_{rp}=50$. The iteration terminates when $\epsilon<10^{-6}$. The reference solutions (i.e. $\rho_{ref}$ and $q_{2,ref}$) are obtained from the GSIS with $N_2=251$, that is, the spatial cell size is about one tenth of the mean free path of gas molecules.}
	\label{fig:Fourier_spatial_error}
\end{figure}

Another important property of the GSIS is that the numerical error caused by the spatial discretization is much reduced when compared to that of the CIS.  From Fig.~\ref{fig:Fourier_spatial_error} we see that when $N_2$ is decreased from 251 to 6, that is, when the spatial cell size is respectively about $1/10$ and 5 times of the mean free path of gas molecules, the relative error in the density profile increases from 0.3\% to 9\%, while that in the heat flux increases from 0.3\% to 16\% in the CIS. However, the relative error in the GSIS always remain within 1\%, even when the cell size is about 5 times larger than the gas mean free path. Note that even when $\delta_{rp}=500$, the heat flux obtained from the GSIS only changes from $3.721\times10^{-3}$ when $N_2=551$ to $3.726\times10^{-3}$ when $N_2=6$. The reason for this excellent performance is that the GSIS is asymptotically preserving the Navier-Stokes limit, while in the CIS the ``numerical'' thermal conductivity may be different to the physical one. Besides, in the CIS, the false convergence, e.g. the non-uniform distribution of heat flux in in Fig.~\ref{fig:Fourier_spatial_error}(b), may be reached when the spatial resolution is not enough. The superior GSIS,  however, does not suffer this problem.

It should be noted that the implicit UGKS~\cite{Zhu2019JCP} and other variants~\cite{yang2018PoF,yang2018PRE} can also produce accurate results when the cell size is much larger than the molecular mean free path. This is achieved through a complex evaluation of the numerical flux at the cell interface to spontaneously treat the molecular streaming and collision. The GSIS, however, does not need complex flux evaluation.

Using the accurate and efficient GSIS, the LBE is solved for different molecular collision models~\eqref{collision_kernel} and the corresponding Knudsen layer functions are obtained. In the numerical simulation, we set the rarefaction parameter to be $\delta_{rp}=60$, so that the distance between two plates is about 60 times as large as the mean free path of gas molecules; thus, the interference between the Knudsen layers near each plate is avoided. In the fast spectral approximation of the linearized Boltzmann collision operator~\eqref{LBE_collision}, the integral with respect to the solid angle $\Omega$ is calculated by the Gauss-Legendre quadrature with $M=6$, see Eq.~(39) in Ref.~\cite{Lei2013}. In the spatial discretization we let
\begin{equation}\label{space_discrete}
x_2=(10-15s+6s^2)s^3, \quad s=(0,1,\cdots,N_s-1)/2(N_s-1)
\end{equation}
with $N_s=200$. The iterations terminate when $\epsilon<10^{-6}$.

\begin{figure}[t]
	\centering
	\includegraphics[scale=0.55,viewport=20 0 510 320,clip=true]{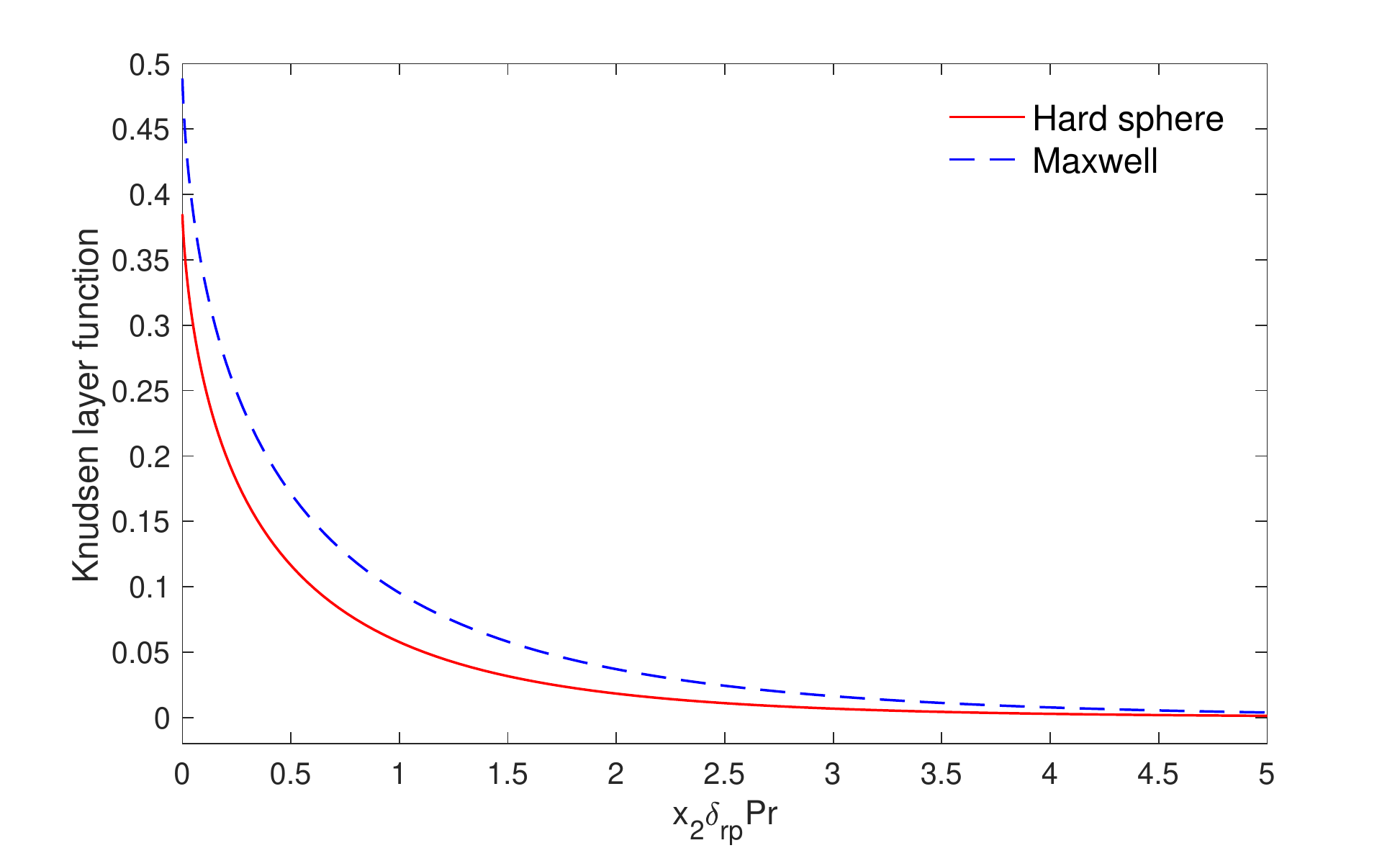}
	\caption{The Knudsen layer function $T_s$ for the temperature profile in the Fourier flow between two parallel plates obtained from GSIS.}
	\label{fig:Fourier_Knudsen_layer}
\end{figure}

When the steady-state solution is obtained, the temperature profile in the bulk region (i.e. $0.4\le{}x_2\le0.5$) is linearly fitted by $T_{NS}=k_1(x_2-1/2)$ in the dimensionless form, where $k_1$ is the coefficient from the least square fitting. Then the KLF is calculated according to the following equation:
	\begin{equation}\label{NS_fit}
	T_s\left({x_2}\delta_{rp}\mathrm{Pr}\right)=\delta_{rp}\mathrm{Pr}\frac{T_{NS}(x_2)-T(x_2)}{k_1},
	\end{equation}
	and the temperature jump coefficient, according to Sharipov's review paper, is calculated as
	\begin{equation}\label{slip_coe}
	{\zeta}_T=\frac{\delta_{rp}}{2}\left( \frac{5}{4\delta_{rp}\mathrm{Pr}|q_2|}-1 \right).
	\end{equation}

The GSIS results for the LBE with the HS and Maxwell molecules reaches the steady-state after 22 and 27 iterations, respectively, and the temperature jump coefficients are respectively 1.892 and 1.954, which do not vary a lot to the collision model. However, the Knudsen layer functions shown in Fig.~\ref{fig:Fourier_Knudsen_layer} has larger difference. It is amazing that the small terms $2\int{(L-L_s)v_iv_j}\mathrm{d}\bm{v}$ in Eq.~\eqref{HoT_sigma} and $\int{(L-L_s)v_i|\bm{v}|^2} \mathrm{d}\bm{v}$ in Eq.~\eqref{HoT_q} significantly affect the Knudsen layer function.

\subsection{Two-dimensional lid-driven cavity flow}

The two-dimensional lid-driven cavity flow is a canonical test for the algorithms of both Navier-Stokes equations and gas-kinetic equations. The flow domain is a square with size of $1\times1$, with the left and right walls locate at $x_1$ = 0 and $x_1=1$, bottom and top walls at $x_2=0$, $x_2=1$. The top wall, i.e., the lid moves in the $x-$direction with a constant velocity of $U_w$, while the other sides are static walls. All of the walls are kept at uniform temperature of $T_0$. To demonstrate the accuracy and efficient of the GSIS, the Shakhov kinetic equation is linearized by choosing $\alpha = U_w/v_m$ in Eq.~\eqref{eq:1}. The boundary conditions are
\begin{equation}
\begin{aligned}
h\left(x_{1}=0, \bm{v}\right)=-2 \sqrt{\pi} f_{eq}\int_{v_{1}<0} v_{1} h\left(x_{1}=0, \bm{v}\right) \mathrm{d}\bm{v}, \quad \text { when } v_{1}>0,\\
h\left(x_{1}=1, \bm{v}\right)=2 \sqrt{\pi} f_{eq}\int_{v_{1}>0} v_{1} h\left(x_{1}=1, \bm{v}\right) \mathrm{d}\bm{v}, \quad \text { when } v_{1}<0,\\
h\left(x_{2}=0, \bm{v}\right)=-2 \sqrt{\pi} f_{eq}\int_{v_{2}<0} v_{2} h\left(x_{2}=0, \bm{v}\right) \mathrm{d}\bm{v}, \quad \text { when } v_{2}>0,\\
h\left(x_{2}=1, \bm{v}\right)=\left[\sqrt{\pi}+ 2v_1+2 \sqrt{\pi} \int_{v_{2}>0} v_{2} h\left(x_{2}=1, \bm{v}\right) \mathrm{d}\bm{v}\right] f_{e q}, \quad \text { when } v_{2}<0.
\end{aligned}
\end{equation}
The problem is solved on non-uniform Cartesian grids, which are discretized by 
\begin{equation}\label{Couette_spatial_grid}
x_{1,2} = (10-15s+6s^2)s^3, \quad s = (0,1,2,\ldots,N_s-1)/\left(N_s-1\right),
\end{equation}
where $N_s$ is the number of grid nodes in both $x_1$ and $x_2$ axis. The linearized Shakhov equation is solved by DVM with the 2nd-order upwind finite-difference scheme, where the distribution functions stored at the centers of grid cells. In the synthetic acceleration step, the continuity equation and the momentum equations in Eq.~\eqref{eq123} are solved using a finite-difference version of the well known Semi-Implicit Method for Pressure Linked Equations (SIMPLE). In each SIMPLE iteration, we solve four discrete diffusive equations (the two velocity components, pressure correction, and temperature) using the Jacobi iteration methods.

When the macroscopic flow variables are solved by SIMPLE algorithm, the velocity distribution function is updated as
\begin{equation*}
h^{(k+1)}(\bm{x},\bm{v})=h^{(k+1/2)}(\bm{x},\bm{v})
+\frac{\delta_{rp}}{\max(10,\delta_{rp})}\left[\lambda_{\rho}(\bm{x})
+2\lambda_{\bm{U}}(\bm{x})\cdot{\bm{v}}
+\lambda_T(\bm{x})\left(|\bm{v}|^2-\frac{3}{2}\right)
\right]f_{eq},
\end{equation*}
because (i) the update of the shear stress and heat flux does not affect the accuracy and efficiency of the GSIS, and (ii) for highly rarefied gas flows, the high-order terms are very large and the macroscopic synthetic equations become stiff near the solid corners due to the small value of $\delta_{rp}$, hence the limiter ${\delta_{rp}}/{\max(10,\delta_{rp})}$ is introduced to increase the numerical stability.

We first test the converging speeds of the CIS and GSIS for the cases of $\delta_{rp} = 0.1$, 1, 10, 100 and 1000. The corresponding spatial grids are non-uniform with $N_s=21$, 21, 21, 41, 61 respectively. For the cases of $\delta_{rp} = 0.1$, 1 and 10, the molecular velocity in both $v_1$ and $v_2$ are discretized by Eq.~\eqref{nonuniform_v}, with $\imath=3$, and $N_v=48$, 48 and 24, respectively. For $v_3$, 24, 24 and 12 uniform points in the range of $[-6,6]$ are used. While for $\delta=100$ and 1000, the 6- and 8-point Gauss-Hermite quadrature nodes are used in all three velocity components. The iterations in both CIS and GSIS are assumed to be converged when
\begin{equation}\label{error_lid}
\epsilon=\iint \left | \frac{|\bm U^{(k+1)}|}{|\bm U^{(k)}|} - 1 \right |\mathrm{d}x_1\mathrm{d}x_2 < 10^{-5}.
\end{equation}

\begin{figure}
	\begin{center}
		\includegraphics[width=0.6\textwidth]{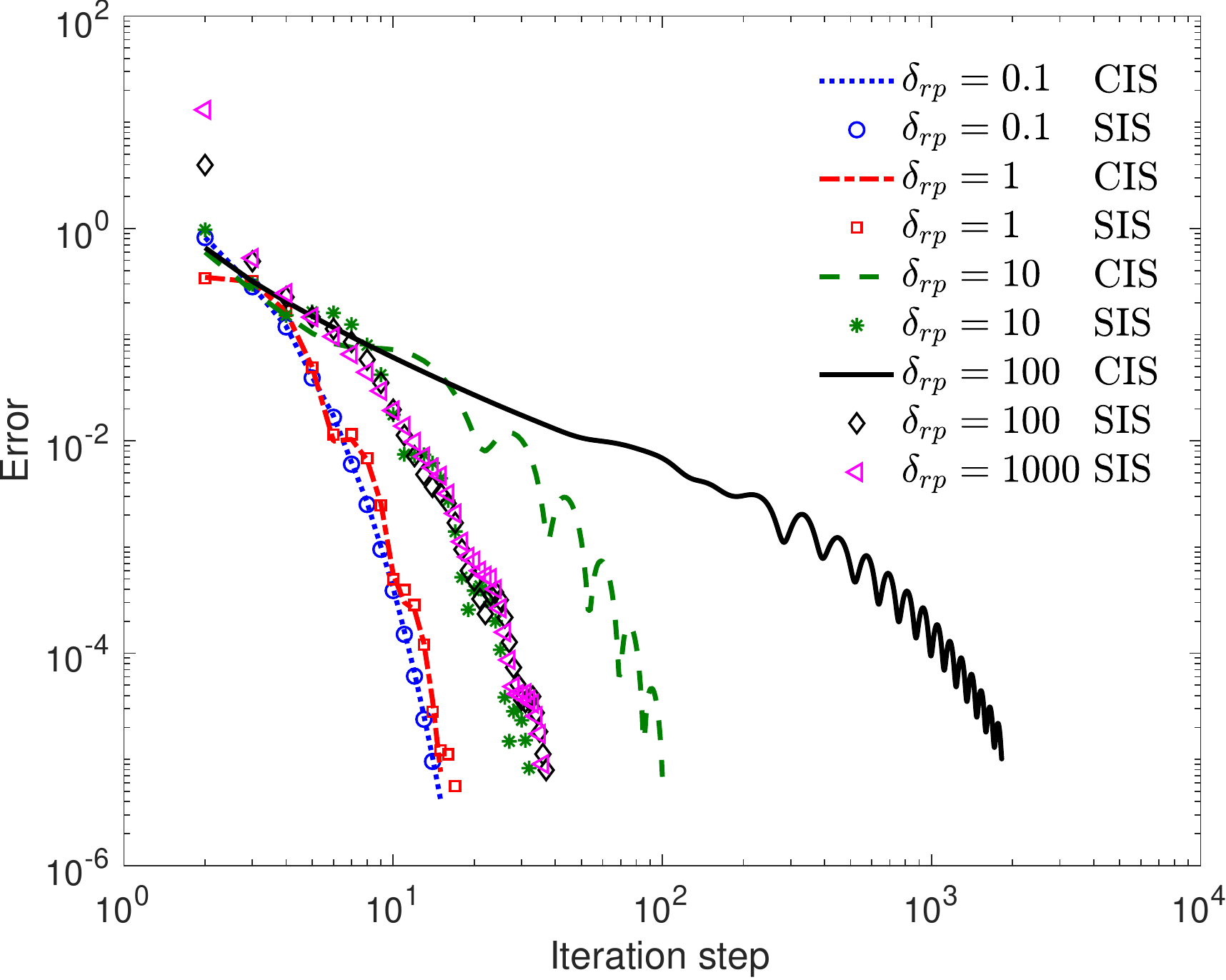}
	\end{center}
	\caption{The decay of error $\epsilon$ as a function of the iteration step in the lid-driven cavity flow described by the linearized Shakhov model.  The iteration is assumed to be converged when $\epsilon$ defined in Eq.~\eqref{error_lid} is less than $10^{-5}$.} %The spatial region is discretized by $N_s=21$, 21, 21, 41 and 61 for the cases of $\delta=0.1$, 1, 10, 100 and 1000, respectively.
	\label{fig:convergence}
\end{figure}

Figure~\ref{fig:convergence} compares the decay of error $\epsilon$ as a function of the number of iteration steps in CIS and GSIS for solutions of flows at different values of rarefaction parameter, while Table~\ref{tab:iteration} summarizes the number of iteration steps and the total CPU time of the calculations with a single threaded Matlab 2018 code on Intel Xeon-E5-2680 v4 CPU. Similar to the test case of the Fourier heat transfer, at small $\delta_{rp}$ (0.1 and 1), the errors in both GSIS and CIS decay with the same rate and converged in less than 20 steps. In the cases of larger $\delta_{rp}$, the iteration step in GSIS slightly increases, but it is less than 40 steps even for the case of $\delta_{rp}= 1000$. In contrast, the convergence of the CIS iteration deteriorates severely as $\delta_{rp}$ increase. The iteration step reaches 1823 for the case of $\delta_{rp} = 100$. Due to the slow convergence of CIS for near continuum flows, the case of $\delta = 1000$ is not simulated.

\begin{table}[htbp]
	\centering
	\caption{Number of iteration steps and CPU time to reach convergence for the lid-driven cavity flow. }
	\begin{tabular}{rllrrrr}
		\hline
		\multicolumn{1}{l}{$\delta$} & $N^2$ & $N_{v_1}N_{v_2}N_{v_3}$    & \multicolumn{2}{c}{Iteration steps} & \multicolumn{2}{c}{Total CPU time (s)} \\ %& Synthetic equations\\
		&       &       & \multicolumn{1}{r}{CIS} & \multicolumn{1}{r}{GSIS} & \multicolumn{1}{r}{CIS} & \multicolumn{1}{r}{GSIS}\\ %&  cost in GSIS \\
		%    \cline{4-7}
		\hline
		0.1   & $20\times 20$ & $48\times48\times 24$ & 14    & 13    &     28.5  & 32.6 \\%& 18.2\%\\
		1   & $20\times 20$ & $48\times48\times 24$& 14    & 16    &     28.2   & 38.4 \\%& 16.1\% \\
		10  & $20\times 20$ & $24\times24\times 12$& 99    & 31    &   121.4    & 47.7 \\ %&20.3\% \\
		100   & $40\times40$ & $16\times16\times 16$ & 1823  & 36    &   3176.1    & 144.1 \\%& 56.4\% \\
		1000  & $60\times60$ & $8\times8\times8$ &     ---   & 36     &   ---   & 492.5 \\%& xxx \\
		\hline
	\end{tabular}%
	\label{tab:iteration}%
\end{table}%

With significantly faster convergence rate, the GSIS takes much less CPU time than the CIS for cases of large $\delta_{rp}$ as shown in Table~\ref{tab:iteration}. Note that although the iteration number is reduced in GSIS, the time for each iteration increases as the cost to solve the synthetic equation is non-negligible. see the last column of Table~\ref{tab:iteration}. This is because the segregated approach of the SIMPLE algorithm can take up to several hundreds of iterations to converge, depending on the value of $\delta_{rp}$. We note that using a coupled algorithm to solve the discretized pressure and velocity components in a single linear equation system would be much faster than the segregated approach, especially for high $\delta_{rp}$ cases, as have been studied in the incompressible CFD theories. For example, in the following section we find that if the kinetic synthetic equations are solved by the DG, the cost of DG for synthetic equations is negligible since pressure, velocity, and temperature are solved simultaneously.

\begin{figure}[t]
\begin{center}
%\subfloat[CIS $N_s=21$]{\includegraphics[width=0.3\textwidth]{CIS20_SIS60.eps}}
%\subfloat[CIS $N_s=41$]{\includegraphics[width=0.3\textwidth]{CIS40_SIS60.eps}}
%\subfloat[CIS $N_s=101$]{\includegraphics[width=0.3\textwidth]{CIS100_SIS60.eps}}\\
%\subfloat[SIS $N_s=21$]{\includegraphics[width=0.3\textwidth]{SIS20_SIS60.eps}}
%\subfloat[SIS $N_s=41$]{\includegraphics[width=0.3\textwidth]{SIS40_SIS60.eps}}
%\subfloat[CIS $N_s=61$]{\includegraphics[width=0.3\textwidth]{SIS60_SIS60.eps}}
\includegraphics[width=0.3\textwidth]{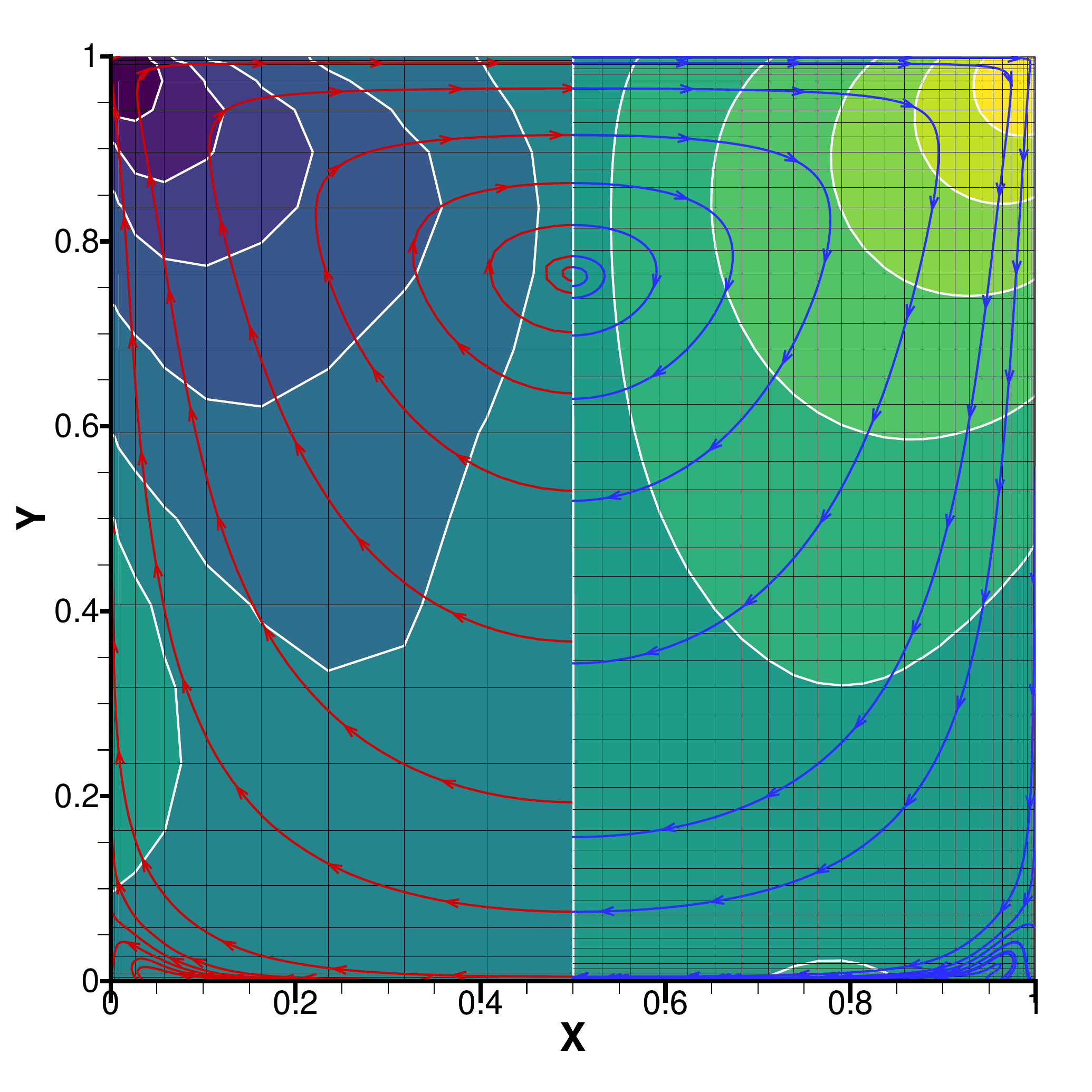}
\includegraphics[width=0.3\textwidth]{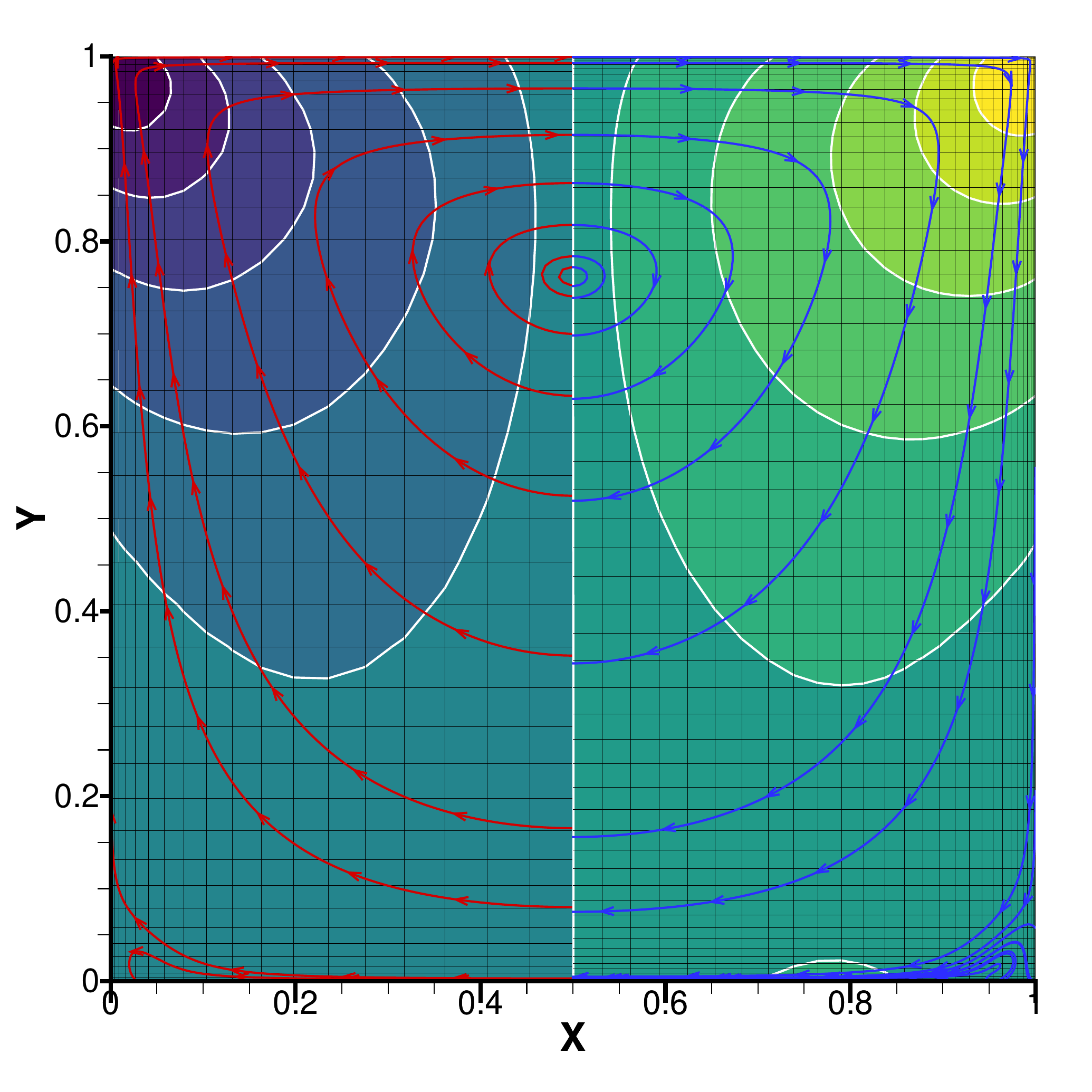}
\includegraphics[width=0.3\textwidth]{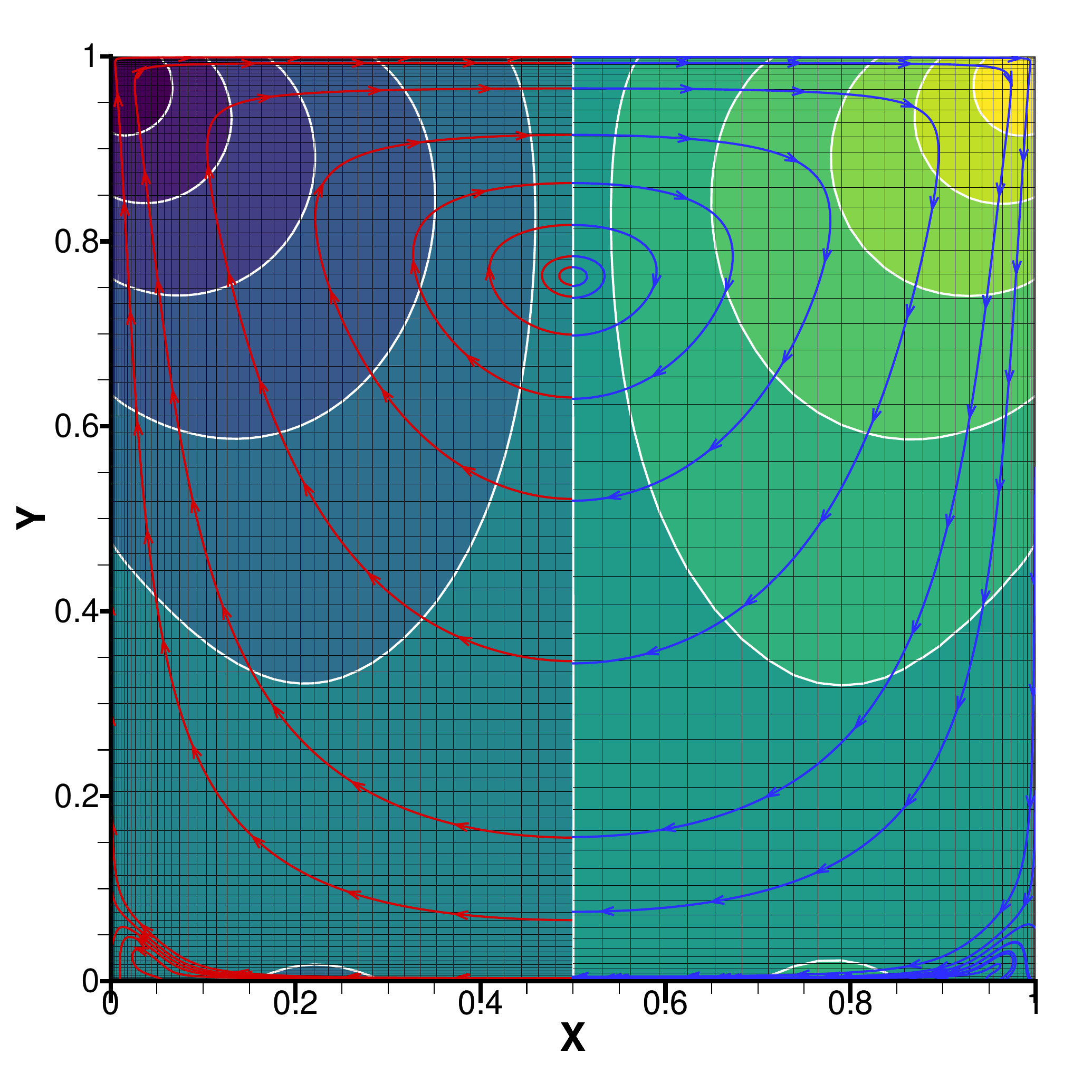}\\
\includegraphics[width=0.3\textwidth]{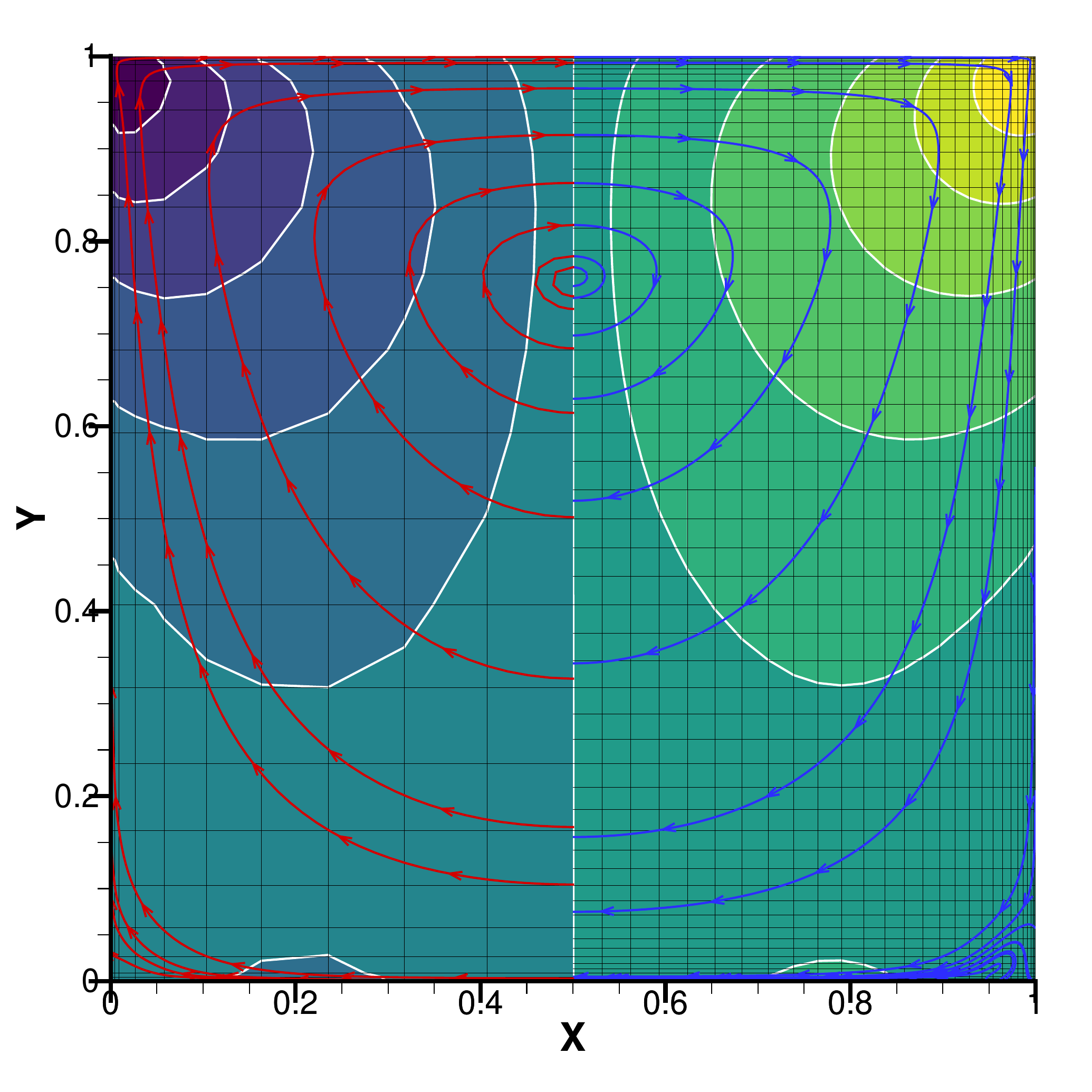}
\includegraphics[width=0.3\textwidth]{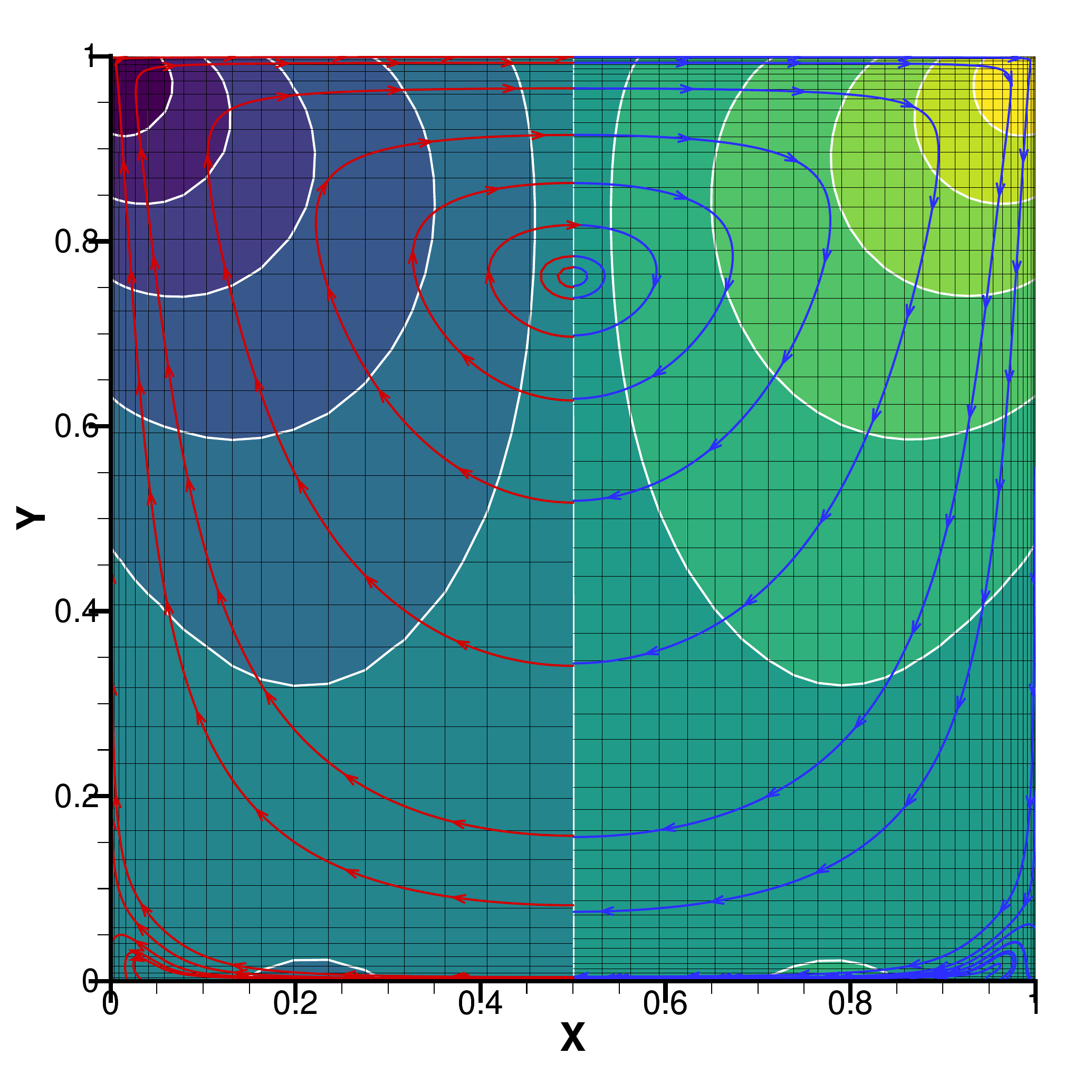}
\includegraphics[width=0.3\textwidth]{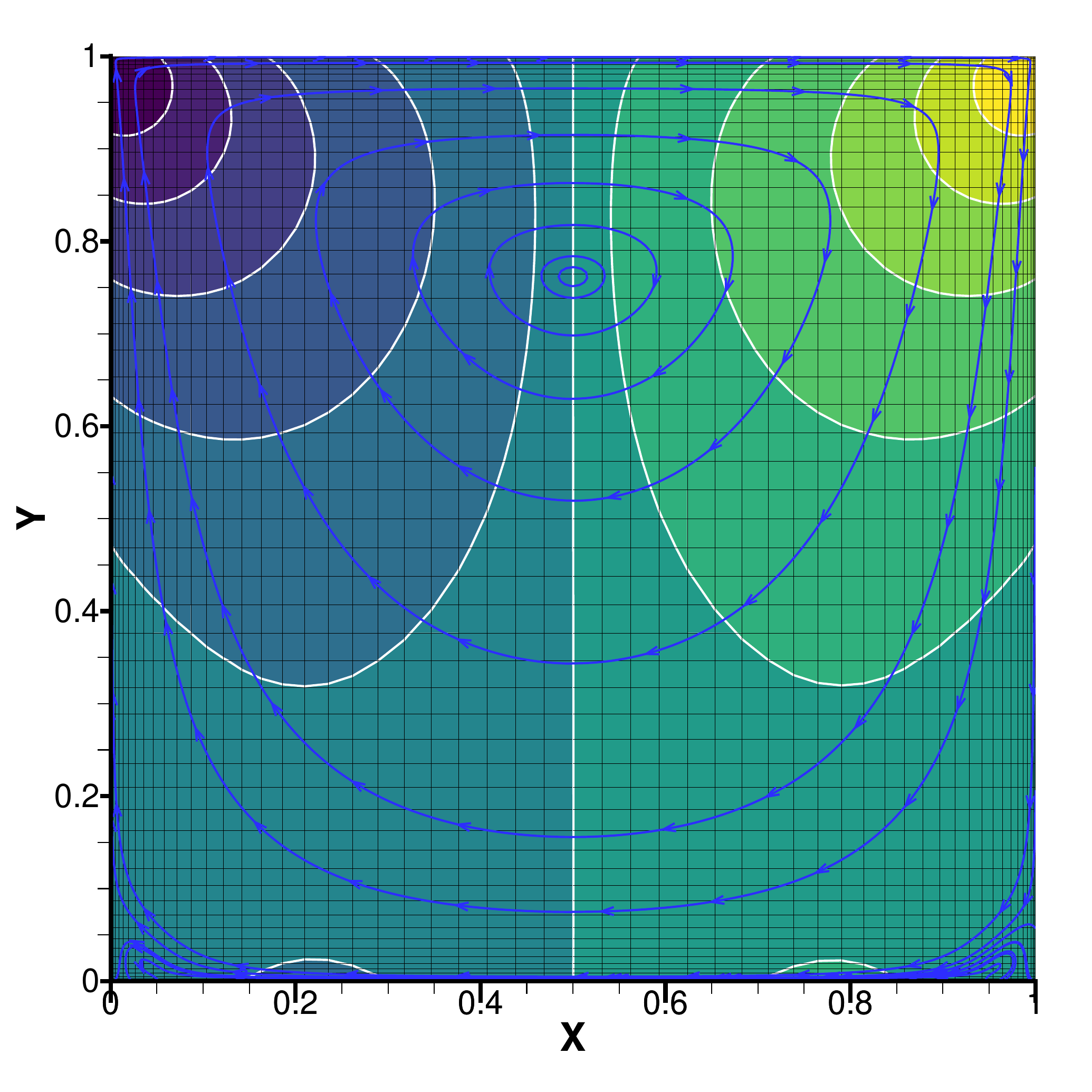}
\end{center}
\caption{Accuracy comparisons between the CIS and GSIS for the lid-driven cavity flow. In each plot, the right half are the reference solution (GSIS results on the $N_s=61$ grid). In the upper rows, the left halves of the plots are CIS solutions on grids of $N_s = 21$, 41 and 101 from left to right.  In the lower rows, the left halves are GSIS solution on grids of $N_s = 21$, 41 and 61 from lest to right. The contour plot is pressure ($\rho+T$), with contour levels of -0.2,  -0.1, -0.05, -0.02, -0.005, 0, 0.005, 0.02, 0.05, 0.1 and 0.2. }
\label{fig:ldc}
\end{figure}

To compare the accuracy of the GSIS with the CIS, we simulated the case of $\delta_{rp} = 100$ with different non-uniform physical grids, including $N_s=21$, 41, 61 and 101. Figure~\ref{fig:ldc} presents the comparisons of the pressure fields and streamlines predicted by both the CIS and GSIS on various grids, in which the reference solutions are taken as the GSIS results on grid of $N_s=61$. The results show that the GSIS solution on the coarsest grid ($N_s=21$) is much more accurate than the CIS counterpart, especially in terms of the pressure field. From Fig.~\ref{fig:ldc}(d) to (f), we can observe that the short contour lines near the bottom wall are accurately captured by the GSIS even on the coarsest mesh, while the CIS can capture them only with the finest mesh ($N_s=100$, see Fig.~\ref{fig:ldc} (c).)

\subsection{Shear-driven flow between two eccentric cylinders}

\begin{figure}[h]
	\begin{center}
		\includegraphics[width=0.3\textwidth]{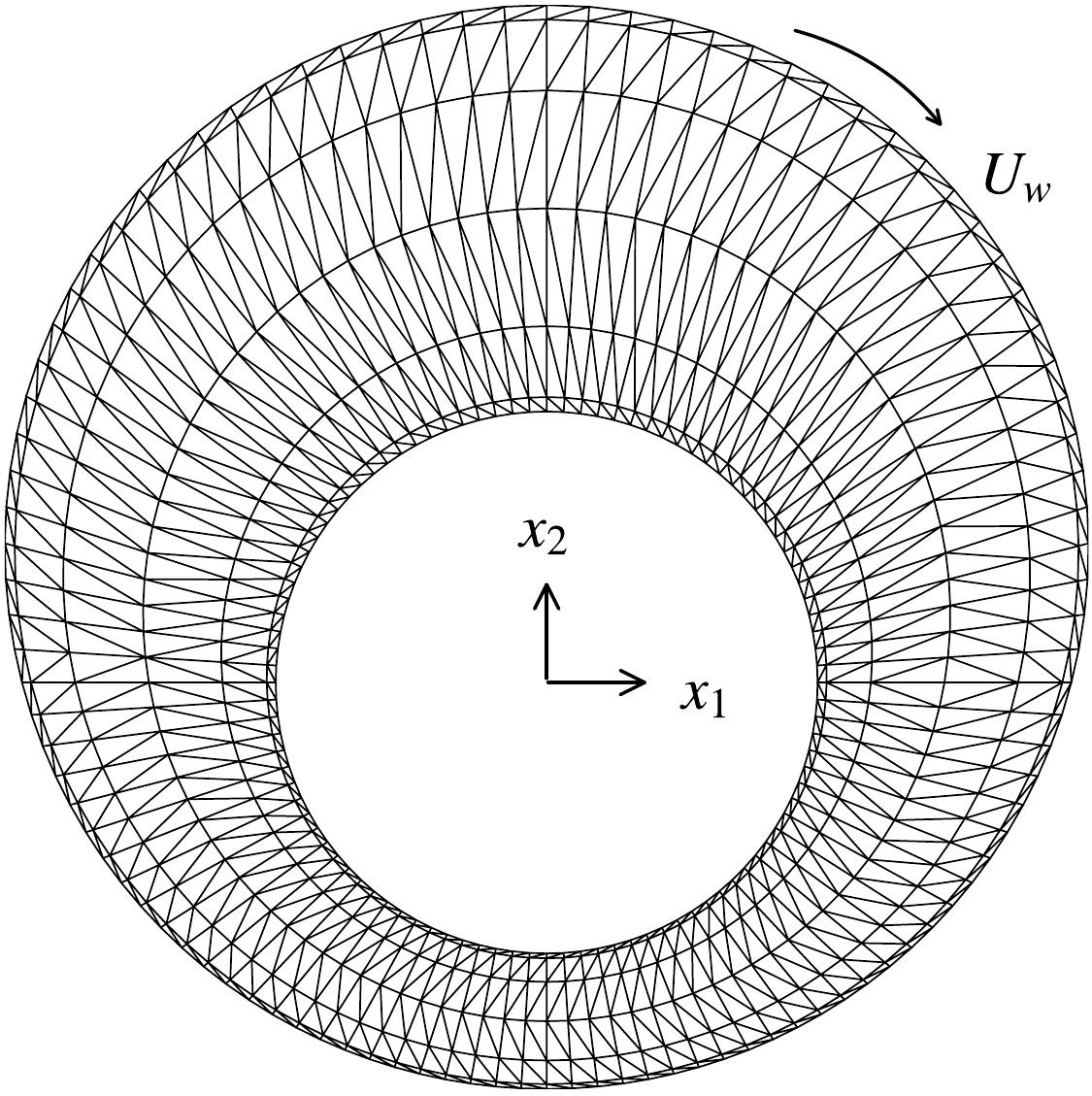}
	\end{center}
	\caption{Schematic of the geometry and structured triangular mesh for shear-driven flow between two eccentric cylinders.}
	\label{CylinderG}
\end{figure}

In this section, we consider a shear-driven gas flow between two noncoaxial cylinders. This test case is used to show that the proposed synthetic iterative scheme can be efficiently implemented through other CFD method rather than the finite difference algorithm to deal with more complicated geometries. As shown in Figure~\ref{CylinderG}, the outer cylinder with a radius of 2 rotates clockwise at a constant speed of $U_w$, while the inner cylinder with a radius of 1 keeps static. The centers of the outer cylinder and inner cylinder are at $\bm x=(0,0.5)$ and the origin, respectively. The cylinders are of a constant temperature $T_0$. It is assumed that $U_w$ is much smaller than the most probable speed $v_m$, thus the gas system can be linearized with $\alpha=U_w/v_m$. The distribution function for the reflected molecules at the outer cylinder is given by

\begin{equation}
h\left(\bm x,\bm v\right)=\left[2\bm{t}_w\cdot\bm{v}-2\sqrt{\pi}\int_{\bm{v}'\cdot\bm{n}_w<0}\bm{v}'\cdot\bm{n}_wh\left(\bm{x},\bm{v}'\right)\mathrm{d}\bm{v}'\right]f_{eq},\quad\text{when}\ \bm{v}\cdot\bm{n}_w>0,
\end{equation}
where $\bm{n}_w$ and $\bm{t}_w$ denote the outward unit normal vector and tangential vector of the solid surface. The boundary condition at the inner cylinder is similar but without the term $\bm{t}_w\cdot\bm{v}$.

Using both the GSIS and CIS, the shear-driven flow is resolved on structured triangular mesh, in which the grid nodes along the radial direction is described by Eq.~\eqref{Couette_spatial_grid}. The high-order DG methods are employed to seek solutions of the linearized Shakhov model equation and the synthetic macroscopic equations in pecewise polynomial spaces of degree of 3. The detailed DG scheme for the gas kinetic equation can be found in~\cite{Su2019IDG}, while the hybridizable DG algorithm to solve the synthetic macroscopic equations is listed in the Appendix. 

\begin{figure}[t]
	\centering
	\includegraphics[width=0.4\textwidth]{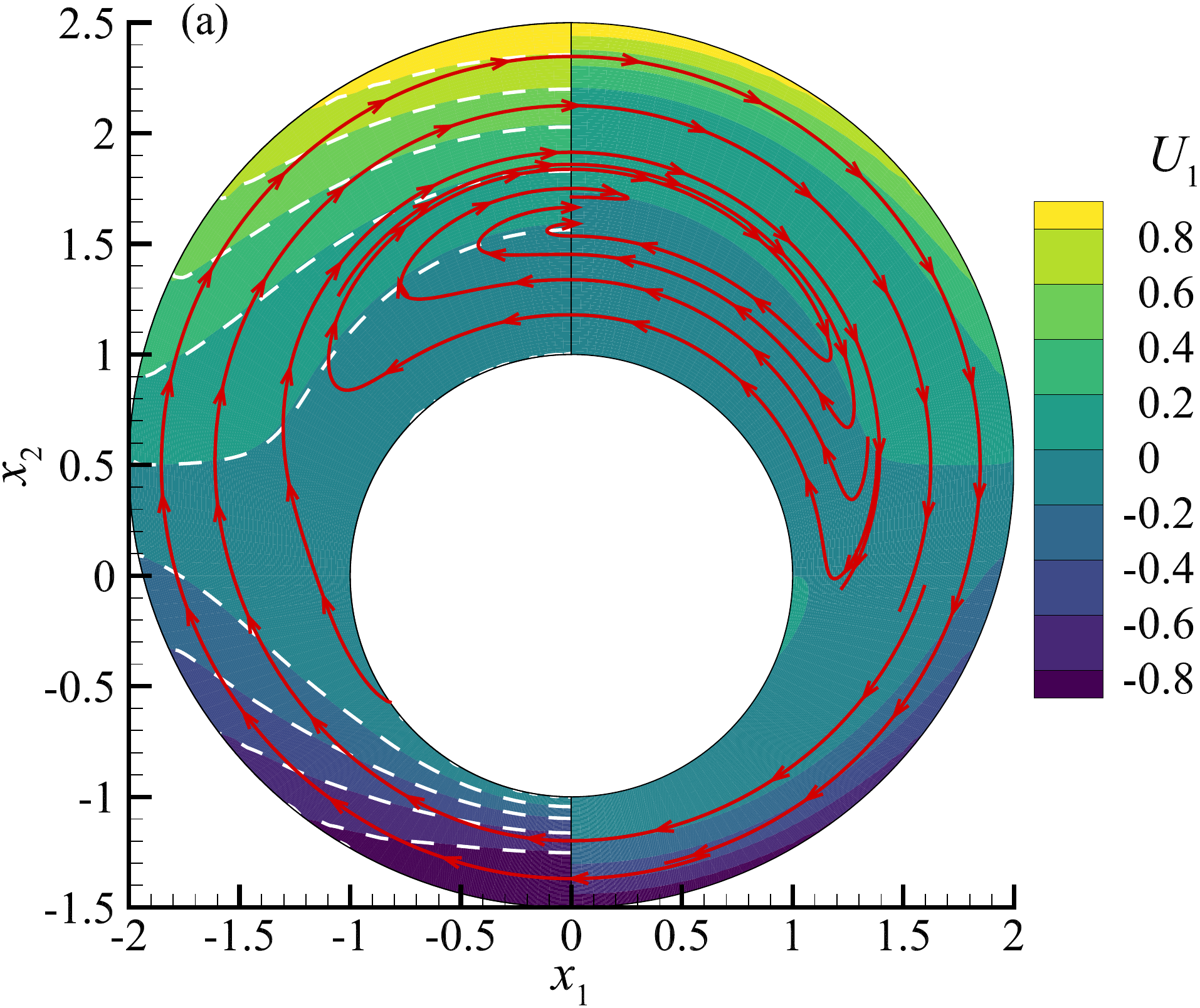}
	\includegraphics[width=0.4\textwidth]{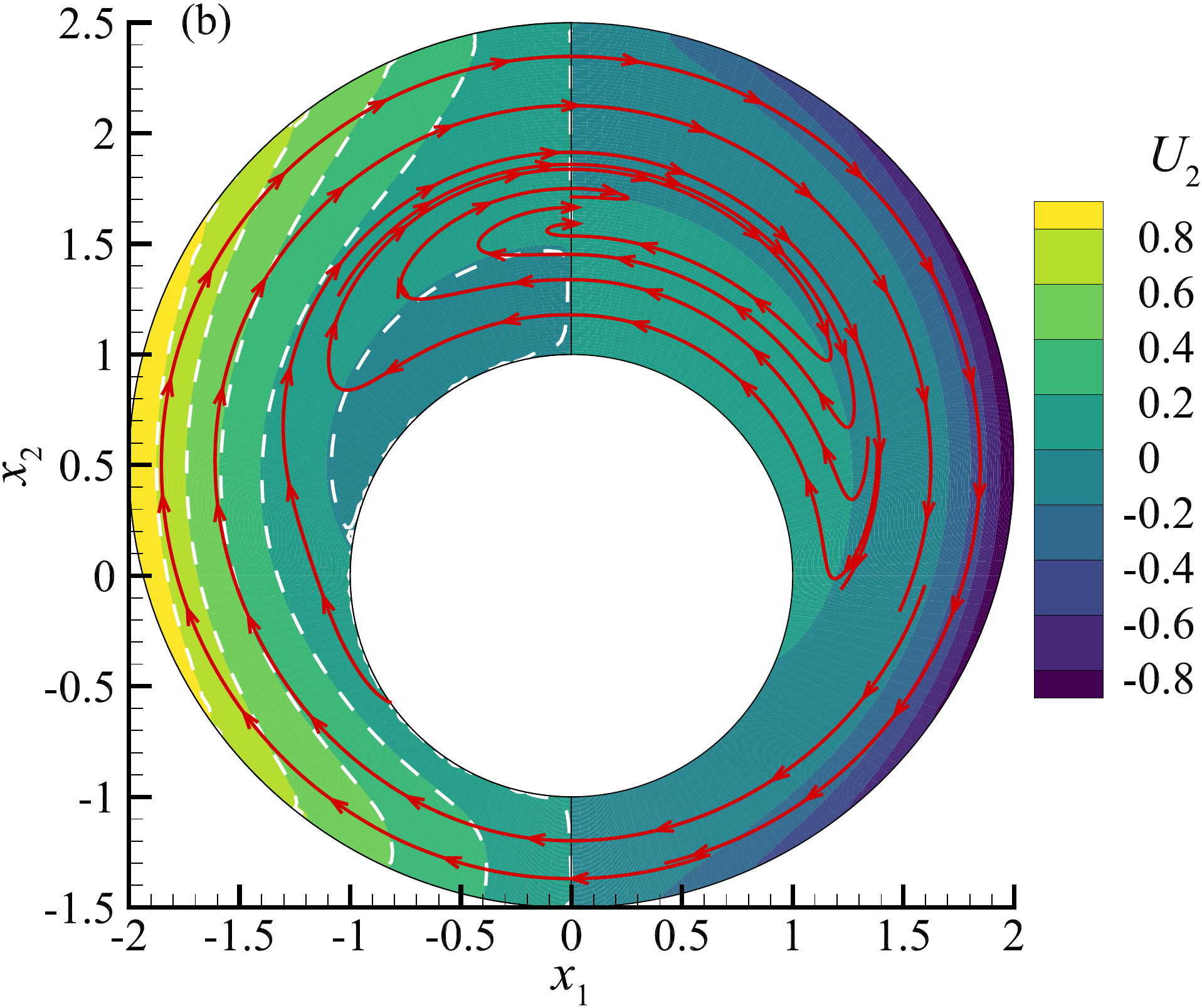}
	\includegraphics[width=0.4\textwidth]{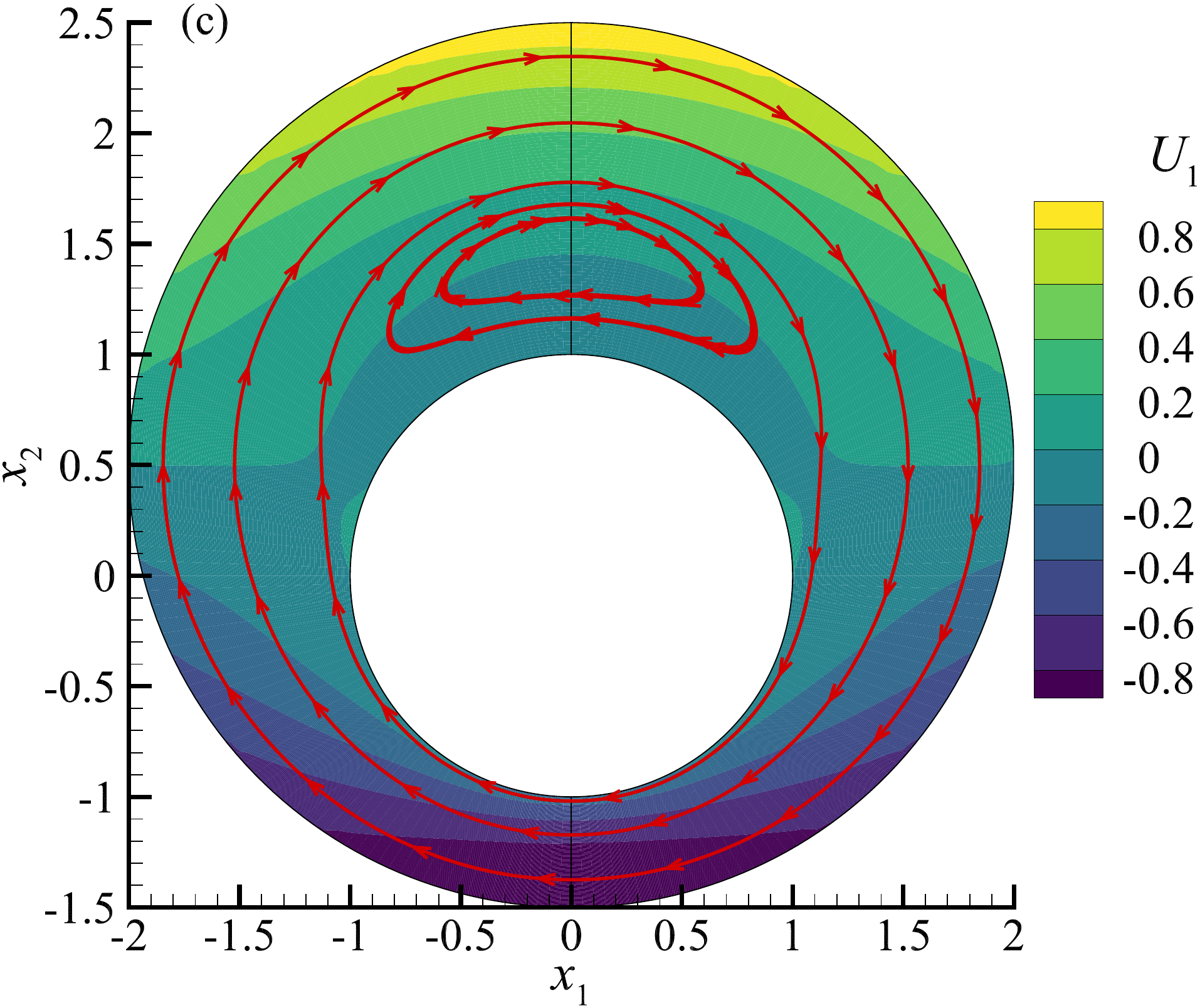}
	\includegraphics[width=0.4\textwidth]{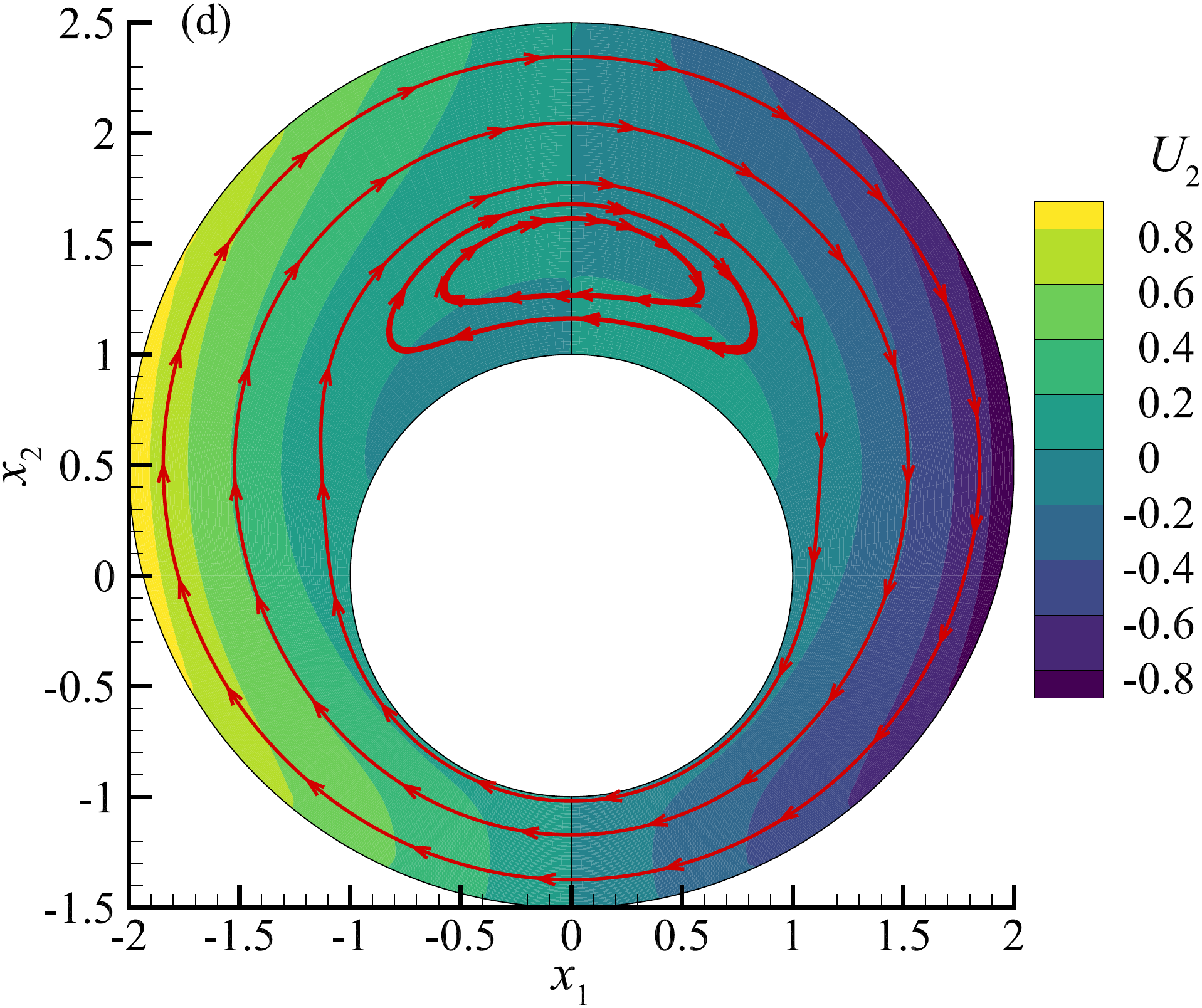}
	\caption{Comparisons of the CIS and GSIS results for shear-driven flow between two eccentric cylinders. (a) Contours of $U_1$ and streamlines at $\delta_{rp}=1000$; (b) Contours of $U_2$ and streamlines at $\delta_{rp}=1000$; (c) Contours of $U_1$ and streamlines at $\delta_{rp}=10$; (d) Contours of $U_2$ and streamlines at $\delta_{rp}=10$. In each sub-figures, the GSIS results are plotted in the left half domain while the CIS ones are illustrated in the right half domain. In (a) and (b) the velocity contours obtained by only solving the Navier-Stokes equations with non-slip velocity boundary are also included, which are indicated by the white dashed lines.
	}
	\label{TwoCylinderV}
\end{figure}

The resultant velocity contours and streamlines are illustrated in Fig.~\ref{TwoCylinderV} for two selected rarefaction parameters $\delta_{rp}=1000$ and 10, in which the GSIS solutions are plotted in the left half domain and the CIS ones are plotted in the right half domain. The results at $\delta_{rp}=1000$ are obtained on 2400 triangles with cell size (characterized by the height of triangle) varying from 3 to 260 times the mean free path of gas molecules. The molecule velocity space is discretized by 8-point Gauss-Hermite quadrature nodes in $v_1$ and $v_2$ and 12 equidistant nodes in the range of $[-4,4]$ in $v_3$. The results at $\delta_{rp}=10$ are obtained on 1600 triangles with cell size varying from 0.1 to 3 times the mean free path of gas molecules. The molecule velocity space is discretized in the domain of $[-4,4]^3$ by 32 non-uniform nodes in $v_1$ and $v_2$ and 24 equidistant nodes in $v_3$. The solutions are believed to be converged when the relative error in velocity magnitude $|\bm{U}|$ between two consecutive iteration steps is less than $10^{-5}$. The streamlines show that, as the gas rotates clockwise from the top to the bottom, due to the shrink of the flow pass, part of the gas near the outer surface is squeezed into the bottom narrow space while the other part of the gas flows back along the surface of the inner cylinder; as a consequence, a vortex appears above the inner cylinder.

Large discrepancies in the velocity contours are observed between the GSIS and CIS results at $\delta_{rp}=1000$. To test the accuracy of both schemes, we also include the results obtained by only solving the Navier-Stokes equations with the non-slip velocity boundary condition, which are illustrated by the white dashed lines in Fig.~\ref{TwoCylinderV}(a) and (b). The GSIS results coincide with the ones from the Navier-Stokes equations, thus the GSIS can asymptotically preserve the Navier-Stokes limit. However, the CIS cannot predict accurate solutions due to the large numerical dissipation on such a coarse mesh, i.e. the maximum cell size is about 260 times of the molecular mean free path.  As the rarefaction parameter decreases to 10, the GSIS and CIS can produce close solutions on the same mesh. 

Consider the rate of convergence to the steady-state solution, the GSIS cost only 26 iterative steps to reach the convergence criterion for both the cases of $\delta_{rp}=1000$ and 10, while the CIS consumes 49454 and 296 steps, respectively. Since compared to that of solving the kinetic equation, the computational consumption for DG to solve the macroscopic equations is negligible, since the number of degrees of freedom for the latter one is much smaller. Therefore, the GSIS can be nearly 2000 and 10 times faster than the CIS when $\delta_{rp}=1000$ and 10, respectively.

\section{Numerical results for periodic oscillation problems}\label{sec:results2}

For linearized problems, if the external force that drives the flow changes periodically in time, then the velocity distribution function can be expressed as~\cite{Sharipov2008Couette,Kalempa2009Sound,Wu2014JFM}: 
\begin{equation}
f=f_{eq}(\bm{v})+
A\Re\left[\exp({i{\mathrm{St}}t})
h(\bm{x},\bm{v})\right],
\end{equation}
where $\Re$ is a real part of a variable and $h$ satisfies the following linearized Boltzmann equation:
\begin{equation}\label{LBE_osci}
i\mathrm{St} h+\bm{v}\cdot\frac{\partial
	{h}}{\partial{\bm{x}}}=L(h,f_{eq}).
\end{equation}
Note that here $i$ is the imaginary unit and $h$ is a complex function, so are the macroscopic quantities defined in Eqs.~\eqref{nuT} and~\eqref{sigmaQ}. These complex values will introduce phase shifts relative to that of the external disturbance. The Strouhal number $\mathrm{St}$ 
\begin{equation}
{\mathrm{St}}=\frac{\varpi{}H}{v_m}
\end{equation}
is the oscillation frequency $\varpi$ normalized by $v_m/H$.

The solutions to these oscillating problems can also be accelerated by the GSIS; the corresponding macroscopic synthetic equations can be derived if we replace $\partial/\partial{t}$ in Sec.~\ref{secIII} by $i\text{St}$. Several numerical examples are given to demonstrate the accuracy and efficient of the present method.

\subsection{Spontaneous Rayleigh-Brillouin scattering}\label{sec:srbs}

In the spontaneous Rayleigh-Brillouin scattering (SRBS), light propagating through the gas is scattered by the thermal motion of gas molecules, where the spectrum of the scattered light contains the information of gas such as temperature, speed, and viscosity. Thus, SRBS provides a non-intrusive way to probe the gas properties in a remote way. Theoretically, the SRBS spectrum can be obtained by solving the LBE~\eqref{LBE} with the initial condition $h(t=0,{x_2},\bm{v})\propto{}\delta({x_2})f_{eq}(\bm{v})$, which represents a density impulse~\cite{Sugawara1968PoF,LeiJFM2015}. To be more specific, the SRBS spectrum is calculated as
\begin{equation}\label{SRBS}
S_s(\delta_{rp},f_s)=\Re\left(\int \hat{h}\mathrm{d}\bm{v}\right),
\end{equation}
where $\hat{h}(\bm{v})$, the Laplace and Fourier transforms of $h$ in the temporal and spatial directions, respectively, satisfies (suppose the scattered light propagates in the $x_2$ direction)
\begin{equation}\label{SRBS_LBE}
2\pi{i}(f_s-v_2)\hat{h}={L}^+(\hat{h})-\nu_{eq}\hat{h}+f_{eq}.
\end{equation}

\begin{figure}[t]
	\centering
	\includegraphics[scale=0.4,viewport=20 0 540 420,clip=true]{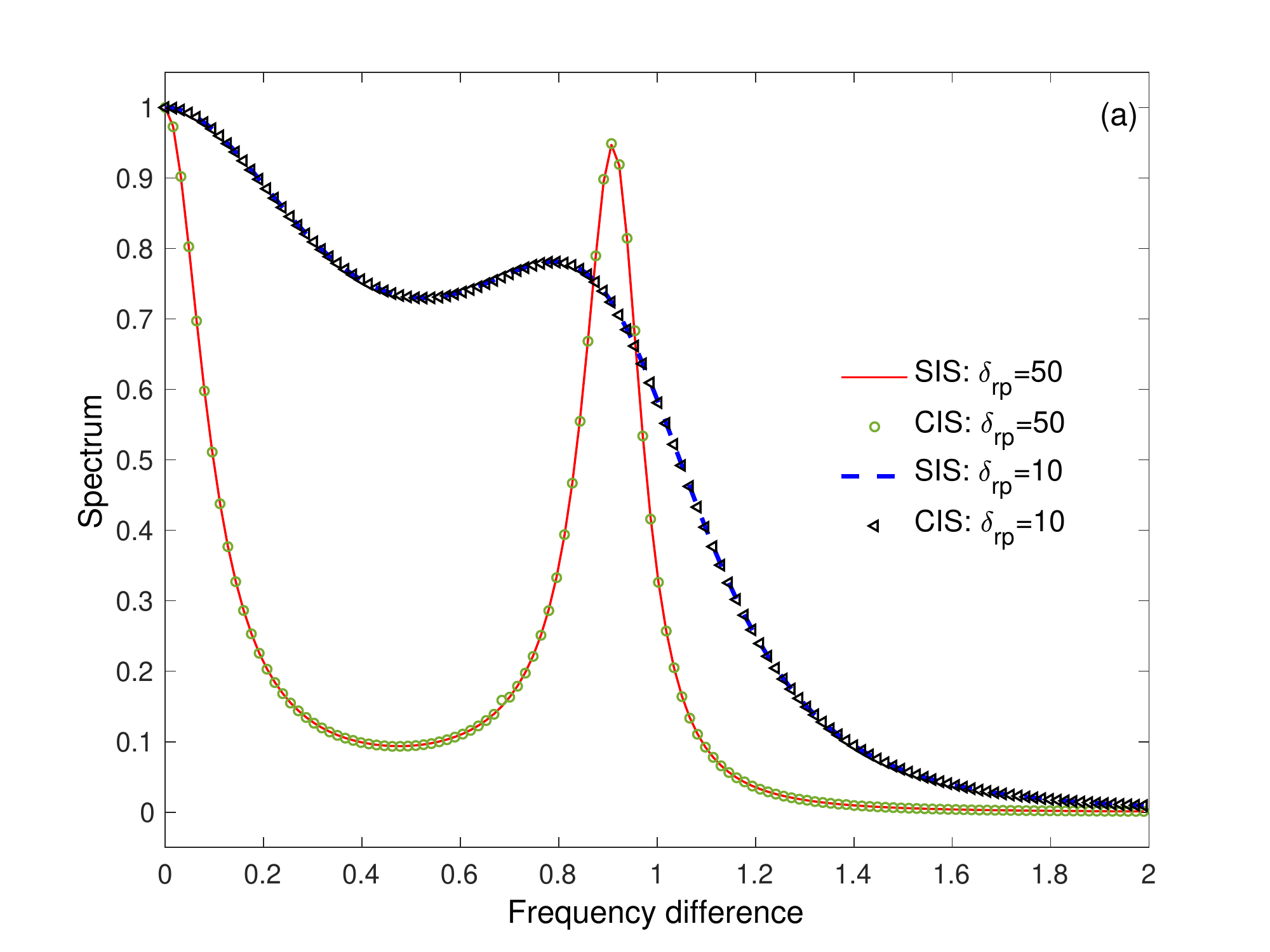}
	\includegraphics[scale=0.4,viewport=20 0 540 420,clip=true]{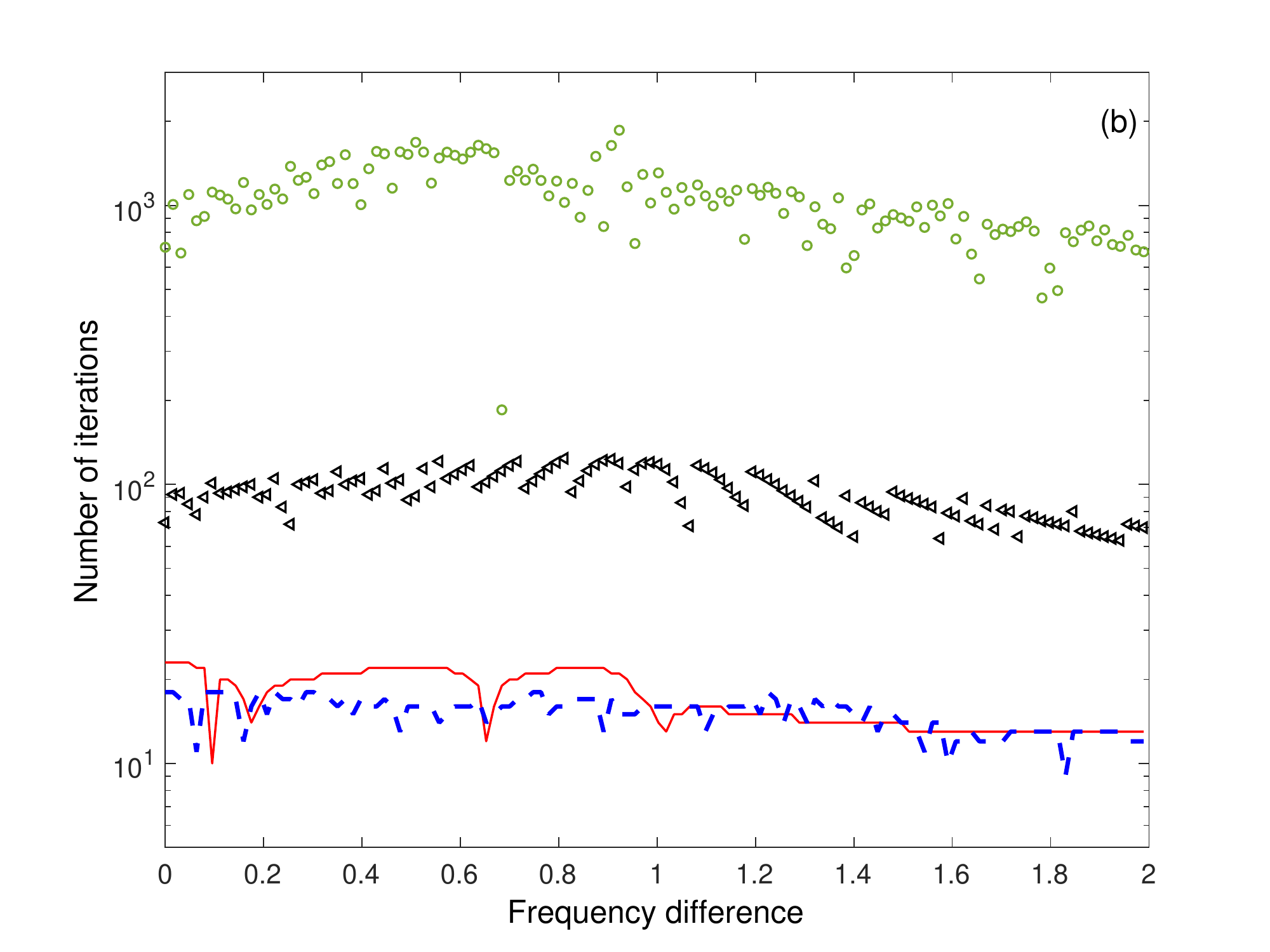}
	\caption{Comparisons of the SRBS spectrum (a) and iteration numbers (b) between the CIS and GSIS when the rarefaction parameter is large. The HS molecular model is used in the LBE. The molecular velocity space $[-6,6]^3$ is discretized by $24\times32\times24$ uniformly distributed points. The solutions are believed to be converged when the relative error in $\hat{\rho}$ between two consecutive iteration steps is less than $10^{-7}$.
	}
	\label{fig:srbs_profile}
\end{figure}

Note that in Eq.~\eqref{SRBS} and~\eqref{SRBS_LBE},   the rarefaction parameter $\delta_{rp}$ is defined when the characteristic flow length $H$ is $\lambda_L/2\sin(\theta_s/2)$, with $\lambda_L$ being the wavelength of laser and $\theta_s$ the angle of light scattering, and $f_s (=\text{St}/2\pi)$ is the frequency shift in the scattering process normalized by the characteristic frequency $v_m/H$, and the hat denotes the Laplace-Fourier transform of the corresponding quantity. Also note that terms in the left-hand-side of Eq.~\eqref{SRBS_LBE} appear because operators $\partial/\partial {t}$ and $\partial/\partial {x_2}$ in Eq.~\eqref{LBE} are replaced by $2i\pi{f_s}$ and $-2i\pi$, respectively. Finally, the source term $f_{eq}$ in Eq.~\eqref{SRBS_LBE} is from the Laplace transform of the initial density impulse. This term will change the first equation in Eq.~\eqref{eq123} to $\frac{\partial {\rho}}{\partial{t}}+\frac{\partial {U_i}}{\partial{x_i}}=1$, while other synthetic equations remain unchanged.

In CIS, the velocity distribution function
is  obtained by solving the following equation iteratively:
\begin{equation}\label{iteration_SRBS}
\hat{h}^{(k+1)}(\bm{v})=\frac{{L}^+(\hat{h}^{(k)})+{f_{eq}(\bm{v})} } {2\pi{}i(f_s-v_2)+\nu_{eq}(\bm{v}) },
\end{equation}
which converges fast when $\delta_{rp}$ is small, but extremely slow when $\delta_{rp}$ is large as the flow enters the near-continuum regimes. 

In the GSIS, the synthetic equations can be obtained by solving the following matrix at the $(k+1)$-th iteration step:
\begin{equation}\label{Grad_sp}
\left[ \begin {array}{cccccc} 
2i\pi{f_s}& -2i\pi& 0 &0 &0
\\ \noalign{\medskip}
-2i\pi &2i\pi{f_s} &-2i\pi &-2i\pi &0
\\ \noalign{\medskip}
0&-2i\pi& 3i\pi{f_s} &0 &-2i\pi
\\ \noalign{\medskip}
0  &-\frac{8}{3}i\pi  & 0 &2i\pi{f_s}+\delta_{rp} &0
\\ \noalign{\medskip}
0  & 0 & -3i\pi{C_q} &0  &2i\pi{f_s}+\frac{2}{3}\delta_{rp}
\end {array} \right]
\left[ \begin {array}{cccccc} \hat{n}^{(k+1)} \\ \hat{U}_2^{(k+1)}
\\ \hat{T}^{(k+1)}  \\ \hat\sigma_{22}^{(k+1)} \\ \hat{q}_2^{(k+1)}
\end {array} \right] =\left[ \begin {array}{cccccc} 1 \\ 0
\\ 0  \\ \text{R}_4 \\ \text{R}_5
\end {array} \right], 
\end{equation}
where $\text{R}_4= 2i\pi{}\text{HoT}_{\sigma_{22}}^{(k+1/2)}+2\int{}(L-L_s)v_2^2\mathrm{d}\bm{v}$ and $\text{R}_5=2i\pi{}\text{HoT}_{q_2}^{(k+1/2)}+\int{}(L-L_s)v_2|\bm{v}|^2\mathrm{d}\bm{v}$.

In the numerical simulation, started from the zero perturbance at each frequency difference, solutions are believed to be converged when the relative error in $\hat{\rho}$ between two consecutive iteration steps is less than $10^{-7}$. Results in Fig.~\ref{fig:srbs_profile}(a) show that GSIS and CIS generate almost the same SRBS spectra, except at $\delta_{rp}=50$ the CIS has a false converged solution (i.e. the discontinuous spectrum) when the frequency difference is around 0.68. As usual, the iteration number in CIS increases significantly with the rarefaction parameter $\delta_{rp}$, while in GSIS this remains nearly unchanged and is far less than that of the CIS. For example, the iteration number of the GSIS is about 10 and 100 times less than that of the CIS when $\delta_{rp}=10$ and 50, respectively.  We have also tested that, even when $\delta_{rp}=500$, converged solutions are obtained within 20 steps in the GSIS for every frequency difference. 

\begin{figure}[t]
	\centering
	\includegraphics[scale=0.6,viewport=20 0 680 390,clip=true]{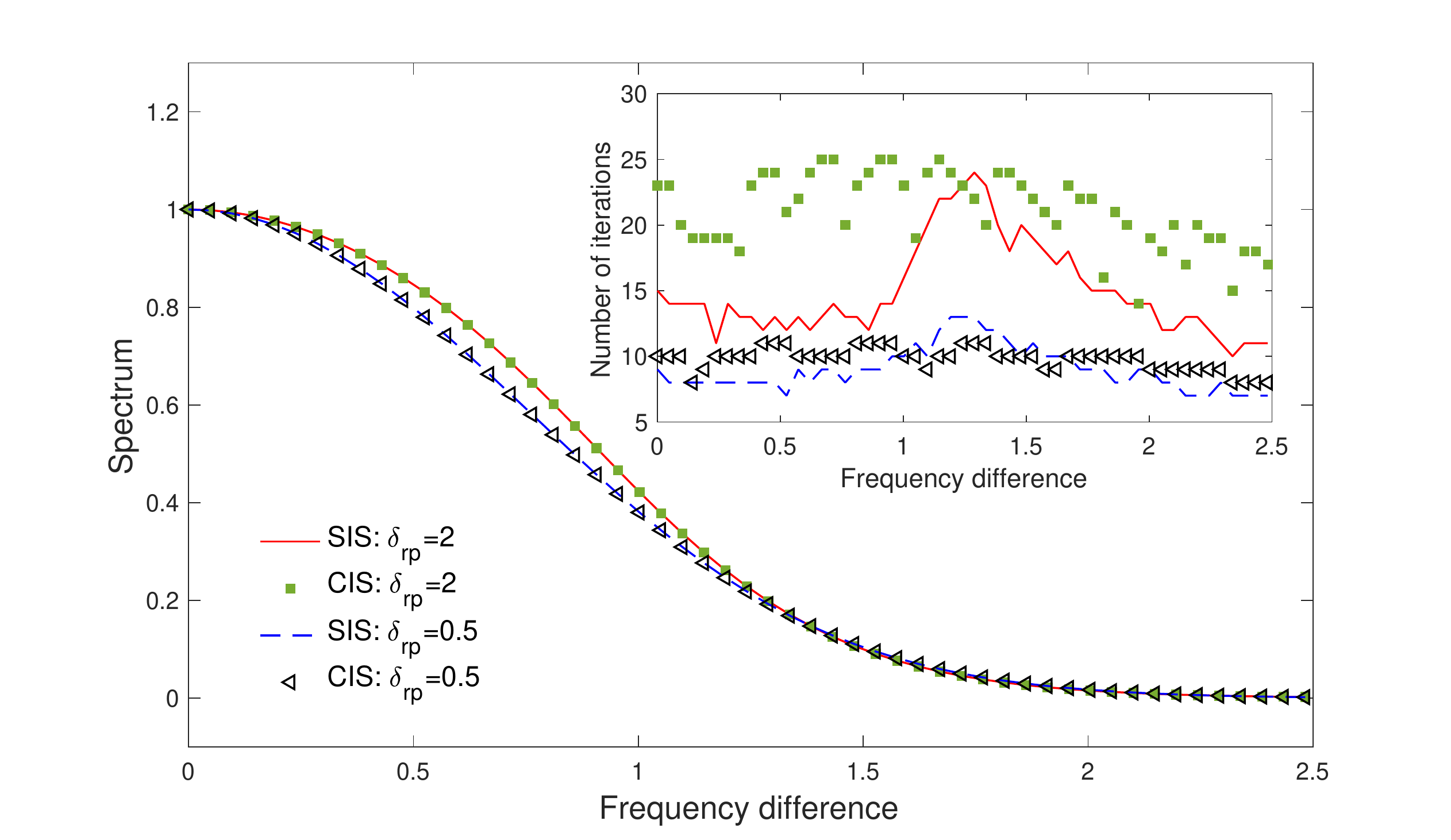}
	\caption{Comparisons of the SRBS spectrum and iteration numbers (inset) between the CIS and GSIS when the rarefaction parameter is small. The Maxwell molecular model is used in the LBE. The molecular velocity space $[-6,6]^3$ is discretized by $24\times192\times24$ uniformly distributed points due to high rarefaction effects. The solutions are believed to be converged when the relative error in $\hat{\rho}$ between two consecutive iteration steps is less than $10^{-7}$.
	}
	\label{fig:srbs_profile2}
\end{figure}

However, when $\delta_{rp}$ is small and $\mathrm{St}$ is large, the GSIS does not converge or even blows up. This is because the eigenvalue of the matrix in Eq.~\eqref{Grad_sp} has large complex values so that any inappropriate initial guess lead to large oscillations that decay rather slow or even blow up. Whereas, physically speaking, the solution should decay fast due to the large rarefaction effect. To remedy this, the small value $\delta_{rp}$ in the left-hand side of Eq.~\eqref{Grad_sp} is replaced by a relative large value $\bar{\delta}_{rp}=\max(\delta_{rp},10)$, while the right-hand side terms are modified correspondingly as
\begin{equation}
\begin{aligned}[b]
\text{R}_4=& 2i\pi{}\text{HoT}_{\sigma_{22}}^{(k+1/2)}+2\int{}(L-L_s)v_2^2\mathrm{d}\bm{v}+(\bar{\delta}_{rp}-\delta_{rp})\hat{\sigma}_{22}^{(k+1/2)}, \\  \text{R}_5=&
2i\pi{}\text{HoT}_{q_2}^{(k+1/2)}+\int{}(L-L_s)v_2|\bm{v}|^2\mathrm{d}\bm{v}+\frac{2}{3}(\bar{\delta}_{rp}-\delta_{rp})\hat{q}_{2}^{(k+1/2)}.
\end{aligned}
\end{equation}

This simple treatment helps to decay non-physical solutions at initial few iteration steps. When the solution of the new system converges, it can be proven that it satisfies Eq.~\eqref{Grad_sp}. Therefore, no approximation is introduced to the converged solution. This point is proven in Fig.~\ref{fig:srbs_profile2}, where the GSIS and CIS solutions agree perfectly with each other, and from the inset we see that the GSIS needs slightly less iteration steps than CIS in most of frequency differences.

\begin{figure}[t]
	\centering
	\includegraphics[scale=0.44,viewport=20 0 540 420,clip=true]{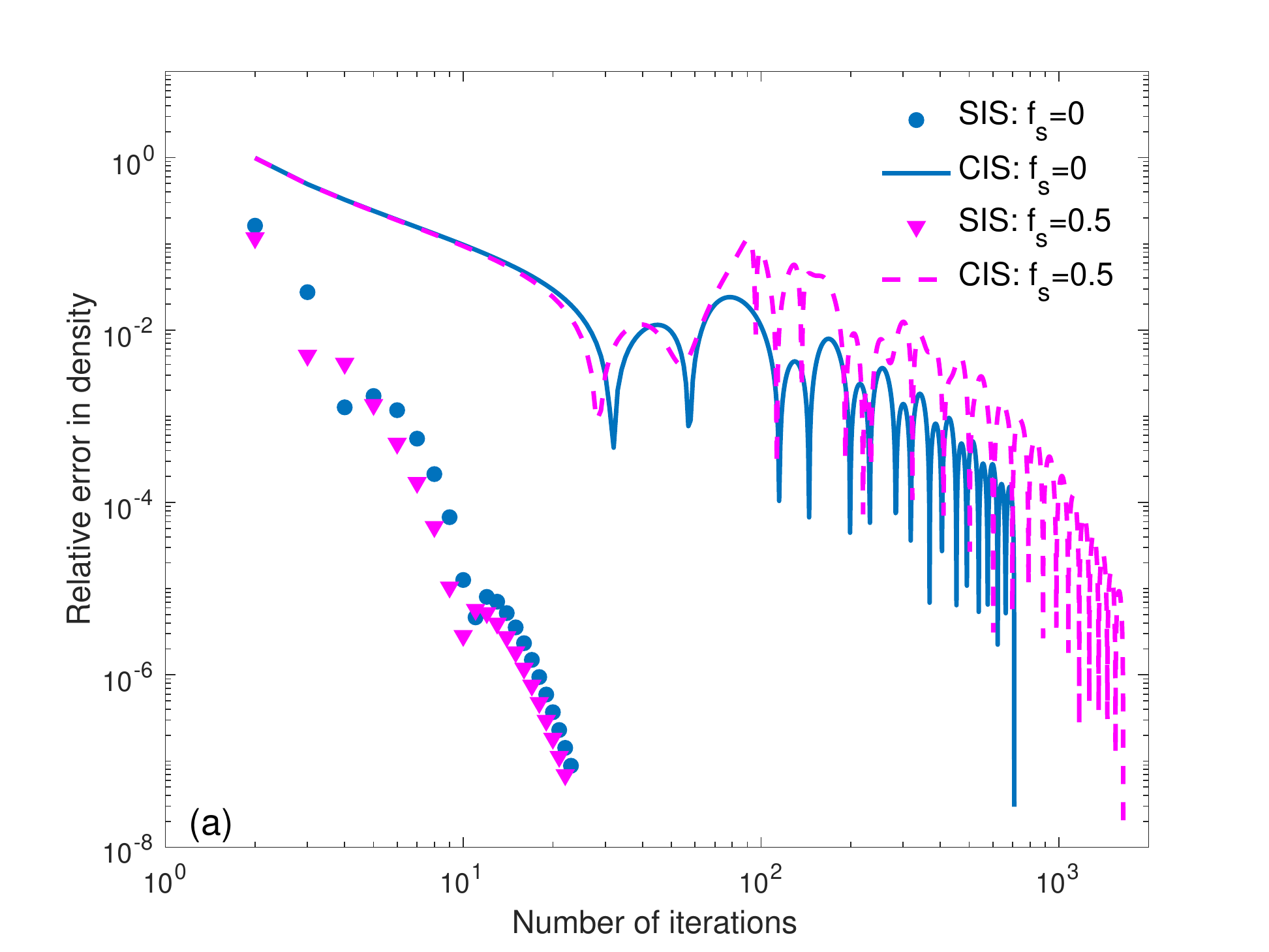}
	\includegraphics[scale=0.44,viewport=20 0 540 420,clip=true]{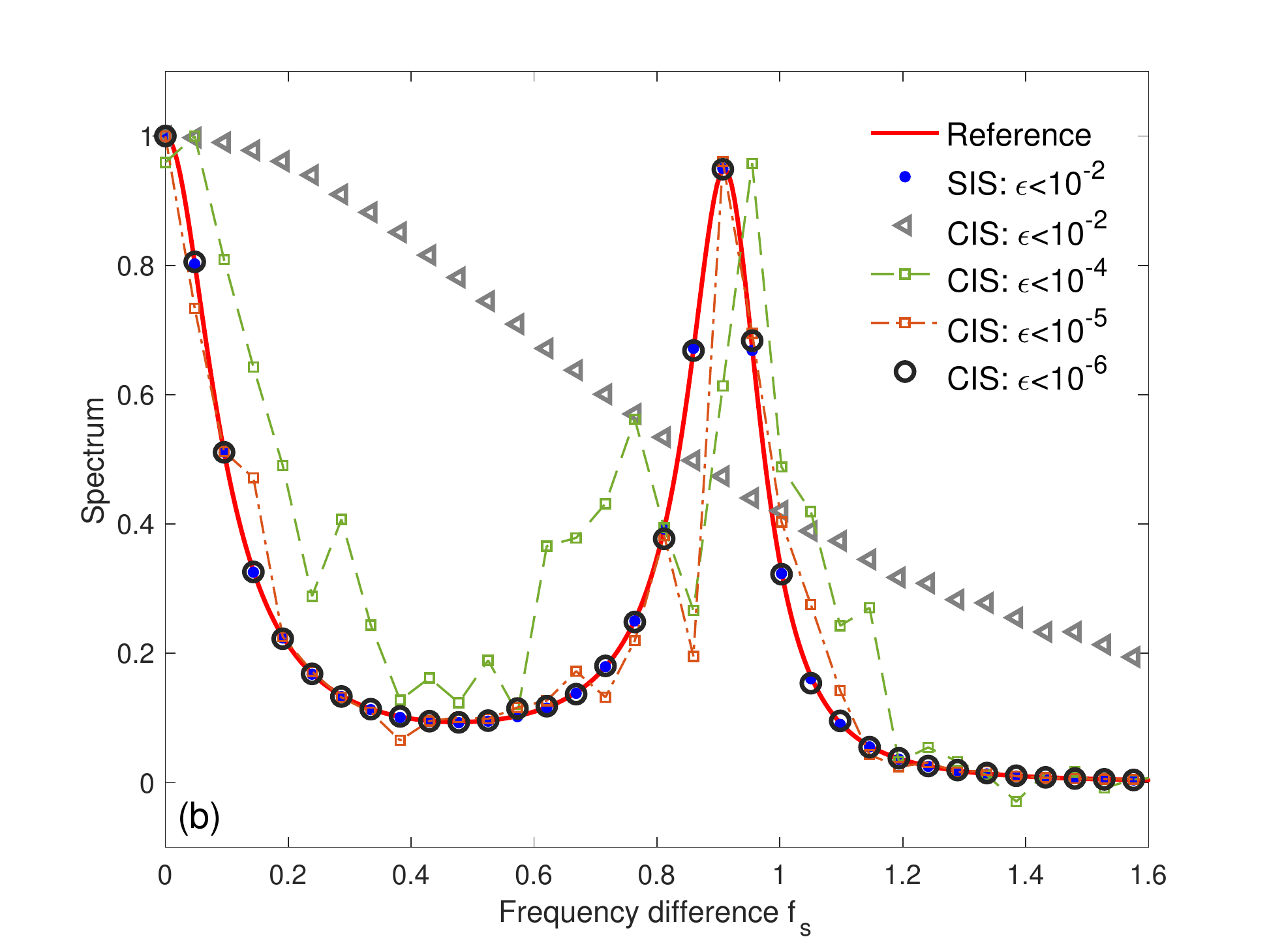}
	\caption{ (a) The decay of the relative error $\epsilon=|\hat{\rho}^{(k+1)}/\hat{\rho}^{(k)}-1|$ between two consecutive iteration steps, and (b) the SRBS spectra obtained at different level of convergence criterion. The reference solution is obtained from the GSIS when $\epsilon=10^{-7}$. The linearized Boltzmann equation with HS molecular model is used, with the rarefaction parameter $\delta_{rp}=50$.}
	\label{fig:srbs_error}
\end{figure}

Another remarkable property of the GSIS is that, at the same level of convergence criterion, the GSIS provides more accurate numerical solutions. One example is given in Fig.~\ref{fig:srbs_error}, where one can see that the relative error between two consecutive iteration steps  $\epsilon=|\hat{\rho}^{(k+1)}/\hat{\rho}^{(k)}-1|$ decays rather fast in the GSIS, while in the CIS it decreases slowly with many oscillations. As a consequence, the GSIS finds the correct spectrum profile even when the relative error in density is less $10^{-2}$, while the CIS can only reach the correct solution when the error is less than $10^{-6}$. This can be explained below. According to the analysis of Adam and Larsen for radiation transfer problem~\cite{DSA2002},  if one stops at the $(k+1)$-th step with
\begin{equation}
 \left|\frac{\hat{\rho}^{(k+1)}}{\hat{\rho}^{(k)}}-1\right|=\epsilon
 \end{equation}
  in the CIS, then the relative difference from the true solution $\hat{\rho}$ is
  \begin{equation}
  \left|\frac{\hat{\rho}}{\hat{\rho}^{(k+1)}}-1\right|\approx\frac{\gamma}{1-\gamma}\epsilon, 
  \end{equation} 
  where $\gamma$ is the spectral radius of the iteration operator. For problem with slow convergence, $\gamma$ is very close to one (see Figure 1 in Ref.~\cite{LeiJCP2017} for the kinetic BGK model equation), which could make the difference from true solution magnified by thousands of times.

\subsection{Oscillatory Couette flow between two parallel plates}\label{Oscillatory_Couette}

Consider the rarefied gas dynamics between two infinite parallel plates with a distance $H$, located at $x_2=0$ and $x_2=1$. Both plates have a temperature $T_0$, the one at $x_2=1$ is stationary, while that at $x_2=0$ oscillating in the $x_1$ direction with the velocity
\begin{equation}
U_{w,1}=\Re\left[U_0\exp({i\mathrm{St}t})\right].
\end{equation}

The Boltzmann equation is linearized by choosing $\alpha=U_0/v_m$ in Eq.~\eqref{LBE_osci}. If we consider the diffuse boundary condition, then we have $h(x_2=0,\bm{v})= 2v_1 f_{eq}$ when  $v_2>0$, and $h(x_2=1,\bm{v})=0$ when $v_2<0$~\cite{Sharipov2008Couette}. The synthetic equations~\eqref{eq123}, \eqref{HoT_sigma}, and~\eqref{HoT_q} can be simplified to 
\begin{equation}\label{Couette2}
	\begin{aligned}[b]
	2i\mathrm{St}U_1+\frac{\partial\sigma_{12}}{\partial {x_2}}=0, \\
	i\mathrm{St}\sigma_{12}+\text{HoT}_{\sigma_{12}}+\frac{\partial U_1}{\partial {x_2}}=-\delta_{rp}\sigma_{12}+2\int{(L-L_s)}v_1v_2\mathrm{d}\bm{v},
	\end{aligned}
\end{equation}
where the moments involving even order of $v_1$ are all zero, and we do not consider the heat flux $q_1$ in this problem as it does not affect the rate of convergence. It is noted that the above equations reduce to the synthetic equation developed in Ref.~\cite{Su2019JCP2} when $\text{St}=0$.

The two equations in Eq.~\eqref{Couette2} can be combined to produce the following diffusion equation for the flow velocity $U_1$ in the $(k+1)$-th iteration step:
\begin{equation}\label{diffusion_Couette0}
2i\mathrm{St}(i\mathrm{St}+\delta_{rp})U_1^{(k+1)}-\frac{\partial^2U_{1}^{(k+1)}}{\partial {x_2}^2}=\frac{\partial}{\partial{x_2}}\left[2\int{(L^{(k)}-L_s^{(k)})}v_1v_2\mathrm{d}\bm{v}-\text{HoT}_{\sigma_{12}}^{(k)}\right].
\end{equation}

In the numerical simulation, the spatial space is discretized by Eq.~\eqref{Couette_spatial_grid}
%as \begin{equation}
%x_2=(10-15s+6s^2)s^3, \quad s=(0,1,\cdots,N_s-1)/(N_s-1)
%\end{equation}
with $N_s=100$. The kinetic equation~\eqref{syn_LBE0} is solved by the second-order upwind scheme, while the derivative in Eq.~\eqref{diffusion_Couette0} is approximated by the central finite difference scheme with 5 stencils, and the resulting linear algebraic system for $U_1$ is solved exactly in the bulk region (i.e. at least three spatial points away from the boundary).

\begin{figure}[t]
	\centering
	\includegraphics[scale=0.6,viewport=10 0 510 400,clip=true]{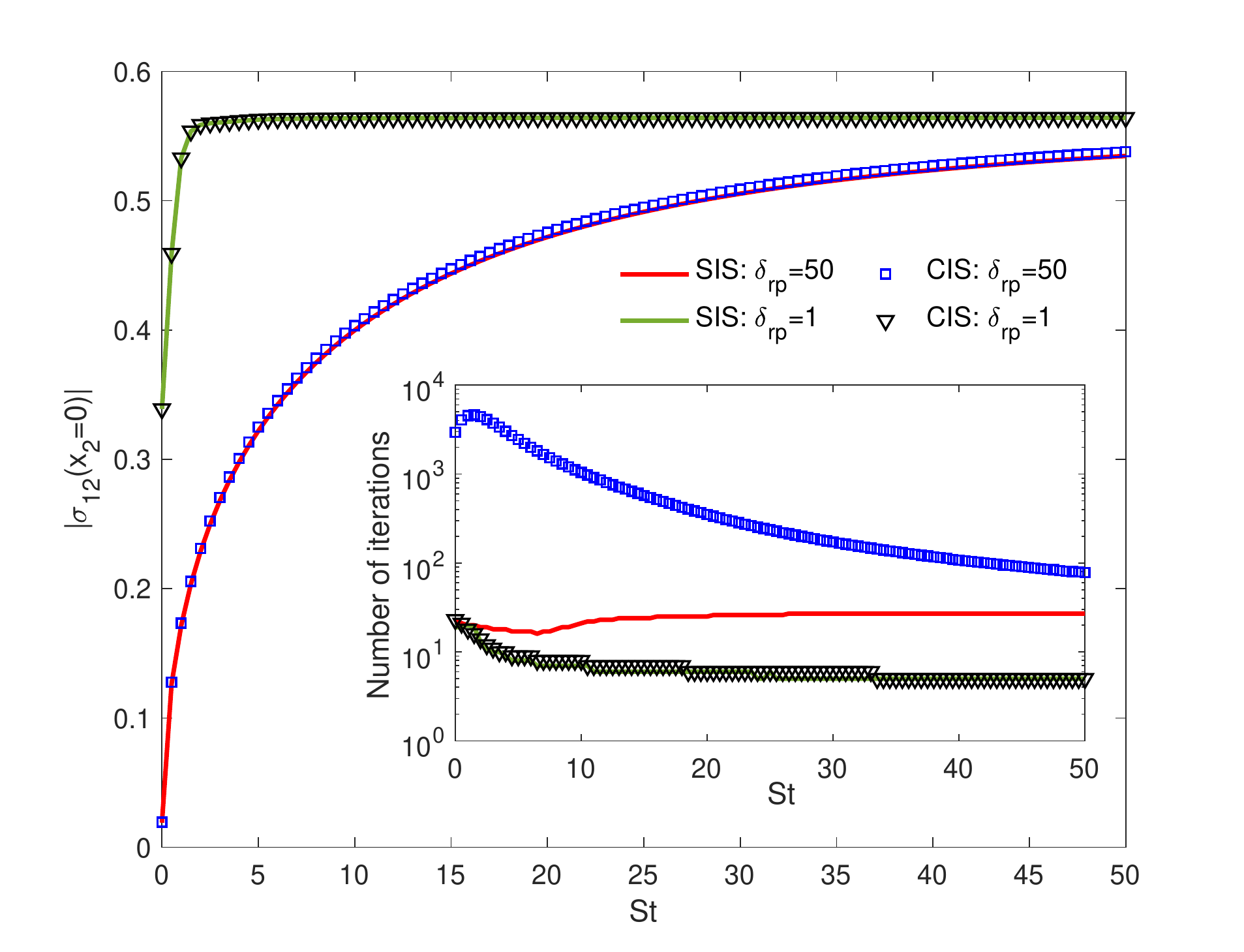}
	\caption{Comparisons of the amplitude of the shear stress exerting on the oscillating plate and iteration numbers (inset) between the CIS and GSIS, for the Oscillating Couette flow. The Shakhov model is solved, where the solution is converged when  $\int{}\left|\frac{U_1^{(k+1)}}{U_1^{(k)}}-1\right|\mathrm{d}x_2<10^{-5}$.  
	}
	\label{fig:Couette_Shear}
\end{figure}
%with the two boundary conditions $U_1^{(k+1)}(x_2=0)=U_1^{(k+1/2)}(x_2=0)$ and $U_1^{(k+1)}(x_2=1)=U_1^{(k+1/2)}(x_2=1)$

The comparison in the accuracy and efficiency between the CIS and GSIS is summarized in Fig.~\ref{fig:Couette_Shear}, where the molecular velocity space is discretized in the same way as that in Sec.~\ref{sec:results1}, but with $N_v=96$ in Eq.~\eqref{nonuniform_v}. The relative difference in the amplitude of the shear stress $\sigma_{12}$ is within 1\%. When the rarefaction parameter is $\delta_{rp}=50$, we see that the number of iterations in the CIS decreases from 30,000 to 100 when Strouhal number increases from 0 to 50. The reason for this reduction can be understood in the following way. The temporal Knudsen number $\text{Kn}_t$, which is defined as the ratio of characteristic oscillation frequency to the mean collision frequency of gas molecules, i.e.
\begin{equation}
\text{Kn}_t=\frac{\varpi}{v_m/\lambda}=\frac{\text{St}}{\delta_{rp}},
\end{equation}
increases with $\text{St}$. Therefore, even when $\delta_{rp}$ is large, that is, when the spatial Knudsen number is small, the large temporal Knudsen number can also make the flow rarefied, and the more rarefied the gas is, the fast the iteration to the steady-state. Even with this effect, the GSIS is still faster than the CIS: only about 20 iterations are needed in the GSIS for each Strouhal number considered.

However, for the GSIS in oscillating problems, there is a problem, like the one encountered in Sec.\ref{sec:srbs}. From Eq.~\eqref{diffusion_Couette0} we see that the eigenvalue of this second-order differential equation is imaginary, which means that when $\delta_{rp}$ is small and $\text{St}$ is large, the solution will change quasi-periodically in the spatial direction with large frequency, whereas physically the solution should decay fast from the oscillating sources as the dissipation is huge due to the large values of both spatial and temporal Knudsen numbers. Mathematically speaking, for highly oscillating solutions, any slight inaccurate boundary conditions will lead to completely different solutions. Therefore, in the numerical simulation, when we solve Eq.~\eqref{diffusion_Couette0} directly, the solution is either wrong or blows up. To fix this problem, again we introduce a relative large value of $\bar{\delta}_{rp}$ to decay the fast oscillation. That is, instead of solving Eq.~\eqref{diffusion_Couette0}, we solve the following diffuse-type equation:
	\begin{equation}\label{diffusion_Couette}
	\begin{aligned}[b]
	2i\mathrm{St}(i\mathrm{St}+\bar{\delta}_{rp})U_1^{(k+1)}-\frac{\partial^2U_{1}^{(k+1)}}{\partial {x_2}^2}=&\frac{\partial}{\partial{x_2}}\left[2\int{(L^{(k)}-L_s^{(k)})}v_1v_2\mathrm{d}\bm{v}-\text{HoT}_{\sigma_{12}}^{(k)}\right] \\
	&+2i\mathrm{St}(\bar{\delta}_{rp}-\delta_{rp})U_1^{(k+1/2)},
	\end{aligned}
	\end{equation}
where 
\begin{equation}\label{delta_bar}
\bar{\delta}_{rp}=\max(\delta_{rp},\text{St}).
\end{equation}  

It can be proven that, when the solution of Eq.~\eqref{diffusion_Couette} converges, Eqs.~\eqref{diffusion_Couette} and~\eqref{diffusion_Couette0} are equivalent. This treatment does not affect the accuracy and efficiency of the GSIS when $\delta_{rp}$ is small, while when $\delta_{rp}$ is large, the solution from the synthetic equations are always stable, and we see in Fig.~\ref{fig:Couette_Shear} that in most cases the GSIS needs slightly less iterations than CIS.

\begin{figure}[t]
	\centering
	\includegraphics[scale=0.44,viewport=10 0 520 400,clip=true]{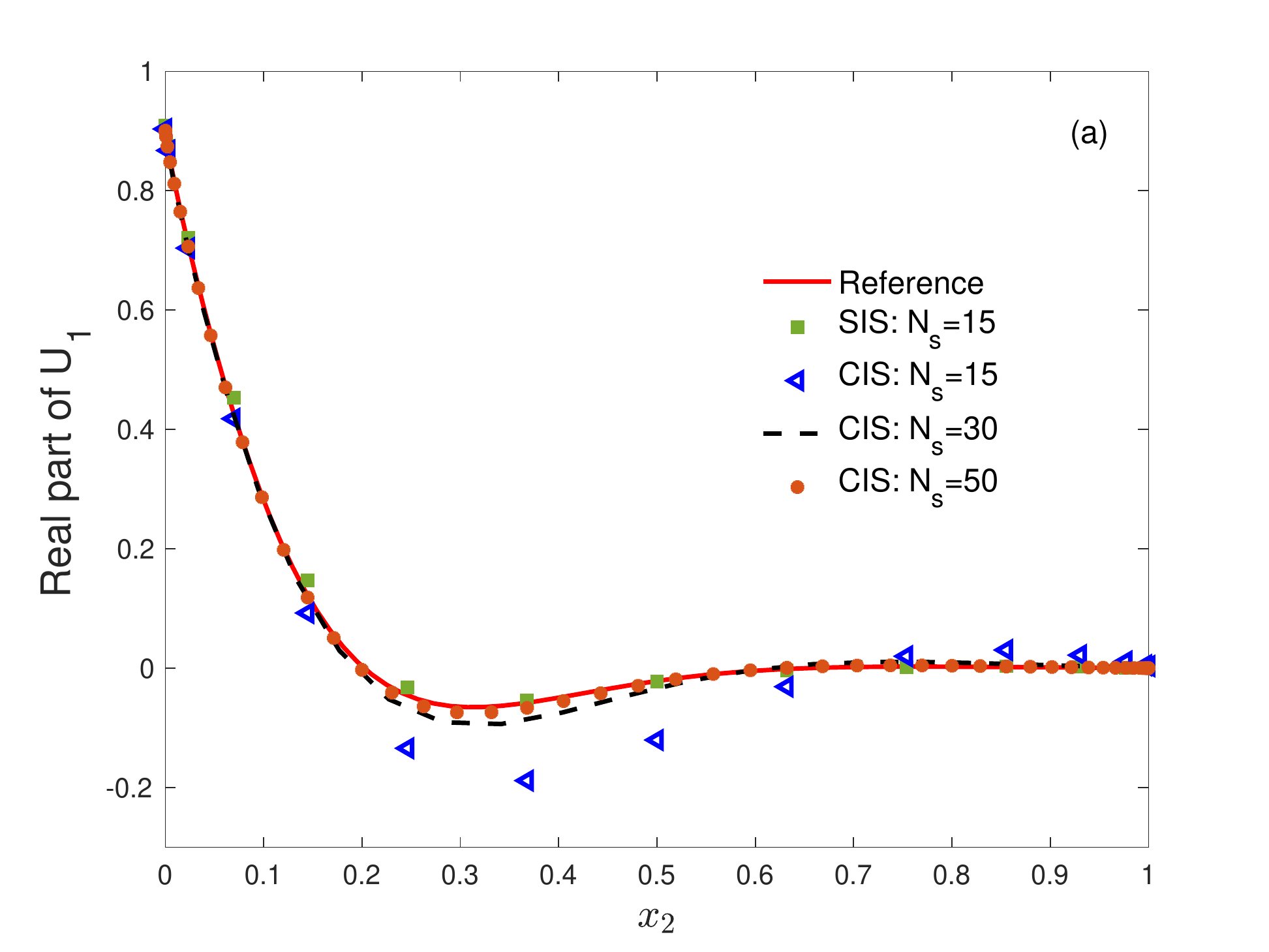}
	\includegraphics[scale=0.44,viewport=10 0 520 400,clip=true]{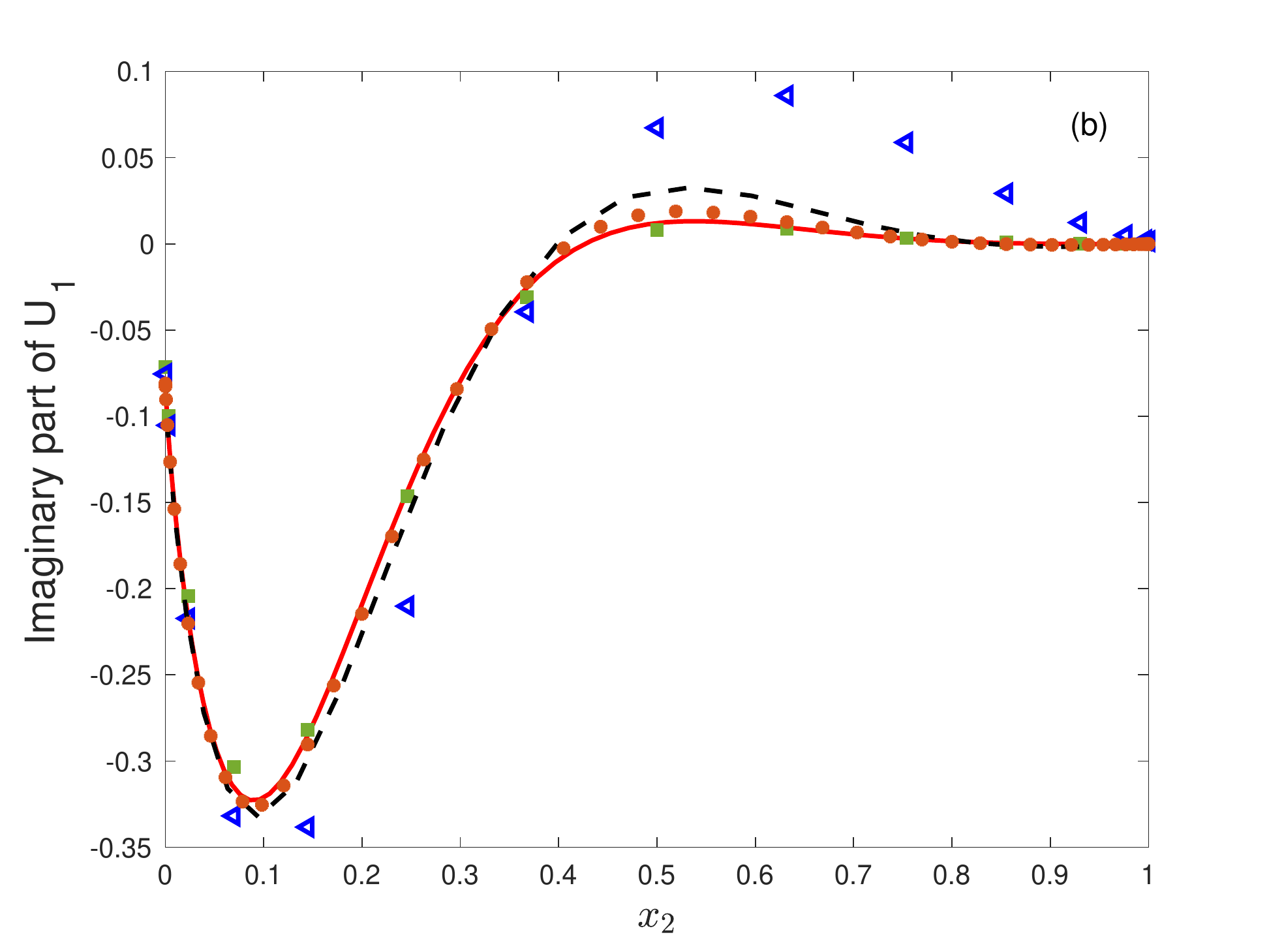}\\
	\includegraphics[scale=0.44,viewport=10 0 520 400,clip=true]{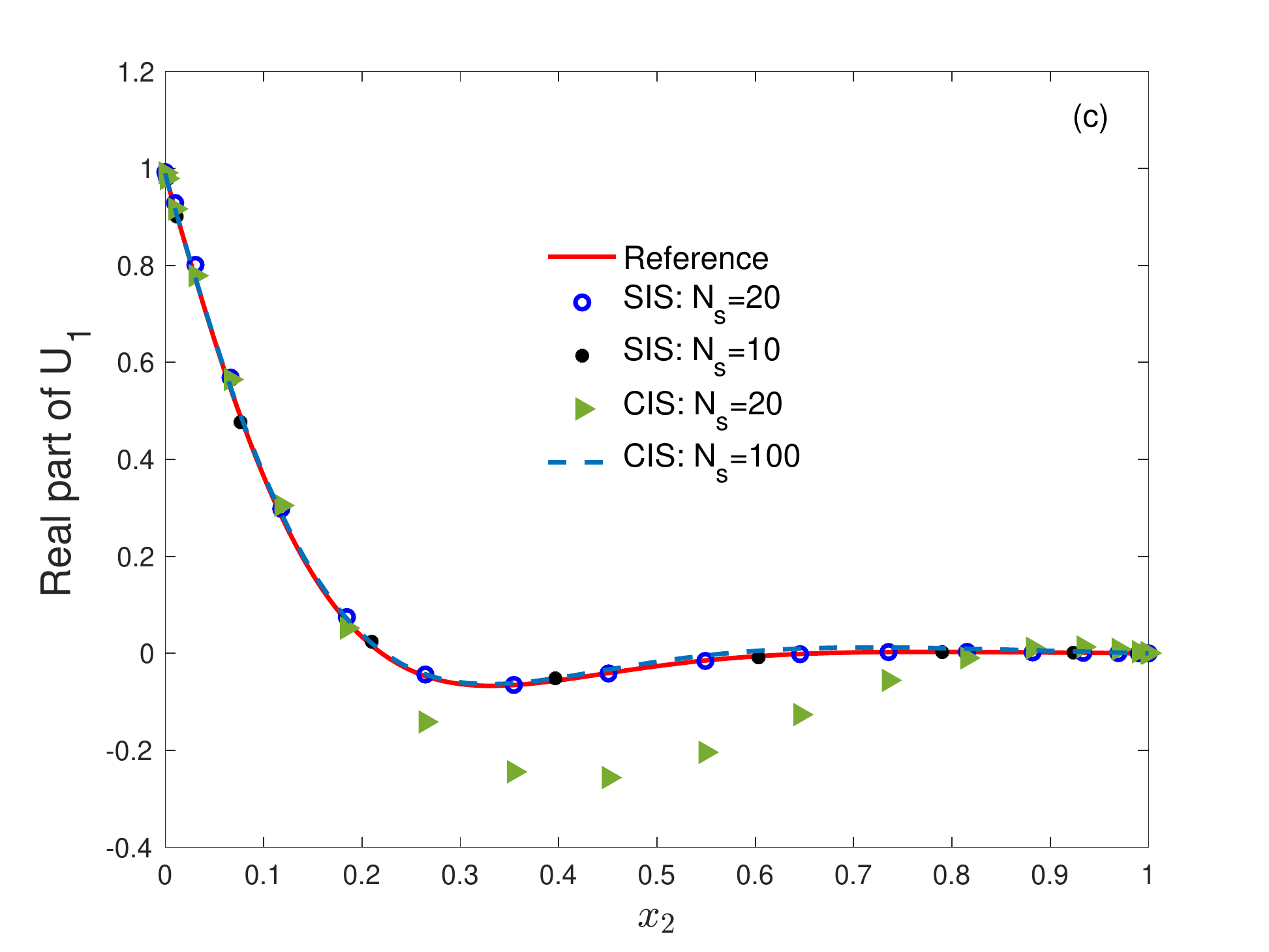}
	\includegraphics[scale=0.44,viewport=10 0 520 400,clip=true]{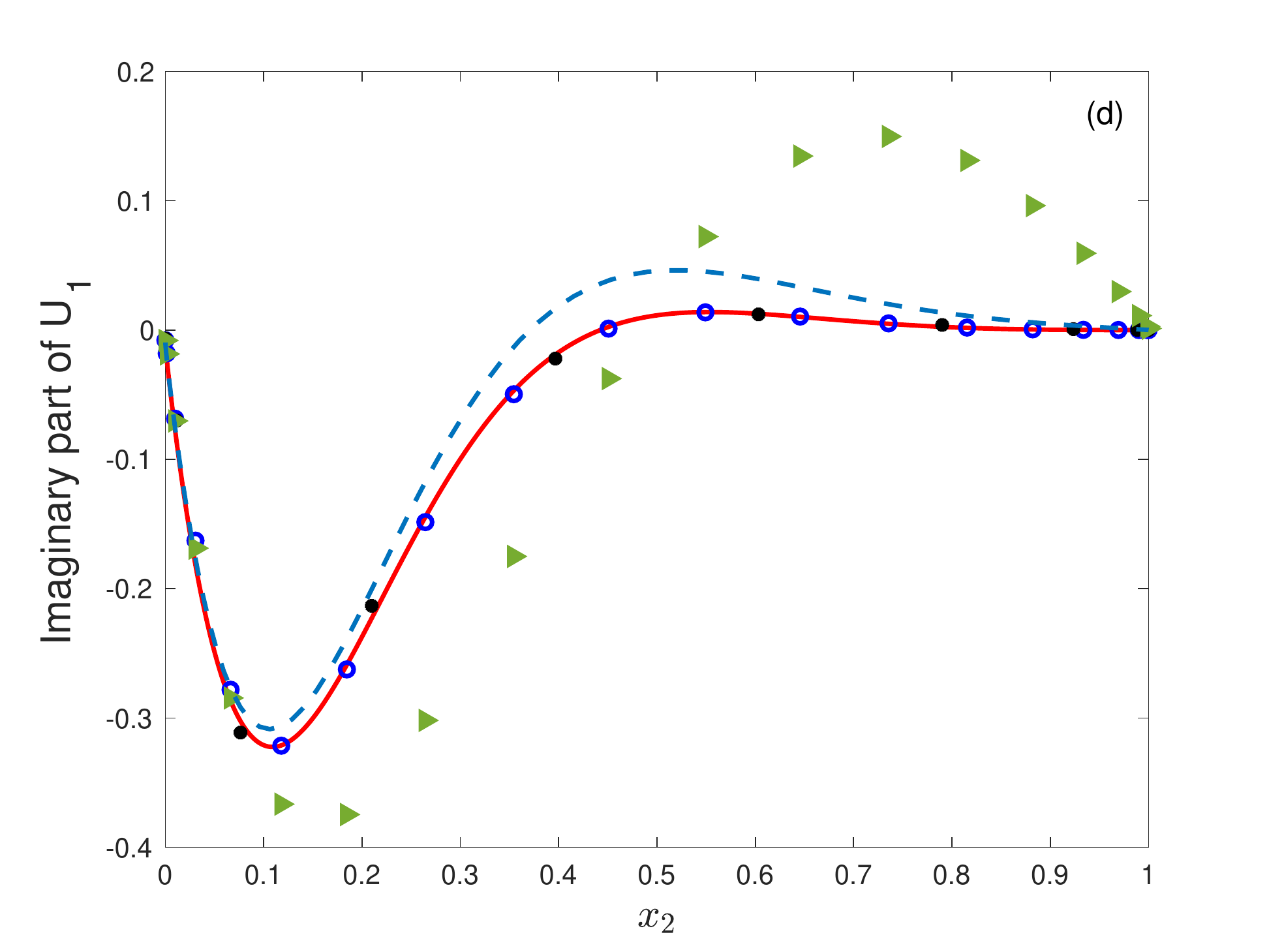}
	\caption{Comparisons of the velocity profiles in the oscillating Couette flow with different spatial discretizations, when (a, b) $\delta_{rp}=50$ and $\text{St}=1$ and (c, d) $\delta_{rp}=500$ and $\text{St}=0.1$. The reference solution is obtained from the GSIS, where the spatial domain is discretized by Eq.~\eqref{Couette_spatial_grid}, with $N_s=100$ when $\delta_{rp}$=50 and $N_s=500$ when $\delta_{rp}=500$. } % the corresponding CIS results overlap with these lines and are not shown here. }
	\label{fig:Couette_velocity}
\end{figure}

In addition to the significant reduction of iteration number, the GSIS needs less spatial grids than that of the CIS. Two examples are given in Fig.~\ref{fig:Couette_velocity}, where one can see that the GSIS can yield accurate results even when the cell sizes are respectively about 6.6 and 50 times of the molecular mean free path, while the CIS has large error due to the strong numerical dissipation.

\begin{figure}[t]
	\centering
	\includegraphics[scale=0.44,viewport=20 0 520 400,clip=true]{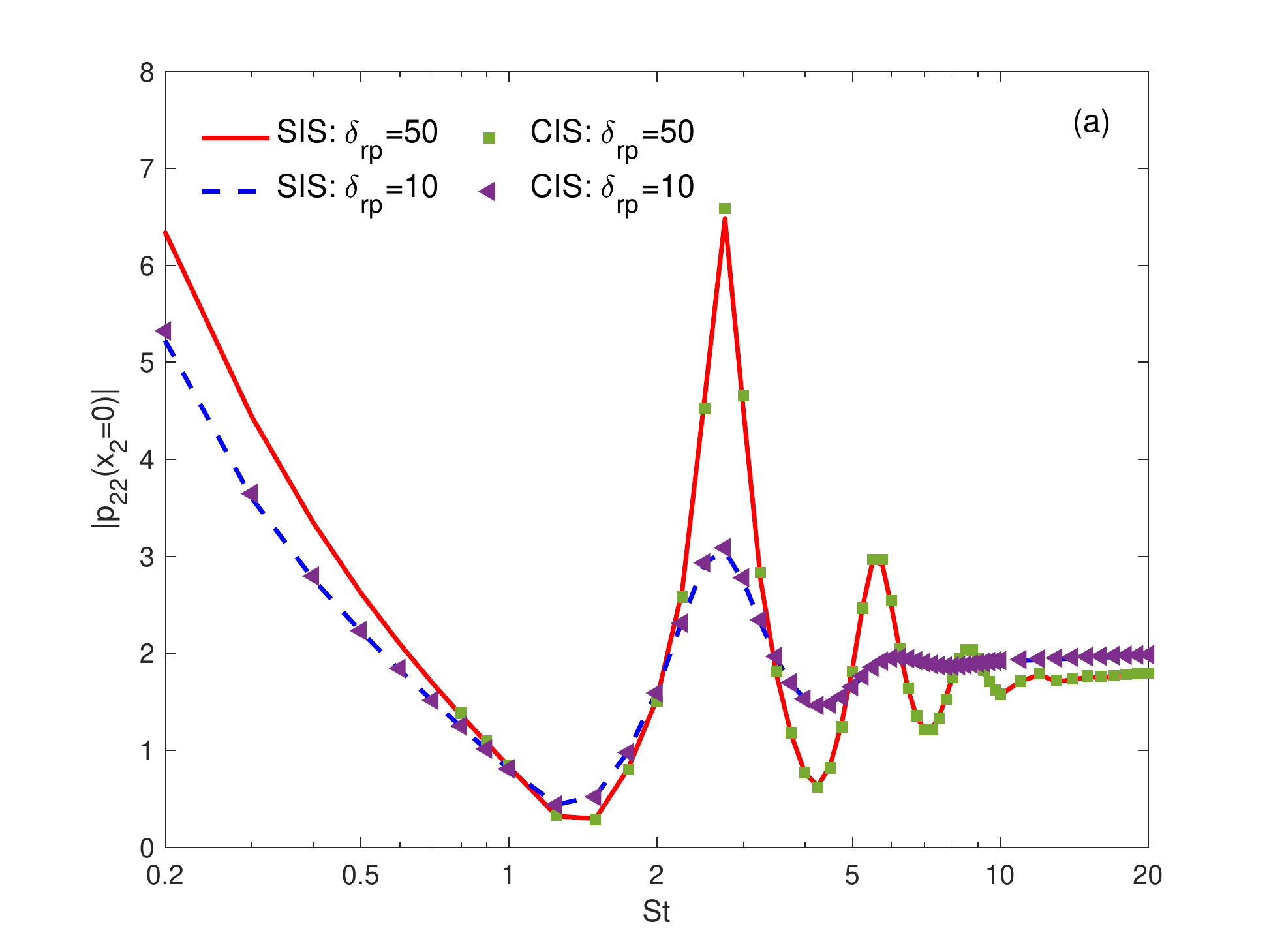}
	\includegraphics[scale=0.44,viewport=20 0 520 400,clip=true]{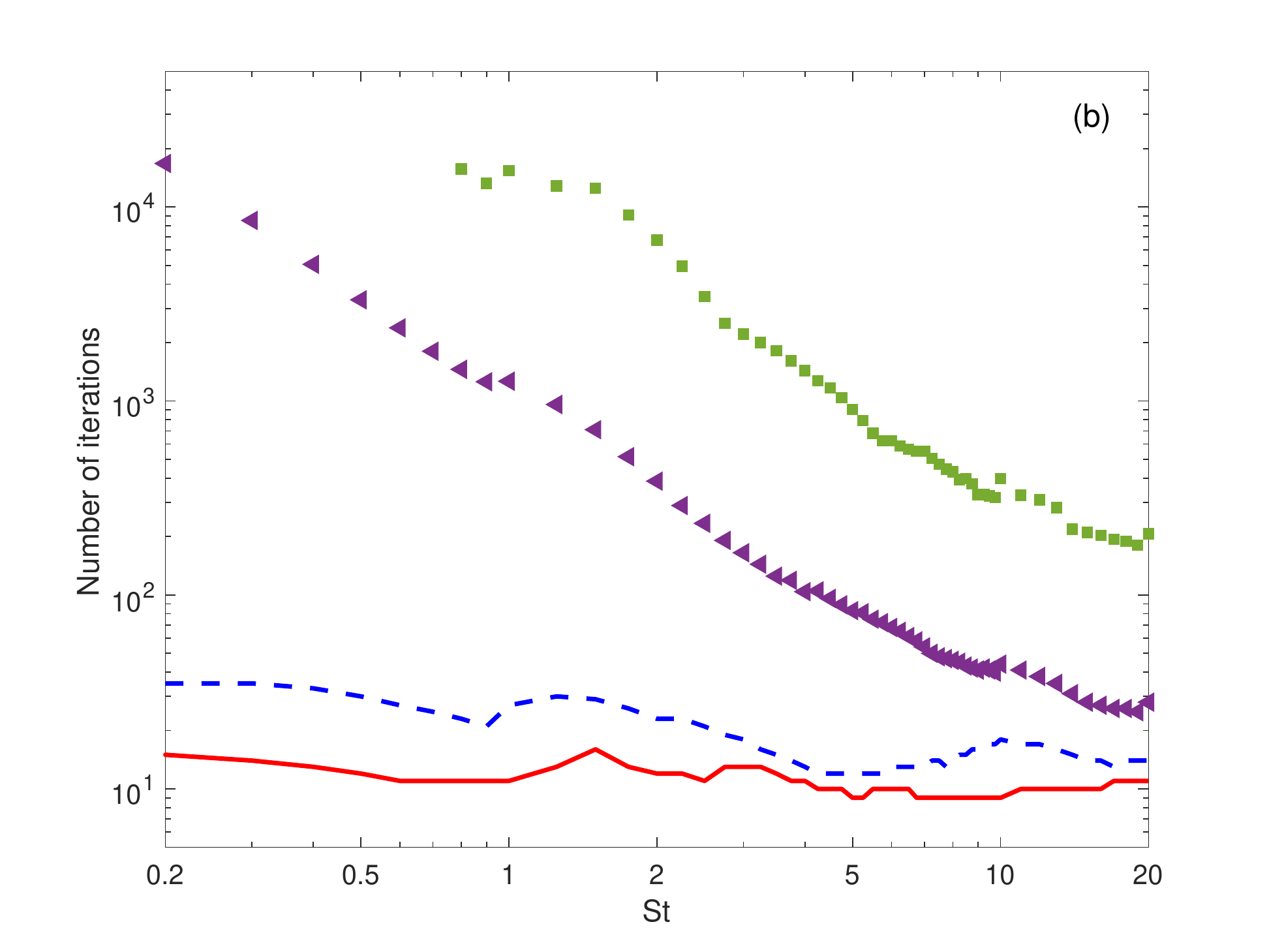}
	\caption{Comparisons of (a) the amplitude of normal pressure exerting on the oscillating plate and iteration numbers (b) between the CIS and GSIS, for the sound propagation problem. The Shakhov model is solved, where the solution is converged when  $\max\left\{
		\int{}\left|\frac{\rho^{(k+1)}}{\rho^{(k)}}-1\right|\mathrm{d}x_2, 
		\int{}\left|\frac{U_2^{(k+1)}}{U_2^{(k)}}-1\right|\mathrm{d}x_2,
		\int{}\left|\frac{T^{(k+1)}}{q^{(k)}}-1\right|\mathrm{d}x_2
		\right\}<10^{-5}$.   }
	\label{fig:sound}
\end{figure}

\subsection{Sound propagation between two parallel plates}

Consider the sound propagation through a gas between two infinite parallel plates with a distance $H$, located at $x_2=0$ and $x_2=1$. The two plates have a temperature $T_0$,   the one at $x_2=1$ is stationary, while that $x_2=0$ oscillating in the $x_2$ direction with the speed $U_{w,2}=\Re\left[U_0\exp({i\mathrm{St}t})\right]$. The Boltzmann equation is linearized by choosing $\alpha=U_0/v_m$ in Eq.~\eqref{LBE_osci}. The boundary conditions are~\cite{Kalempa2009Sound}
\begin{equation}
\begin{aligned}
h(x_2=0,\bm{v})=& \left[\sqrt{\pi}+2v_2-2\sqrt{\pi}\int_{v_2<0} v_2h(x_2=0,\bm{v})\mathrm{d}\bm{v}\right] f_{eq}, \ \  \text{when~}  {v_2>0}, \\
h(x_2=1,\bm{v})=& 2\sqrt{\pi}f_{eq}\int_{v_2<0} v_2h(x_2=1,\bm{v})\mathrm{d}\bm{v} , \ \  \text{when~}  {v_2<0}.
\end{aligned}
\end{equation}

The synthetic equations~\eqref{eq123}, \eqref{HoT_sigma}, and~\eqref{HoT_q} can be simplified to 
\begin{eqnarray}
i\text{St}\rho+\frac{\partial{U_2}}{\partial{x_2}}=0, \label{sound_n} \\
2i\text{St}U_2+\frac{\partial{\rho}}{\partial{x_2}}+\frac{\partial{T}}{\partial{x_2}}+\frac{\partial{\sigma_{22}}}{\partial{x_2}}=0, \\
\frac{3}{2}i\text{St}T+\frac{\partial{q_2}}{\partial{x_2}}+\frac{\partial{U_2}}{\partial{x_2}}=0,\\
i\text{St}\sigma_{22}+\text{HoT}_{\sigma_{22}}+\frac{4}{3}\frac{\partial U_2}{\partial {x_2}}=-\delta_{rp}\sigma_{22}+2\int{(L-L_s)}\left(v_2^2-\frac{|\bm{v}|^2}{3}\right)\mathrm{d}\bm{v}, \label{sound_sigma} \\
i\text{St}q_{2}+\text{HoT}_{q_2}+\frac{3C_q}{2}\frac{\partial T}{\partial {x_2}}=-\frac{2}{3}\delta_{rp}q_{2}+\int{(L-L_s)}v_2|\bm{v}|^2\mathrm{d}\bm{v}. \label{sound_q}
\end{eqnarray}

These synthetic equations can be combined to form two diffusion equations for the flow velocity $U_2$ and temperature $T$. To quickly decay the non-physical oscillations when $\delta_{rp}$ is small and $\text{St}$ is large, in numerical iterations we set
\begin{equation}
\begin{aligned}[b]
\delta_{rp}\sigma_{22}^{(k+1)}=&\bar{\delta}_{rp}\sigma_{22}^{(k+1)}+(\delta_{rp}-\bar{\delta}_{rp})\sigma_{22}^{(k+1/2)},\\
\delta_{rp}q_{2}^{(k+1)}=&\bar{\delta}_{rp}q_{2}^{(k+1)}+(\delta_{rp}-\bar{\delta}_{rp})q_{2}^{(k+1/2)},
\end{aligned}
\end{equation}
where $\bar{\delta}_{rp}$ is given in Eq.~\eqref{delta_bar}. When $U_2$ and $T$ are solved, the perturbed density, shear stress and heat flux can be solved from Eqs.~\eqref{sound_n}, \eqref{sound_sigma}, and~\eqref{sound_q}.

\begin{figure}[t]
	\centering
	\includegraphics[scale=0.5,viewport=50 20 810 590,clip=true]{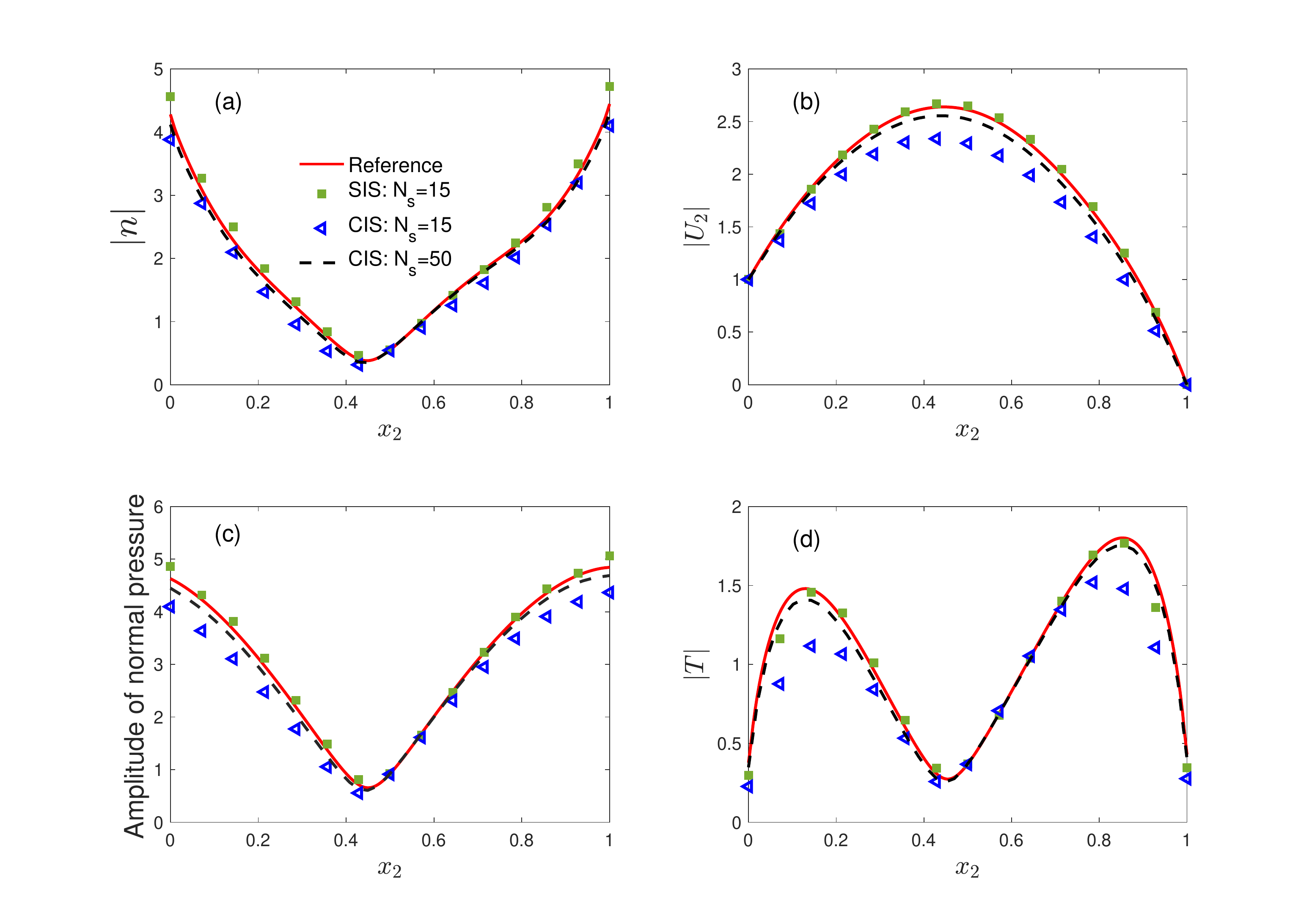}
	\caption{Profiles of macroscopic quantities in the sound propagation problem with different spatial discretizations, when $\delta_{rp}=50$ and $\text{St}=2.5$. The reference solution is obtained from the GSIS, where the spatial domain is discretized by $N_s=200$ uniform grids; the corresponding CIS results overlap with these lines and are not shown here. The normal pressure is defined as $P_{22}=2\int{v_2^2}h\mathrm{d}\bm{v}$. }
	\label{fig:Sound_velocity}
\end{figure}

Typical numerical results are shown in Fig.~\ref{fig:sound} when the spatial region $x_2\in[0,1]$ is discretized by 200 uniformly-distributed points, while the velocity grids are the same as that used in Sec.~\ref{Oscillatory_Couette}. For the CIS, it is very hard to find the converged solution when the Strouhal number $\text{St}$ is small, where the iteration number scales roughly as $\text{St}^{-1.5}$. However, this problem does not exist in the GSIS, as the Strouhal number has little effect on the number of iterations. The effect of spatial resolution on the fidelity of the solution is demonstrated in Fig.~\ref{fig:Sound_velocity} when $\text{St}=2.5$, where the sound waves between two plates resonance. It is seen that the GSIS needs less spatial grids than CIS. Again, this example proves the accuracy and efficiency of the GSIS.

\subsection{Two-dimensional oscillatory Couette flow}

\begin{figure}[pt]
	\centering
	\includegraphics[scale=0.3,viewport=50 -10 560 500,clip=true]{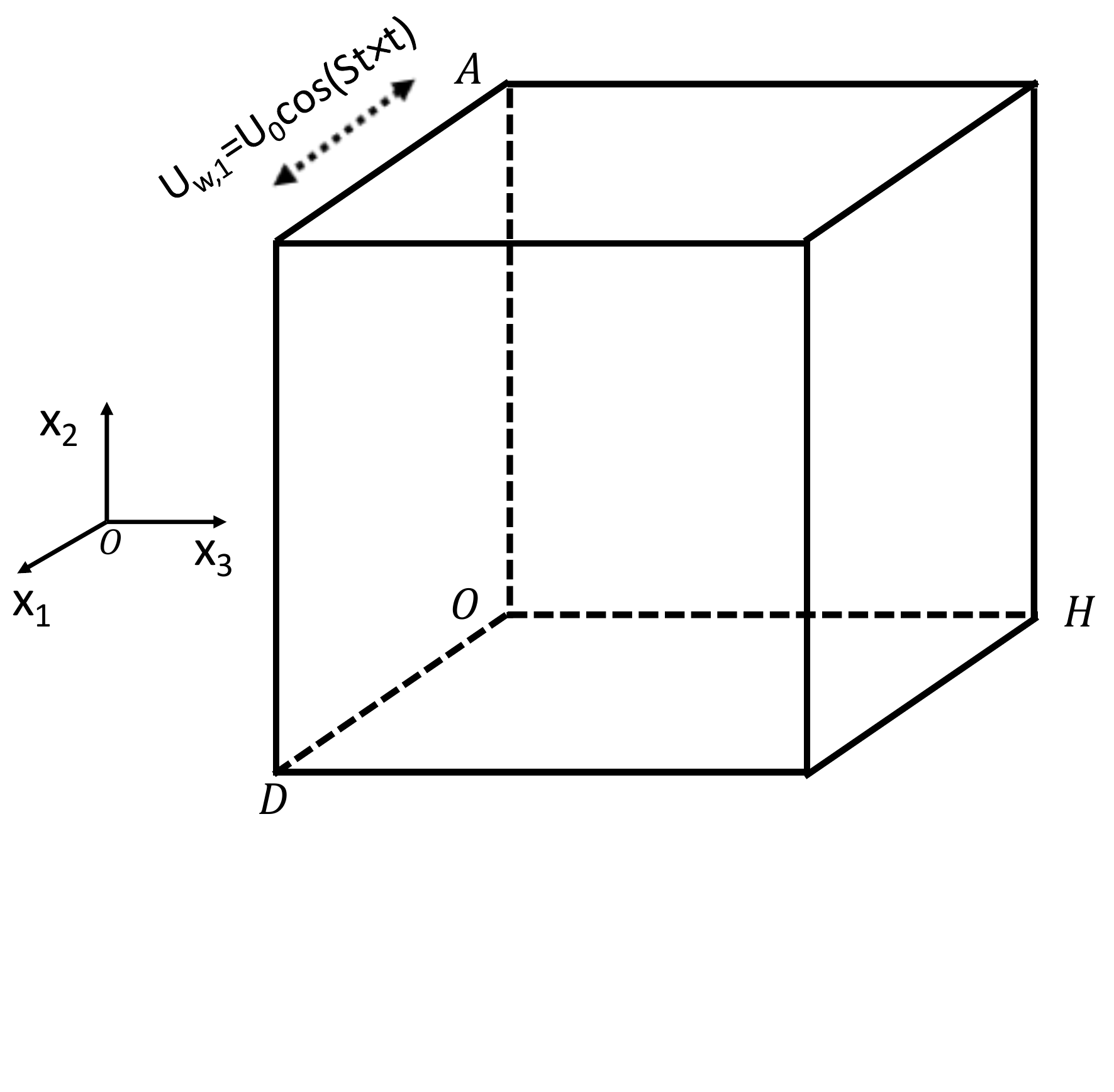}
	\includegraphics[scale=0.45,viewport=30 10 550 400,clip=true]{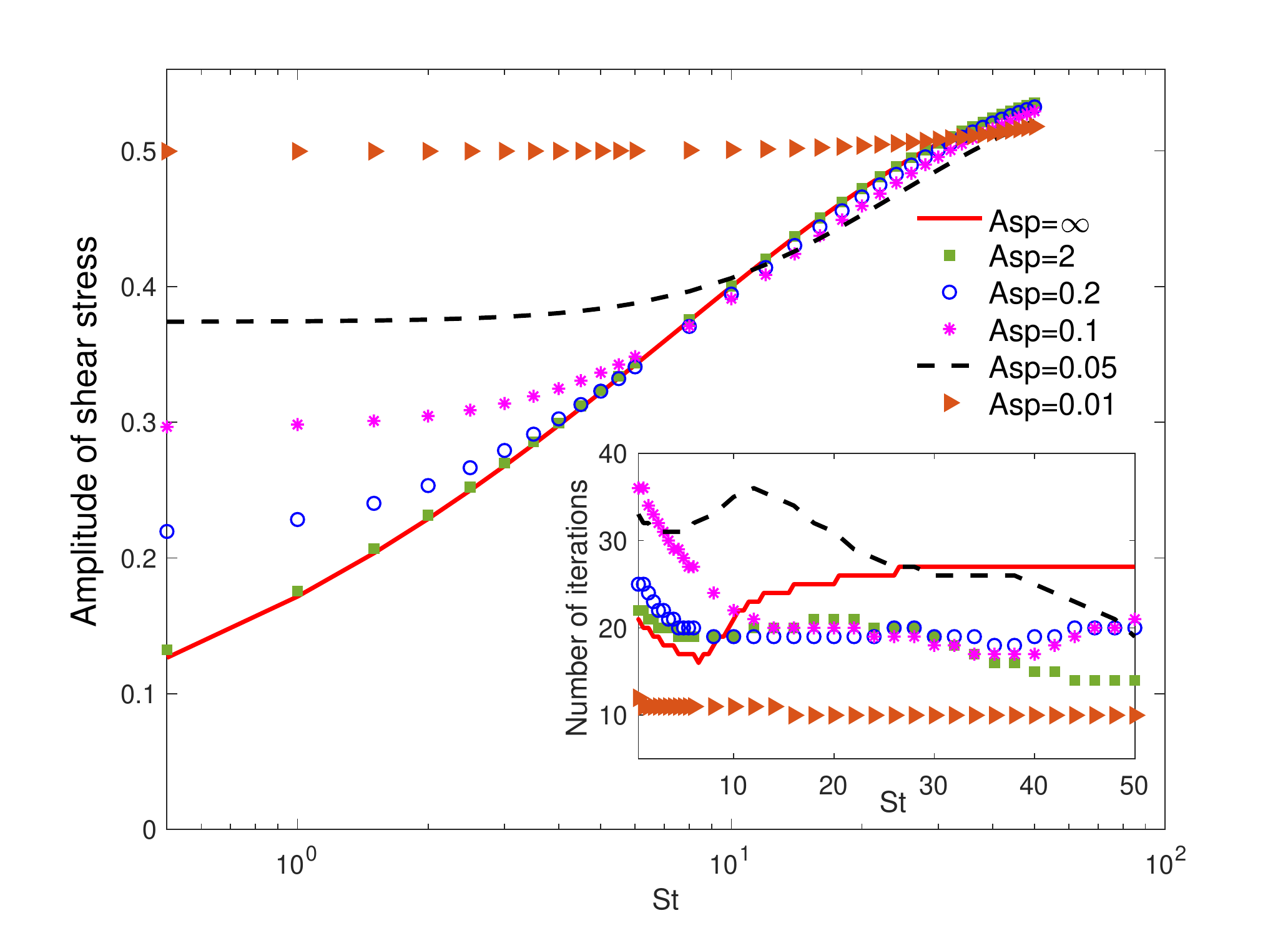}\\
	\includegraphics[scale=0.48,viewport=50 150 770 445,clip=true]{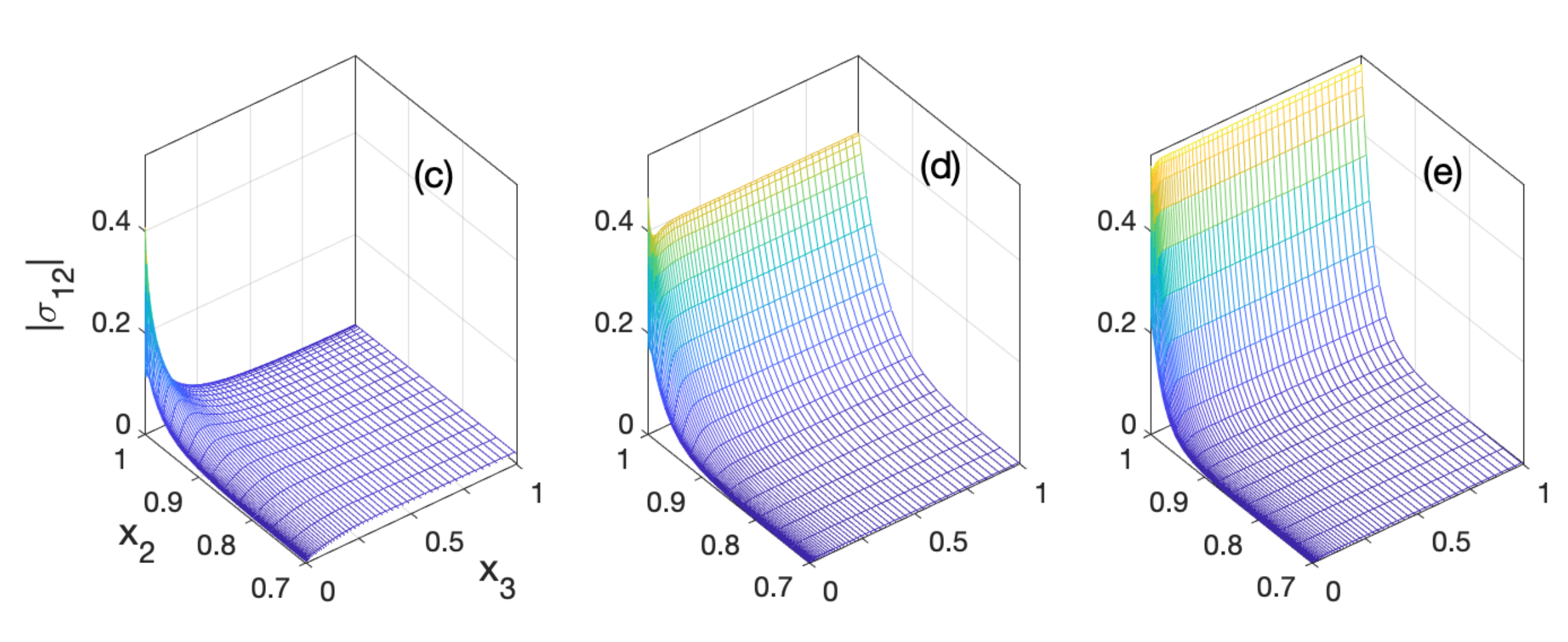}
	\includegraphics[scale=0.48,viewport=50 150 770 445,clip=true]{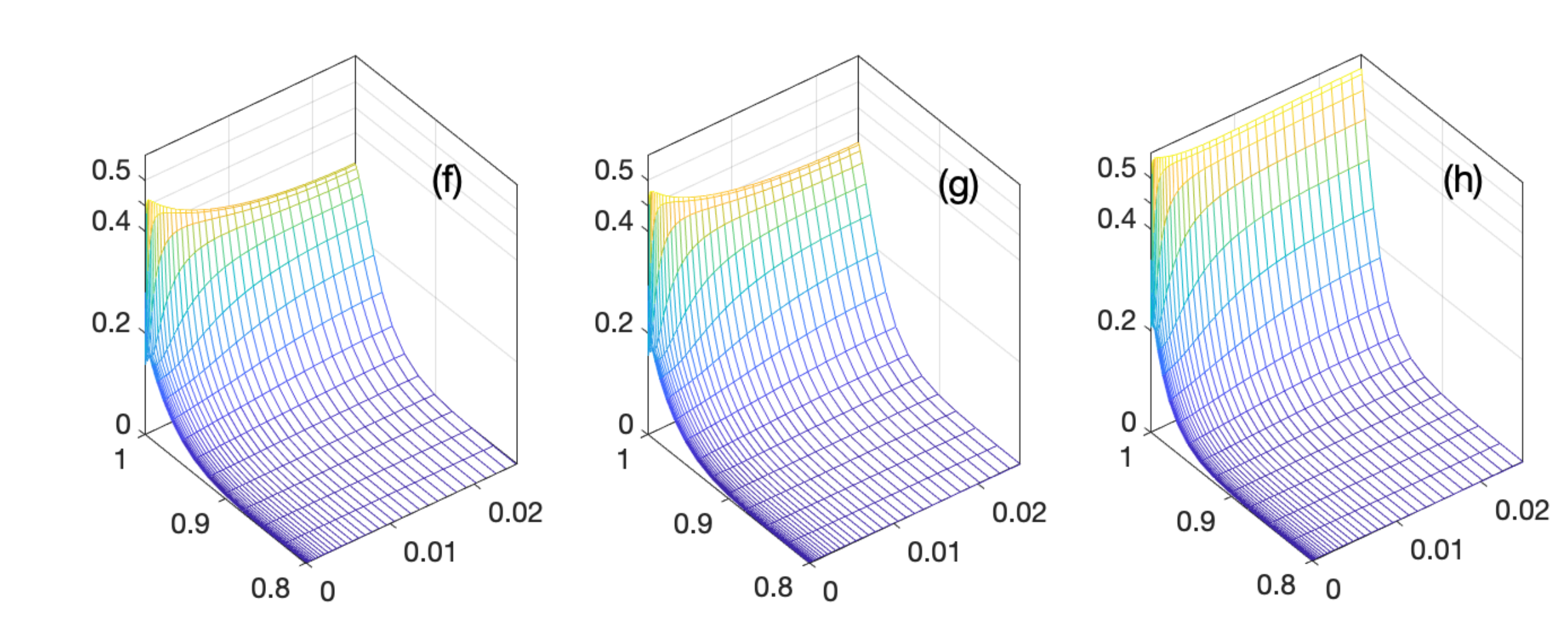}

	\caption{ (a) Schematic of the oscillatory flow in a 3D rectangular cavity, where `O' is the origin of the coordinate. The top lid oscillates in the $x_1$ direction periodically. (b)
	The amplitude of shear force exerting on the oscillating lid that is normalized by the aspect ration $\text{Asp}=OH/OA$, see Eq.~\eqref{average_force}; Inset shows the iteration number when the relative error in $U_1$ between two consecutive iteration is less than $10^{-5}$. (c, d, e) The distribution of shear stress when $\text{St}=0$, 10, and 50, respectively, and $\text{Asp}=2$. (f, g, h) Same as (c, d, e), respectively, but with $\text{Asp}=0.05$. The linearized Shakhov model is used  with $\delta_{rp}=50$ in all cases. }
	\label{fig:couette2d}
\end{figure}

Finally we consider the oscillatory flow in a three-dimensional cavity shown in Fig.~\ref{fig:couette2d}(a). We assume the side length OD is much larger than OH and OA, so that the problem is quasi two-dimensional. The characteristic length $H$ is chosen as the side length OA, and the aspect ratio is defined as $\text{Asp}=OH/OA$. If $\text{Asp}=\infty$, the problem is just the oscillatory Couette flow between two parallel plates studied in Sec.~\ref{Oscillatory_Couette}. This problem is interesting because it displays a counter-intuitive phenomenon that the shear force exerting on the oscillating lid in two-dimensional cavity could be even smaller than that of the one-dimensional Couette flow~\cite{Wu2014JFM}. The full three-dimensional oscillatory flow was studied in Ref.~\cite{Wang2018PoF}, but not all the parameter region are covered, for example, the case with OA much larger than OA and OH.

The synthetic equations~\eqref{eq123}, \eqref{HoT_sigma}, and~\eqref{HoT_q} can be simplified to 
\begin{equation}\label{Couette2D}
\begin{aligned}[b]
2i\text{St}U_1+\frac{\partial\sigma_{12}}{\partial {x_2}}+\frac{\partial\sigma_{13}}{\partial {x_3}}=0, \\
i\text{St}\sigma_{12}+\text{HoT}_{\sigma_{12}}+\frac{\partial U_1}{\partial {x_2}}=-\delta_{rp}\sigma_{12}+2\int{(L-L_s)}v_1v_2\mathrm{d}\bm{v},\\
i\text{St}\sigma_{13}+\text{HoT}_{\sigma_{13}}+\frac{\partial U_1}{\partial {x_3}}=-\delta_{rp}\sigma_{13}+2\int{(L-L_s)}v_1v_3\mathrm{d}\bm{v},
\end{aligned}
\end{equation}
which leads to the following diffusion-type equation for the flow velocity $U_1$ that is solved in an stable iterative manner:
\begin{equation}\label{diffusion_Couette2D}
\begin{aligned}[b]
2i\text{St}(i\text{St}+\bar{\delta}_{rp})U_1^{(k+1)}-\left(\frac{\partial^2}{\partial {x_2}^2}+\frac{\partial^2}{\partial{x_3}^2}\right)U_{1}^{(k+1)}=\text{Source}+2i\text{St}(\bar{\delta}_{rp}-\delta_{rp})U_1^{(k+1/2)},
\end{aligned}
\end{equation}
where $\bar{\delta}_{rp}$ is given in Eq.~\eqref{delta_bar}, and 
\begin{equation}
\begin{aligned}[b]
\text{Source}=&\frac{\partial}{\partial{x_2}}\left[2\int{(L^{(k)}-L_s^{(k)})}v_1v_2\mathrm{d}\bm{v}-\text{HoT}^{(k+1/2)}_{\sigma_{12}}\right]\\
&+
\frac{\partial}{\partial{x_3}}\left[2\int{(L^{(k)}-L_s^{(k)})}v_1v_3\mathrm{d}\bm{v}-\text{HoT}^{(k+1/2)}_{\sigma_{13}}\right].
\end{aligned}
\end{equation}

In numerical simulations, the molecular velocities $v_2$ and $v_3$ are discretized non-uniformly according to Eq.~\eqref{nonuniform_v} with 48 points in each direction, while $v_1$ is truncated into the region of $[-6,6]$ and discretized by 24 uniformly-distributed points. Due to the symmetry $h(x_2,x_3,v_1,v_2,v_3)=h(x_2,\text{Asp}-x_3,v_1,v_2,-v3)$, we only consider the domain where $0\le{}x_2\le1$ and $0\le{}x_3\le\text{Asp}/2$, which are discretized by 50 and 60 points according to Eqs.~\eqref{Couette_spatial_grid} and~\eqref{space_discrete}, respectively. The velocity distribution function entering the domain from the stationary walls is zero, while that from the oscillating wall is $2v_1f_{eq}$. The diffusion-type equation~\eqref{diffusion_Couette2D} is approximated by the central finite difference with 5 stencils, which are solved exactly by rewriting it in the matrix form. From the inset of Fig.~\ref{fig:couette2d}(b) we see that the GSIS is very efficient as converged solutions are obtained within 40 iterations.

We are interested in how the average shear force exerting on the oscillating lid change with the normalized oscillation frequency $\text{St}$. Hence $\sigma_{13}$ is not considered here as it is anti-symmetric along the line $x_3=\text{Asp}/2$ so its overall contribution to the friction is zero. The amplitude of the average shear force on the oscillating lid is defined as
\begin{equation}\label{average_force}
\bar{\sigma}_{12}=\frac{2|\int_0^{\text{Asp}/2}\sigma_{12}(x_2=1)\mathrm{d}x_3|}{\text{Asp}},
\end{equation}  
which is shown in Fig.~\ref{fig:couette2d}(b) for different aspect ratios of the cavity over a wide range of the oscillation frequency, when $\delta_{rp}=50$. It can be seen when $\text{Asp}=2$, the average shear force is the same as that of $\text{Asp}=\infty$, except that it is slightly larger when $\text{St}$ is small. This is seen more clearly in Fig.~\ref{fig:couette2d}(c) that the two lateral walls, i.e. the left and right walls in Fig.~\ref{fig:couette2d}(a), increases the shear stress from a nearly small constant to a high rise near the left top corner. When $\text{St}$ increases, the shear stress quickly decay from the oscillating lid to the zero value at the bottom surface, and its value at the oscillating lid is nearly uniform, see Fig.~\ref{fig:couette2d}(d) and (e). As the aspect ratio of the cavity reduces, the average shear force increases when $\text{St}$ is small, see Fig.~\ref{fig:couette2d}(f) for an example; this is easy to understood as the lateral walls increase the total friction according to our daily life experience. However, from Fig.~\ref{fig:couette2d}(f) and (g) we can see that the shear stress quickly saturates, as the increase of oscillation frequency only slightly increases the shear stress at the lid, such that the average shear stress on the lid remains nearly constant over a wide range of $\text{St}$; and the smaller the aspect ratio is, the wider this region is. This may be useful to design a micro-electro-mechanical system where the shear force remains constant in a certain wide range of oscillation frequency. It is  this efficient algorithm we are able to find this new phenomenon which are missed in Ref.~\cite{Wang2018PoF}. Another counter-intuitive thing is that, when $\text{St}$ is large, the average shear force at small values of cavity aspect ratio is slightly smaller than that of the one-dimensional cavity, although the relative difference is within 5\%.

\section{Conclusions and outlooks}\label{sec:summary}

In summary, we have developed a general synthetic iteration scheme to find the steady-state solution of the linearized Boltzmann equation efficiently and accurately. Various numerical results have demonstrated that our scheme is able to find the converged solution within about 20 iterations at any Knudsen number, due to the fact that the synthetic macroscopic equations not only asymptotically preserves the Navier-Stokes limit in the framework of Chapman-Enskog expansion, but also explicitly contains the constitutive laws for the shear stress and heat flux at the first order approximation in the the Knudsen number to the linearized Boltzmann equation. As a consequence, accurate solutions that are not contaminated by large numerical dissipation and accumulated error, can be obtained when the cell size is much larger than the mean free path of gas molecules. Moreover, the numerical error in the synthetic iteration scheme decays very fast and the convergence criterion can be set at a much smaller value than the conventional iteration scheme. These factors enables our synthetic iteration scheme to find the steady-state solution in 10-ish iterations.

This paper provides a framework to solve the general rarefied gas flow problems. The advantages and future works are highlighted below:
\begin{enumerate}
	
	\item Compared to implicit UGKS~\cite{Zhu2019JCP} and it variants~\cite{yang2018PoF,yang2018PRE}, we conclude that in order to develop efficient multiscale numerical schemes, macroscopic equations must be solved together with the Boltzmann or kinetic model equations. While in Refs.~\cite{Zhu2019JCP,yang2018PoF,yang2018PRE} only five equations from the conservation law are used so that complex flux evaluation across the cell interface must be adopted to asymptotically preserve the Navier-Stokes limit, our scheme needs no complex flux evaluation as the Navier-Stokes equations are recovered explicitly. Thus, the numerical implementation is much easy and the convergence to steady-state solution is much faster. More importantly, our scheme does not depend on the specific form of the collision operator, while that in Refs.~\cite{Zhu2019JCP,yang2018PoF,yang2018PRE} relies only on the BGK-type kinetic equations to enable exact evaluation of numerical flux.

	\item  Since the limitation on the cell size is removed and fast convergence is enabled, the present synthetic iteration scheme can be directly applied to low-variance~\cite{Radtke2009PRE,Radtke2011} and even frequency-domain~\cite{Ladiges2015JCP} DSMC that solves the linearized Boltzmann/kinetic model equations to improve the computational efficiency, especially in the near-continuum flow regime. 
	%The basic idea is first seen in Ref.~\cite{Degond2011} with the purpose of reducing statistical error in transient flows, but it will be more efficient when the steady-state solution is sought each time we solve the time-independent macroscopic equations.

	\item The present work can be extended to multi-species and compressible flow easily. The key is to construct macroscopic equations which recovers the compressible Navier-Stokes equation to the first order of Knudsen number. As a matter of fact, the Grad 13 moment equations~\cite{Grad1949,henning} can be directly used if the high-order velocity moments are calculated from the numerical solution of the Boltzmann equation, rather than closed by making assumption on the form of velocity distribution function. Actually the authors have implemented the synthetic iteration scheme for nonlinear Fourier heat transfer, and started from the global equilibrium distribution converged solution at arbitrary Knudsen number is found within 20 iterations.

	\item 
	It is noted that recently the gas-kinetic wave-particle (UGKWP) method, which uses the essential idea of UGKS that the streaming and collision should be treated spontaneously, has been applied in the framework of DSMC to remove the constraint on the cell size when the Knudsen number is small~\cite{Liu2018arXiv,zhu2019arXiv}; the BGK kinetic model is solved and the complex and time-consuming particle sorting is used to enable the asymptotically preserving property. We believe that the synthetic iteration scheme can also be applied to DSMC to remove the limitation on cell size and boost convergence, and the advantage is clear: it relies on no specific collision operator so that can be extended naturally to multi-species flows and even flows involving chemical reactions.

%	\item
%	\lei{Ease/remove the limitation on the cell size in DSMC and boost convergence, as indicated in Fig.~\ref{fig:Fourier_spatial_error}; using the same coupling idea in UGKWP but with no partial sorting; more efficient; relies on no assumption on the detailed collision model so can be extended to chemical reaction naturally; reform the openFOAM in collaboration with Craig White? }
	
\end{enumerate}

With these new development implemented, it is foreseen that in the near future that the problem of numerical simulation of multiscale rarefied gas flows will be solved completely. Also, the same idea can be applied to other kinetic equations such as the Enskog equation for dense gases dynamics with applications to shale gas extraction and non-equilibrium evaporation and condensation~\cite{Lei2015Enskog,Wu2016JFM,Frezzotti2005}.

\appendix

\section*{Appendix}
\renewcommand{\theequation}{A.\arabic{equation}}

Here, some details to solve the synthetic macroscopic equations using the high-order hybridizable discontinuous Galerkin (HDG) method~\cite{Cockburn2010} on arbitrary triangular mesh are presented. The steady-state governing equations can be written in the following mixed form as a system of first-order equations

\begin{equation}
\begin{aligned}
\nabla\cdot\left[\bm{\mathcal{G}}_\text{c}+\bm{\mathcal{G}}_\text{d}\right]=0,\\
\bm{L}-\nabla\bm u-\bm{\Pi}=0,\\
\bm{E}-\nabla T-\bm{\Theta}=0,
\end{aligned}
\end{equation}
where
\begin{equation}
\begin{aligned}
\bm{\mathcal{G}_\text{c}}=\left[\begin{array}{c}
\bm U\\p\bm I\\\bm 0
\end{array}
\right],\quad
\bm{\mathcal{G}_\text{d}}=\left[\begin{array}{c}
\bm 0\\-\frac{1}{\delta_{rp}}\left(\bm L+\bm L^{\text{T}}-\frac{2}{3}\mathrm{tr}\left(\bm L\right)\bm I\right)\\
-\frac{5}{4\delta_{rp}\mathrm{Pr}}\bm E
\end{array}
\right],\\
\bm{\Pi}=\left[\begin{array}{cc}
\text{HoT}_{\sigma_{11}}+\frac{1}{2}\text{HoT}_{\sigma_{22}} & \frac{1}{2}\text{HoT}_{\sigma_{12}}\\
\frac{1}{2}\text{HoT}_{\sigma_{12}} & \frac{1}{2}\text{HoT}_{\sigma_{11}}+\text{HoT}_{\sigma_{22}}
\end{array}
\right],\quad\bm{\Theta}=\left[\begin{array}{c}
\frac{4}{5}\text{HoT}_{q_{1}} \\
\frac{4}{5}\text{HoT}_{q_{2}} 
\end{array}
\right]
\end{aligned}
\end{equation}
with $\bm{I}$ being the identity matrix. The auxiliary variables $\bm{L}$ and $\bm{E}$ are introduced to approximate the combination of the velocity gradient $\nabla\bm{U}$, temperature gradient $\nabla T$ and the high-order moments. Then, the stress tensor and heat flux are evaluated as
\begin{equation}\label{sq}
\sigma_{ij}=-\frac{1}{\delta_{rp}}\left(L_{ij}+L_{ji}-\frac{2}{3}L_{kk}\delta_{ij}\right),\quad q_i=-\frac{5}{4\delta_{rp}\mathrm{Pr}}E_i
\end{equation}

Let $\Delta\in\mathbb{R}^2$ be an two-dimensional domain with boundary $\partial\Delta$ in the $x_1-x_2$ plane. Then, $\Delta$ is partitioned in $M$ disjoint regular triangles $\Delta_i$: $\Delta=\cup^{M}_i\Delta_i$. The boundaries $\partial\Delta_i$ of the triangles define a group of $N$ faces $\Gamma_c$: $\Gamma=\cup^{M}_i\{\partial\Delta_i\}=\cup^{N}_c\{\Gamma_c\}$. For HDG discretization, two types of discontinuous finite element approximation space, one for solutions within $\Delta_i$ and the other for traces of solution on $\Gamma_c$, are defined as

\begin{equation}
\begin{aligned}[b]
\mathcal{V}=\{\varphi:\ \varphi|_{\Delta_i}\in\mathcal{P}^k(\Delta_i),\ \forall\ \Delta_i\subset\Delta\},\\
\mathcal{W}=\{\psi:\ \psi|_{\Gamma_c}\in\mathcal{P}^k(\Gamma_c),\ \forall\ \Gamma_c\subset\Gamma\},
\end{aligned}
\end{equation}
where $\mathcal{P}^k(D)$ denotes the space of $k-$th order polynomials on a domain $D$.

The HDG method solves the system in two steps. First, a global problem is set up to determine the traces of the flow properties $\hat{\bm{Q}}=\left[\hat{p},\hat{\bm{U}},\hat{T}\right]$ on the faces $\Gamma$. Then, a local problem with $\hat{\bm{Q}}$ as the boundary condition on $\partial\Delta_i$ is solved element-by-element to obtain the solutions for the flow properties $\bm{Q}=\left[p,\bm{Q},T\right]$, as well as the ones for the auxiliary variables $\bm{L}$ and $\bm{E}$. Generally speaking, when moving from the interior of the triangle element $\Delta_i$ to its boundary $\partial\Delta_i$, the traces defines what the values of field variables on the boundary should be. In the HDG method, it is assumed that the traces are singled-valued on each face. 

We introduce the notations $\left(a,b\right)_D=\int_{D\in\mathbb{R}^2}\left(a\odot b\right)\mathrm{d}x_1\mathrm{d}x_2$ and $\langle a,b\rangle_D=\int_{D\in\mathbb{R}^1}\left(a\odot b\right)\mathrm{d}\Gamma$, where $\odot$ can be either the dot product $\cdot$ or tensor product $\otimes$. The local problem is stated as: find $\left(\bm{Q},\bm{L},\bm{E}\right)\in\left[\mathcal{V}\right]^4\times\left[\mathcal{V}\right]^4\times\left[\mathcal{V}\right]^2$ such that

\begin{equation}\label{local}
\begin{aligned}
-\left(\bm{\mathcal{G}_\text{c}}+\bm{\mathcal{G}_\text{d}},\nabla\bm{r}\right)_{\Delta_i}+\langle\hat{\bm{\mathcal{F}}}\cdot\bm{n},\bm{r}\rangle_{\partial\Delta_i}=0,\\
\left(\bm{L},\bm{w}\right)_{\Delta_i}+\left(\bm{U},\nabla\cdot\bm{w}\right)_{\Delta_i}-\langle\hat{\bm{U}},\bm{w}\cdot\bm{n}\rangle_{\partial\Delta_i}=\left(\bm{w},\bm{\Pi}\right)_{\Delta_i}\\
\left(\bm{E},\bm{z}\right)_{\Delta_i}+\left(T,\nabla\cdot\bm{z}\right)_{\Delta_i}-\langle\hat{T},\bm{z}\cdot\bm{n}\rangle_{\partial\Delta_i}=\left(\bm{z},\bm{\Theta}\right)_{\Delta_i}
\end{aligned}
\end{equation}  
for all $\left(\bm{r},\bm{w},\bm{z}\right)\in\left[\mathcal{V}\right]^4\times\left[\mathcal{V}\right]^4\times\left[\mathcal{V}\right]^2$. The numerical flux $\hat{\bm{\mathcal{F}}}\cdot\bm{n}$ is defined as~\cite{Peraire2010}

\begin{equation}
\hat{\bm{\mathcal{F}}}\cdot\bm{n}=\left[\begin{array}{c}
\bm{U}\\
\hat{p}\bm{I}-\frac{1}{\delta_{rp}}\left(\bm{L}+\bm{L}^{\mathrm{T}}-\frac{2}{3}\mathrm{tr}\left(\bm L\right)\bm{I}\right)\\
-\frac{5}{4\delta_{rp}\mathrm{Pr}}\bm{E}

\end{array}\right]\cdot\bm{n}+\left[\begin{array}{ccc}
\tau & &\\
& \frac{\tau}{\delta_{rp}} &\\
& & \frac{5\tau}{4\delta_{rp}\mathrm{Pr}}
\end{array}
\right]\left[\begin{array}{c}
p-\hat{p}\\
\bm{U}-\hat{\bm U}\\
T-\hat{T}
\end{array}\right].
\end{equation}
Here $\bm{n}$ being the outward unit normal vector of $\partial\Delta_i$. $\tau$ is the stabilization parameter that have important effects on the accuracy and convergence of the HDG method. In this work, we chosen $\tau=1/H_\text{min}$, with $H_\text{min}$ the minimum height of the triangles $\Delta_i$. 

The global problem is set up by enforcing the continuity of the numerical flux over all the interior faces. It is stated as: find $\hat{\bm{Q}}\in\left[\mathcal{W}\right]^4$ such that
\begin{equation}\label{Global1}
\langle\bm{\left(\hat{\mathcal{F}}}\cdot\bm{n}\right)^+,\bm{\psi}\rangle_{\Gamma_c}+\langle\bm{\left(\hat{\mathcal{F}}}\cdot\bm{n}\right)^-,\bm{\psi}\rangle_{\Gamma_c}=0,\quad\text{on}\ \Gamma_c\in\Gamma\backslash\partial\Delta,
\end{equation}
for all $\bm{\psi}\in\left[\mathcal{W}\right]^4$. Here the superscripts $\pm$ denote the numerical fluxes obtained from the triangles on both sides of the face. Note that the traces on boundary faces are calculated as
\begin{equation}\label{Global2}
\langle\hat{\bm Q}-\bm{Q}^+_{VDF},\bm{\psi}\rangle_{\Gamma_c},\quad\text{on}\ \Gamma_c\in\Gamma\cap\partial\Delta,
\end{equation}
where $\bm{Q}^+_{VDF}$ is the field solutions directly calculated from the approximated velocity distribution function (see Eq.~\eqref{MP}) within the triangle where the boundary face $\Gamma_c$ belongs to. 

By assembling the local problem~\eqref{local} and global problem~\eqref{Global1} and~\eqref{Global2} over all the triangles and faces, we can obtain a matrix system 

\begin{equation}
\left[\begin{array}{cccc}
A_Q & A_L & A_E & A_{\hat{Q}}\\
B_Q & B_L & B_E & B_{\hat{Q}}\\
C_Q & C_L & C_E & C_{\hat{Q}}\\
D_Q & D_L & D_E & D_{\hat{Q}}\\
\end{array}
\right]\left[\begin{array}{c}
\mathbb{Q}\\
\mathbb{L}\\
\mathbb{E}\\
\hat{\mathbb{Q}}
\end{array}
\right]=\left[\begin{array}{c}
S_{Q}\\
S_{L}\\
S_{E}\\
S_{\hat{Q}}
\end{array}
\right]
\end{equation}
where $\mathbb{Q}$, $\mathbb{L}$, $\mathbb{E}$ and $\hat{\mathbb{Q}}$ are the vectors of degrees of freedom of the flow properties $\bm{Q}$, the auxiliary variables $\bm{L}$ and $\bm{E}$, and the trace of the flow properties $\hat{\bm{Q}}$, respectively. Note that the degrees of freedom for $\bm{Q}$, $\bm{L}$ and $\bm{E}$ are grouped together and ordered element-by-element, and the corresponding coefficient matrix $\left[A_Q,A_L,A_E;B_Q,B_L,B_E;C_Q,C_L,C_E\right]$ has block-diagonal structure. Therefore, we can eliminate $\bm{Q}$, $\bm{L}$ and $\bm{E}$ to obtained a reduced linear system involving only $\hat{\mathbb{Q}}$. Once $\hat{\mathbb{Q}}$ is determined, $\bm{Q}$, $\bm{L}$ and $\bm{E}$ are reconstructed corresponding to the local problem~\eqref{local} in an element-wise fashion, while the stress tensor and heat flux are calculated as Eq.~\eqref{sq}.

\section*{Acknowledgments}

This work is supported in the UK by the Engineering and Physical Sciences Research Council under grant EP/R041938/1. L.~Zhu acknowledges the financial support of European Union’s Horizon 2020 Research and Innovation Programme under the Marie Skłodowska-Curie grant agreement number 793007.

\section*{References}
\bibliography{bibnew}
\end{document}